\documentclass[10pt]{iopart}

\DeclareMathAlphabet{\mathpzc}{OT1}{pzc}{m}{it}
\usepackage{graphicx}
\usepackage{cancel}
\usepackage{cite}  
\usepackage{color}
\usepackage{epstopdf}
\usepackage{hyperref}
\usepackage{soul} 
\usepackage{xifthen}
\usepackage{ulem} 

\expandafter\let\csname equation*\endcsname\relax
\expandafter\let\csname endequation*\endcsname\relax
\usepackage{amsmath}
\usepackage{amssymb}

\newcommand{\Eqref}[1]{\mbox{equation\hspace{0.25em}\eqref{#1}}}
\newcommand{\Eqsref}[1]{\mbox{equations\hspace{0.25em}\eqref{#1}}}
\newcommand{\figref}[1]{\mbox{figure\hspace{0.25em}\ref{#1}}}
\newcommand{\Figref}[1]{\mbox{Figure\hspace{0.25em}\ref{#1}}}

\newcommand{\secref}[1]{\mbox{section\hspace{0.25em}\ref{#1}}}

\newcommand{\refcite}[1]{\mbox{Ref.\hspace{0.25em}\cite{#1}}}

\definecolor{gray}{RGB}{200,200,200}

\newcommand{\abs}[1]{|#1|}

\newcommand{\vect}{\boldsymbol}

\newcommand{\diff}{\text{d}}
\newcommand{\difffrac}[2]{\frac{\diff #1}{\diff #2}}

\newcommand{\set}[1]{\{{#1}\}}
\newcommand{\takenat}[2]{\left.{#1}\right|_{#2}}
\newcommand{\mean}[1]{\langle #1 \rangle}

\newcommand{\cinf}{{c}_\infty}  
\newcommand{\ceqbase}{c^{(0)}_{\rm out}}
\newcommand{\ceqbaseIn}{c^{(0)}_{\rm in}}

\newcommand{\supSat}{\varepsilon} 
\newcommand{\supSatEq}{\supSat_{\rm eq}} 

\newcommand{\kb}{k_{\rm B}} 
\newcommand{\width}{w} 
\newcommand{\cbase}{c^{(0)}} 
\newcommand{\ceq}{c^{\rm eq}}  
\newcommand{\xd}{x_{\rm c}} 
\newcommand{\Vsys}{V} 
\newcommand{\Vd}{V_\mathrm{d}}  
\newcommand{\Rstab}{R_{\rm stab}}  
\newcommand{\steady}[1]{#1_*}  

\newcommand{\A}{c_{ A}} 
\newcommand{\B}{c_{ B}}
\newcommand{\R}{c_{ R}}

\newcommand{\Rl}{c_{ R}^{-}} 
\newcommand{\Rr}{c_{ R}^{+}} 
\newcommand{\Al}{c_{ A}^{-}} 
\newcommand{\Ar}{c_{ A}^{+}} 

\newcommand{\Fr}{F^{+}} 
\newcommand{\Fl}{F^{-}}

\newcommand{\Rhom}{\bar {c}_R }  
\newcommand{\Ahom}{\bar {c}_A }  

\newcommand{\al}{\alpha} 
\newcommand{\be}{\beta} 

\newcommand{\fe}{f}
\newcommand{\fetot}{f_\mathrm{tot}}

\newcommand{\sLoc}{s}
\newcommand{\sTot}{s_\mathrm{tot}}
\newcommand{\sBase}{\Gamma}
\newcommand{\s}[1][]{\ifthenelse{\isempty{#1}}{s}{s^{(#1)}}}
\newcommand{\sF}[1][]{\ifthenelse{\isempty{#1}}{s_{\rightarrow}}{s^{(#1)}_{\rightarrow}}}
\newcommand{\sB}[1][]{\ifthenelse{\isempty{#1}}{s_{\leftarrow}}{s^{(#1)}_{\leftarrow}}}
\newcommand{\cBase}{\cbase}

\newcommand{\cOut}{c}
\newcommand{\cc}{c_{\rm c}}

\newcommand{\cEqIn}{\ceq_{\rm in}}
\newcommand{\cEqOut}{\ceq_{\rm out}}
\newcommand{\cBaseIn}{\cbase_{\rm in}}
\newcommand{\cBaseOut}{\cbase_{\rm out}}
\newcommand{\cInfty}{c_\infty}
\newcommand{\DIn}{D_{\rm in}}
\newcommand{\DOut}{D_{\rm out}}
\newcommand{\kIn}{k_{\rm in}}
\newcommand{\kOut}{k_{\rm out}}
\newcommand{\sBaseIn}{\sBase_{\rm in}}
\newcommand{\sBaseOut}{\sBase_{\rm out}}
\newcommand{\sBaseInMax}{\sBaseIn^{\rm max}}
\newcommand{\FluxIn}{J_{\rm in}}
\newcommand{\FluxOut}{J_{\rm out}}
\newcommand{\sFluxIn}{S_{\rm in}}

\newcommand{\DiffLenIn}{\ell_{\rm in}}
\newcommand{\DiffLenOut}{\ell_{\rm out}}

\newcommand{\mob}[1][]{\ifthenelse{\isempty{#1}}{\Lambda}{\Lambda_{\rm r}^{(#1)}}}
\newcommand{\mobTot}{\mob[\mathrm{tot}]}
\newcommand{\mobR}{\mob_{\rm r}}
\newcommand{\cInt}{c_\mathrm{I}}
\newcommand{\Rcrit}{R_\mathrm{c}}
\newcommand{\RsA}{R^{(1)}}
\newcommand{\RsB}{R^{(2)}}
\newcommand{\etaM}{\eta_\mathrm{m}}
\newcommand{\etaR}{\eta_R}

\newcommand{\ii}{{\rm i}}

\newcommand{\bbr}{\vect{r}}

\newcommand{\bj}{\vect{j}}
\newcommand{\bk}{\vect{q}}

\newcommand{\sep}{ \ \ \ , \ \ \ }

\newcommand{\beq}{\begin{equation}}
\newcommand{\eeq}{\end{equation}}
\newcommand{\beqn}{\begin{eqnarray}}
\newcommand{\eeqn}{\end{eqnarray}}
\newcommand{\pp}{\partial}
\newcommand{\dd}{{\rm d}}

\newcommand{\eq}{equation~}

\newcommand{\eg}{e.g.,\ }
\newcommand{\ie}{i.e.,\ }

\newcommand{\la}{\langle}

\newcommand{\ra}{\rangle}

\numberwithin{equation}{section} 
\numberwithin{figure}{section} 

\begin{document}

\title[Physics of Active Emulsions]{Physics of Active Emulsions}

\author{Christoph A.\ Weber$^{1,2,3,*}$, David Zwicker$^{1,4,*}$, Frank J\"{u}licher$^{1,2}$ and Chiu Fan Lee$^{5}$}

\address{$^1$ Max Planck Institute for the Physics of Complex Systems,
N\"{o}thnitzer Str.~38, 01187 Dresden,
Germany} 

\address{$^2$ Center for Systems Biology Dresden, CSBD, Dresden, Germany}

\address{$^3$ Paulson School of Engineering and Applied Sciences, Harvard University, Cambridge, MA 02138, USA}

\address{$^4$ Max Planck Institute for Dynamics and Self-Organization, Am Fa{\ss}berg 17, 37077 G\"{o}ttingen, Germany}

\address{$^5$ Department of Bioengineering, Imperial College London, South Kensington Campus, London SW7 2AZ, U.K.}

\address{$^*$ equal contribution}

\ead{weber@pks.mpg.de, david.zwicker@ds.mpg.de, julicher@pks.mpg.de, c.lee@imperial.ac.uk}
\vspace{10pt}

\begin{indented}
\item[] Version of draft:  \today 
\end{indented}

\begin{abstract}
Phase separating systems that are maintained away from thermodynamic equilibrium 
via molecular processes represent a class of active systems, which we call \textit{active emulsions}.
These systems are driven by external energy input for example provided by an external fuel reservoir. 
The external energy input gives rise to novel phenomena that are not present in passive systems.
For instance, concentration gradients  can spatially organise emulsions and cause novel droplet  size distributions.
Another example are active droplets that are subject to chemical reactions such that their nucleation and size can be controlled and they can spontaneously divide. 
In this review we discuss the physics of phase separation and emulsions 
and show how the concepts that governs such phenomena  can be extended to capture the physics of active emulsions.  
This physics  is relevant to the spatial organisation of the biochemistry in living cells, for the development novel applications in chemical engineering and models for the origin of life.
\end{abstract}

\pacs{47.55.D-, 05.70.Ln, 47.10.ab, 82.40.Qt, 82.60.Hc, 83.80.Tc, 87.10.Ca, 87.15.R-, 87.16.Uv}
%
\vspace{1pc}
\noindent{\it Keywords}: 
active emulsions, 
active droplets,
liquid phase separation, 
droplet ripening in concentration gradients, 
positioning of droplets,
driven chemical reactions in emulsions, 
suppression of Ostwald-ripening,
division of droplets,
shape deformations and instabilities of droplets.

\ioptwocol


\section{Introduction: From passive to active emulsions}

The formation and dynamics of condensed phases such as droplets 
represent ubiquitous phenomena that we all experience in our daily life~\cite{syrbe1998polymer}. Examples are droplets condensing on the leaves of flowers and trees due to the supersaturated fog in the morning of some autumn day, or the ``ouzo effect'', where the oil droplets in Ouzo grow in size and cloud the liquid.
These transitions from a homogeneous mixture to a system with coexisting phases can be controlled by temperature or by changing the composition of the mixture.
The physical conditions under which a mixture phase separates are well-understood. 
The interactions that favour the accumulation of components of the same type must exceed the entropic tendency of the system to remain mixed~\cite{clerk1875dynamical,flory1942thermodynamics,huggins42}.
After drops have been nucleated, they undergo 
a ripening dynamics. 
Droplets either fuse, or 
 larger droplets grow at the expense of smaller ones which then disappear.
 The latter phenomenon is referred to as Ostwald ripening~\cite{ostwald1897studien}.
During the ripening dynamics the droplet size distribution continuously broadens.
In the case of Ostwald ripening 
 the droplet size distribution exhibits an universal scaling in time $t$ with the mean droplet size $\propto t^{1/3}$, which was derived by Lifschitz and Slyozov~\cite{Lifshitz_Slyozov_61,wagner61,Bray_Review_1994}. 
However, at large times-scales, droplet growth saturates and ripening stops as there is only one droplet left in the system.
This condensed droplet stably coexists with a surrounding minority phase of lower  solute concentration. 

The behaviour of droplets can change in more complex environments.
For instance, liquid condensed phases can interact with surfaces by   wetting~\cite{cahn1977critical,moldover1980interface,pohl1982wetting}.
Droplets embedded in a gel matrix interact such that the droplet size can be tuned by changing the mechanical properties of the gel~\cite{Dufresne2017arXiv170900500S}.
Droplets can also behave differently in transient systems, which have not yet reached equilibrium.
For instance, surfactants exchanging with the surrounding solvent can induce Marangoni flows, which can propel droplets~\cite{Herminghaus2014Interfacial,Izri2014Self-Propulsion,maass2016swimming,Seemann2016Self-Propelled,Jin2017Chemotaxis} and even lead to spontaneous division~\cite{Caschera2013Oil}.
Generally, more complex behavior can be expected when multiple phases of different composition come in contact and exchange material~\cite{sellier2011self,jehannin2015periodic,cira2015vapour,Tan2016,Li2018a}.

Liquid condensed phases are also influenced  by external ``forces'' such as gravitation,  concentration or temperature gradients, magnetic or electric fields~\cite{loewen_review, Lee_tempgrad_2004}. 
In industrial manufacturing and processing, temperature gradients and sedimentation are explicitly taken advantage of, e.g., for extracting  crude oil~\cite{kokal2005crude}.
Moreover, concentration~\cite{matsuyama1999formation, matsuyama2000structure} and temperature gradients~\cite{caneba1985polymera,caneba1985polymerb} are commonly  used to assemble synthetic membranes for electro-optical devices.
Recently,  the equations governing the ripening dynamics of droplets derived by Lifschitz and Slyozov have been extended to account for the presence of a concentration gradient~\cite{Lee_2013, NJP_Weber_ChiuFan_Juelicher}. It has  been theoretically  shown that  droplets  can be positioned along the concentration gradient and that the universal scaling in time of droplet ripening breaks down.

Liquid condensed phases are also ideal compartments to organise chemical reactions  since they can enrich specific chemical components. 
In particular, chemical agents mix rapidly in small droplets, 
 and thus  control the timing of chemical reactions~\cite{song2006reactions, Stroberg2018}.
 The interface of droplets can serve to concentrate reactants, resulting in a speed up of the reaction~\cite{fallah_prl14}.
Liquid droplets in the presence of chemical reactions 
can even propel across a solid substrate~\cite{john2005self,yabunaka2012self, seemann2016self,maass2016swimming} or
 flow past a chemically patterned substrate leading to unique morphological instabilities~\cite{Kuksenok2003}.
They also provide model systems to study interactions of pattern forming systems.
For example, droplets containing agents undergoing oscillatory Belousov-Zhabotinsky reactions have been considered as coupled phase oscillators~\cite{delgado2011coupled,vanag2011excitatory}.
Recently, it had been suggested that phase separated liquid-like compartments composted of oppositely charged molecules, called coacervates, could be ideal seeds for prebiotic life~\cite{vieregg2016polynucleotides}.
In particular, RNA catalysis is viable within these coacervate droplets and they even provide a mechanism for length selection of RNA~\cite{drobot2018compartmentalized}.
In all these systems, the droplet material does not participate in the chemical reaction.

If the phase separated material undergoes a chemical reaction itself, new physical behaviours can emerge. 
In passive systems, where phase separation and chemical reactions are in thermal equilibrium, coexisting phases cannot be stable. 
These systems settle in a homogeneous state that corresponds to the free energy minimum~\cite{krapivsky_b10}. 
Conversely, if the chemical reaction is driven away from equilibrium,  phase separation can be maintained. 
Under some conditions an arrest of the ripening dynamics has been observed in numerical simulations~\cite{Puri1994,Glotzer1994,Glotzer1994a, Glotzer1995, Glotzer1995a, Verdasca1995, Christensen1996, Toxvaerd1996, Motoyama1996, Motoyama1997, Huo2004, Travasso2005, Kuksenok2006, Li2010} and in experiments where chemical reactions are induced by light~\cite{Harada1996, Tran-Cong1996,Tran-Cong1997,Ohta1998,Tran-Cong1999,Tran-Cong1999a, Sutton2004, TranCongMiyata2017} or proceed spontaneously~\cite{Tanaka1992, Inoue1995, Williams1997, Chan1996, Chan1997, Girard1998, Kataoka1998, Locatelli2006}.
Recently, it has been shown that the chemical reaction between the droplet material can indeed suppress Oswald ripening leading to  an emulsion composed of drops of identical size~\cite{Zwicker2014, Zwicker2015, Wurtz2018}.
Driven chemical reactions can initiate droplet formation as a response to external stimuli~\cite{Wurtz2018a} and they can even trigger the division of droplets~\cite{Zwicker2017}.
Recent experiments on droplets in the presence of non-equilibrium turnover reactions showed the assembly of supramolecular structures with a tuneable lifetime~\cite{tena2017non} and serve as a model for molecular selection of reaction products and their assemblies~\cite{tena2018non}.
These examples highlight the rich phenomenology emerging from the interplay of phase separation with chemical reactions.

Phase separated systems in the presence of external forces and chemical reactions actively driven away from thermal equilibrium are a novel class of physical systems.
 Many physical properties known from  passive emulsions are altered and  novel phenomena occur.
Since energy input is typically necessary to 
generate external forces, and to drive chemical reactions away from equilibrium, we refer to these systems as \textit{active emulsions}.

An intriguing  example where the physics of these active emulsions is relevant are living cells. 
Biological function inside cells is attained by the spatial-temporal organisation of biomolecules and the control of their chemical reactions. 
For this purpose the interior of the cell is divided into compartments, referred to as organelles.
Each organelle has a chemical identity due to a distinct composition of functional biomolecules. 
Some organelles, such as mitochondria, are surrounded by membranes that are permeated by active channels regulating chemical potential differences across the membrane~\cite{ernster1981mitochondria}. 
However, there are also  organelles that do not posses any membrane; they are called non-membrane bound compartments or biomolecular condensates~\cite{brangwynne2011soft, Hyman2011, hyman2014liquid, brangwynne2015polymer,banani2017biomolecular,Cliffs_last_rev}. 
To maintain their chemical identity in the absence of a membrane, 
it has been suggested that these compartments are liquid droplets
formed by liquid-liquid phase separation~\cite{sear_softmatter07,Brangwynne_2009, Lee_2013}. 
Recently, many organelles have been found with properties reminiscent of liquid droplets~\cite{banani2017biomolecular, Woodruff2018, Boeynaems2018a}.
Examples include mRNA-protein-condensates~\cite{Brangwynne_2009,  saha2016polar}, RNA polymerase clusters~\cite{Boehning2018, Lu2018}, centrosomes~\cite{Zwicker2014, woodruff2017centrosome}, and multiple nuclear subcompartments~\cite{brangwynne2011active, feric2016coexisting}. 
These findings suggest that the cytoplasm can be regarded as a multi-component emulsion hosting a large variety  of  coexisting  phases, each of distinct composition~\cite{sear_prl03,jacobs_biophysj17,Ditlev2018,Mao2018}.
 In contrast to passive emulsions, cellular droplets exist in the non-equilibrium environment of living cells.
The associated continuous dissipation of energy can be used to drive chemical reactions or
generate concentration gradients of molecular species.
Both processes can affect the dynamics and stability of these active droplets and cause behaviours not observed for passive droplets~\cite{Cliffs_last_rev, Cates2018}.


Here we review recent theoretical approaches used to describe 
droplets and emulsions under conditions that deviate from 
passive phase separating systems.
Our review starts with an introduction to passive phase separation and dynamics of emulsion undergoing Ostwald ripening (section~\ref{sect:passive_phase_sep}).
In section~\ref{sect:phase_sep_in_gradients}, we discuss phase separation and dynamics of droplets in the presences of external forces and concentrations gradients, while section~\ref{sect:phase_sep_with_turn_over} considers the dynamics and stability of droplets under the influence of chemical reactions that are driven away from thermal equilibrium.


	\section{Liquid-liquid phase separation of binary mixtures}\label{sect:passive_phase_sep}


Phase separation refers to the spontaneous partitioning of a system into subsystems with distinct macroscopic properties. 	 
Examples include the cellular compartments mentioned in the introduction, but also many everyday phenomena that range from fog in the morning to oil droplet formation in salad dressings. 
In this section, we will discuss the physical principles and derive the equations describing the dynamics of phase separation and the ripening of droplets in liquid emulsions.

\subsection{Statistical mechanics of a binary mixture}
\label{sect:PS}\label{sect:bin_phasesep}\label{sec:stat_mech}

We start by considering a binary, incompressible mixture consisting of two types of molecules on a lattice with $M$ sites. 
Each lattice site is occupied by either molecule $A$ or  $B$, with $N_A$ and $N_B$ representing their total numbers in the system, so $N_A + N_B = M$.
The system is at thermal equilibrium with a heat bath at temperature $T$. The thermodynamic properties of the system are thus dictated by the partition function~\cite{balian2007microphysics1} 
\begin{equation}
	\label{eq:Z}
	Z = \sum_{\Omega} \exp\left(-\frac{H(\sigma_1, ..., \sigma_M)}{k_{\rm B}T}\right)
	\ ,
\end{equation}
where the Hamiltonian $H(\sigma_1, ..., \sigma_M)$ denotes the energy of a particular arrangement~$\sigma_1, ..., \sigma_M$ of the molecules on the lattice and $k_{\rm B}$ is the Boltzmann constant. 
Here, we encode the arrangements using a binary variable~$\sigma_n$, where $\sigma_n=1$ if the lattice site $n$ is occupied by molecule $A$ and $\sigma_n=0$ if it is occupied by $B$. 
$\Omega$ refers to the set of all possible arrangements considering that the molecules $A$ are indistinguishable from each other; the same applies to molecules $B$~\cite{frenkel_molphys14}. 
For simplicity, we only consider nearest neighbour interactions, which are described by the following Hamiltonian~\cite{safran_book}
\begin{multline}
	H(\{\sigma \}) = \frac12 \sum_{( m,n )} \Bigl(
		e_{AA} \sigma_m \sigma_n 
		 + e_{BB} (1-\sigma_m) (1- \sigma_n)
		\\[-5pt]
		+ e_{AB}\bigl[\sigma_m (1-\sigma_n) + \sigma_n (1-\sigma_m)\bigr]
	\Bigr) \, ,
	\label{eq:H}
\end{multline}
where the summation is over all nearest neighbour pairs $(m,n)$ on the lattice and the factor~$\frac12$ avoids the double counting of interaction pairs.
Here, the interaction parameters $e_{ij}$ determine what particle types tend to be next to each other.
For instance, if $e_{AA} < 0$, two $A$ molecules on neighbouring sites lower the total energy, making this configuration more probable. 
In general, these interaction parameters can arise from various physical interactions that may include dipolar and van der Waals interactions, screened electrostatic interactions between charged molecular groups or entropy-driven hydrophobic interactions~\cite{israelachvili2011intermolecular,safran_book,rubinstein2003polymer}.

\subsubsection{Thermodynamics of a homogeneous mixture.}\label{thermod_mixture}

Within the canonical ensemble
a homogeneous binary mixture of volume $V$ can be characterised by the Helmholtz free energy~$F= E -TS$, which combines the internal energy $E$ and the entropy $S$ of a system.
This free energy can be expressed by the partition function~\eqref{eq:Z}~\cite{yeomans1992statistical,balian2007microphysics1,balian2007microphysics2}
\begin{equation}
	F(T, V, N_A,N_B)=-\kb T \ln Z(T, V, N_A,N_B) \;.
	\label{eqn:free_energy_Z}
\end{equation}
Derivatives of the free energy $F$ are related to thermodynamic quantities 
that are relevant in our discussion of phase separation.
In particular, the entropy is given as $S=-\partial F/\partial T|_{V,N_A,N_B}$,
the pressure is $p=-\partial F/\partial V|_{T,N_A,N_B}$ and the chemical potentials read $\mu_A=\partial F/\partial N_A|_{T,V,N_B}$
and $\mu_B=\partial F/\partial N_B|_{T,V,N_A}$.

For simplicity, we focus on an incompressible binary system of constant volume $V=\nu M$ and constant molecular volume $\nu$ of the two components.
In this case, adding an $A$ molecule to the system corresponds to removing a $B$ molecule.
Consequently, the relevant thermodynamic quantities  are the {\it exchange chemical potential} $\bar\mu$ and the {\it osmotic pressure} $\Pi$~\cite{callen_b85,safran_book}:
\begin{subequations}
\begin{align}
	\label{eq:def_chem}
	\bar\mu 
		&= \left. \frac{\pp F}{\pp N_A} \right|_{T,V}
		=-  \left. \frac{\pp F}{\pp N_B} \right|_{T,V}
		= \nu  \left. \frac{\pp f}{\pp \phi} \right|_{T}	
\, ,
\\
	\label{eq:def_p}
	\Pi &= -\left. \frac{\pp F}{\pp V} \right|_{T,N_A}
		=  - f + \phi \, \left. \frac{\pp f}{\pp \phi} \right|_{T}	
	\, ,
\end{align}
\end{subequations}
where the number of lattice sites~$M$ is slaved to the total volume by $M=V/\nu$. 
Here, we used the homogeneity of the system, $F=V f(\phi)$, where $f(\phi)$ is the free energy density as a function of the volume fraction $\phi=N_A \nu/V$ of $A$ molecules.

The homogeneous state for a given volume is a stable thermodynamic state if it corresponds to a minimum of the free energy $F$.
This requires that the curvature of the free energy density
as a function of volume fraction is convex, i.e., $f''(\phi) \ge 0$.
The link between stability and curvature of the free energy density can be understood qualitatively:
conservation of molecule numbers implies that raising the volume fraction in one spatial region requires lowering it in another.
If the free energy density is convex, any such perturbation increases the overall free energy.
This can be shown rigorously by considering 
 spatially inhomogeneous perturbations that conserve molecule numbers.
 We will discuss this approach after introducing the free energy functional in section~\ref{sec:GL}.

The stability of the homogenous state can
also be shown using an ensemble where the particle number $N_A$ is fixed and 
the volume $V$ can change. 
This ensemble is governed by the thermodynamic potential  $ G(N_A,\Pi)=F(N_A,V) + V \Pi$, where $\Pi$ is the osmotic pressure given by \Eqref{eq:def_p} and the volume $V=\partial G /\partial \Pi$. 
A homogeneous state with the osmotic pressure
$\Pi$ is stable if the free energy $G$ as a function of $\Pi$ is concave, $\partial^2 G/\partial \Pi^2 <0$. 
The concavity of $G$ with respect to variations of the osmotic pressure can be seen by writing $\partial^2 G/\partial \Pi^2  = \partial V /\partial \Pi = -
V \kappa$, where $\kappa$ is the osmotic compressibility $\kappa = - V^{-1} \partial V /\partial \Pi$.
For the homogeneous state to be stable, the osmotic pressure should increase as the volume decreases, i.e., $\kappa>0$, to push the system back to its thermodynamic state after a perturbation in volume.
This condition is satisfied if the free energy density is convex,
$f''(\phi)>0$, since $\kappa=(\phi^2 f''(\phi))^{-1}$. 
In the thermodynamic limit,  
ensembles become equivalent and thus the convexity of the free energy density 
determines 
the thermodynamic stability of the homogeneous state,  
not only in the ensemble $(N_A,\Pi)$ where the osmotic pressure is imposed but also in the ensemble $(N_A,V)$ where the volume is fixed.

\subsubsection{Mean field free energy density of an incompressible mixture.}
To determine the relevant thermodynamic quantities for phase separation, we need to evaluate the free energy and the partition function given in \Eqref{eqn:free_energy_Z}.
Since this is generally difficult, we discuss a  {\it mean-field approximation} for
the homogeneous case of the incompressible binary mixture on a lattice 
characterised by the Hamiltonian given in~\Eqref{eq:H}. 
This approximation neglects the spatial correlations between the molecules.
Within the mean field approximation, the probability that lattice site~$n$ is occupied by $A$, is  given by $\la \sigma_n \ra= N_A/M$, where $\la \ldots \ra$ denotes the average in the canonical ensemble
and $\phi = N_A/M$ is the volume fraction of $A$ molecules  in the system.
Due to incompressibility the probability of the site being occupied by $B$ is $N_B/M=1-\phi$. 
The partition function hence is
\begin{equation}
	\label{eq:part_function_mean_field}
	Z \simeq \left| \Omega \right|
	\exp\left(- \frac{E(\phi)}{k_{\rm B}T}\right)
\end{equation}
with the internal energy given as 
\begin{equation}\label{eq_mf_energy}
	E(\phi)=\frac{zM}{2}\left[e_{AA} \phi^2 +2e_{AB} \phi (1-\phi) +e_{BB} (1-\phi)^2\right]
	\, ,
\end{equation}
where $z$ is the number of neighbours per lattice site (e.g., $z=6$ for a cubic lattice), and $zM/2$ is the total number of distinct nearest neighbours. 
The number~$\abs{\Omega}$ of all possible arrangements on the lattice appearing in \Eqref{eq:part_function_mean_field},
\begin{equation}
	\left| \Omega \right| = {M \choose N_A} = {M \choose N_B} = \frac{M !}{N_A ! N_B!}
	\label{eqn:possible_arrangements}
	\, ,
\end{equation}
 determines the  entropy $S= \kb \ln \left| \Omega \right|$ for the incompressible binary mixture on a lattice, which is also referred to as mixing entropy. 
 Using \Eqsref{eqn:free_energy_Z}, we obtain the free energy density
\begin{align}
	\fe(\phi) 
	&\simeq
	\frac{z}{2\nu} \bigl[ e_{AA} \phi^2 + 2 e_{AB} \phi(1-\phi) +e_{BB} (1-\phi)^2 \bigr] 
\notag\\
	&\quad + \frac{k_{\rm B}T}{\nu}  \bigl[ \phi\ln \phi +(1-\phi) \ln (1-\phi)\bigr] \, ,
\label{eq:free_en}
\end{align}
where we have used  Stirling's approximation, $\ln{N!} \simeq N\ln N -N$, to evaluate the factorials.
The free energy density can also be written as $\fe(\phi)  = \phi \fe(1) + (1-\phi)\fe(0) + \fe_{\rm mix}(\phi)$, which separates the contribution of the pure systems from the \textit{free energy of mixing}~\cite{safran_book,rubinstein2003polymer},
\begin{equation}
	\label{eq:Fmix}
	\fe_{\rm mix}(\phi) =
		\frac{k_{\rm B}T}{\nu} \bigl[ \phi\ln \phi +(1-\phi) \ln (1-\phi) + \chi \phi(1-\phi) \bigr]
\end{equation}
where 
\begin{equation}
	\label{eq:chi}
	\chi=\frac{z}{2 k_{\rm B}T} \left(2e_{AB}-e_{AA}-e_{BB}\right)
\end{equation}
is the Flory-Huggins interaction parameter \cite{flory1942thermodynamics,huggins42}.  
$\fe_{\rm mix}$ captures the competition between the mixing entropy $S=-k_{\rm B} (V/\nu) \left[\phi\ln \phi +(1-\phi) \ln (1-\phi) \right]$ and the molecular interactions characterised by the single parameter~$\chi$.
In the next section, we will see that both the free energy density (\eq (\ref{eq:free_en})) and the free energy density of mixing (equation~\eqref{eq:Fmix}) lead to the same phase separation  equilibrium.
However, the difference will become apparent when we discuss chemical reactions in section~\ref{sect:phase_sep_with_turn_over}.

The free energy density $f_{\rm mix}$ is  a symmetric function with respect to $\phi=\frac12$.
This symmetry stems from considering equal molecular volumes of components $A$ and $B$ and  the subtraction of the free energy before mixing.
Conversely, if the molecules $A$ and $B$  have different molecular volumes $n_A \nu$ and $n_B \nu$, the free energy of mixing is not symmetric ~\cite{flory1942thermodynamics,huggins42}:
\begin{align}
	\label{eq:Fmix_asym}
	\tilde{\fe}_{\rm mix}(\phi) &=
		\frac{k_{\rm B}T}{\nu} \bigg[ \frac{\phi}{n_A} \ln \phi +\frac{(1-\phi)}{n_B} \ln (1-\phi) 
\notag \\ & \quad
		+ \chi \phi(1-\phi) \bigg] \, ,
\end{align}
where $n_A$ and $n_B$ denote the non-dimensional molecular size in 
multiples of the volume $\nu$ of a single lattice site.

Homogeneous states 
governed by the free energy density $\fe(\phi)$ 
are only stable when the free energy density is convex, \mbox{$\fe''(\phi)>0$}; see section~\ref{thermod_mixture}. 
For the expression given in \Eqref{eq:free_en} this is the case for all~$\phi$ in the absence of interactions ($e_{ij}=0$ for $i,j=A,B$)
 and when entropic effects dominate.
In the presence of interactions however, the free energy density of the homogeneous system can become concave ($\fe''(\phi)<0$) within a range of volume fractions $\phi$, see \figref{fig:ch2_1}(a). 
Within this range, the homogeneous state is not stable,
implying that the thermodynamic equilibrium state is inhomogeneous.

\subsubsection{Phase coexistence.}
\label{sect:maxwell_contruct_homog}

The simplest inhomogeneous state corresponds to two subsystems of different volume fractions, also referred to as phases.
The associated free energy can be written as
\begin{equation}
	\label{eq:freeen}
	F  \simeq V_1 \fe(\phi_1)+V_2 \fe\left(\phi_2\right) \, , 
\end{equation}
where $\phi_\alpha$ and $V_\alpha$ denote the volume fraction and volume of phase $\alpha$, with $\alpha=1,2$. 
The  incompressibility assumption combined with conservation of particles implies   $V_1+V_2=V$ and $V_1\phi_1+V_2\phi_2=V\phi$. 
Consequently, there are only two independent variables in the free energy above, e.g., $\phi_1$ and $V_1$.
In \Eqref{eq:freeen} we neglected the energetic contribution of the interface region that separates the two phases.
This is valid in the thermodynamic limit where the system and the volumes of the phases are infinitely large, so the energetic contribution of the interface is negligible relative to the contribution of the phases.
 We will have to refine \Eqref{eq:freeen} when discussing finite systems in section~\ref{sec:interface}.

	\begin{figure}[tb]
	\begin{centering}
		\includegraphics[width=1.0\columnwidth]{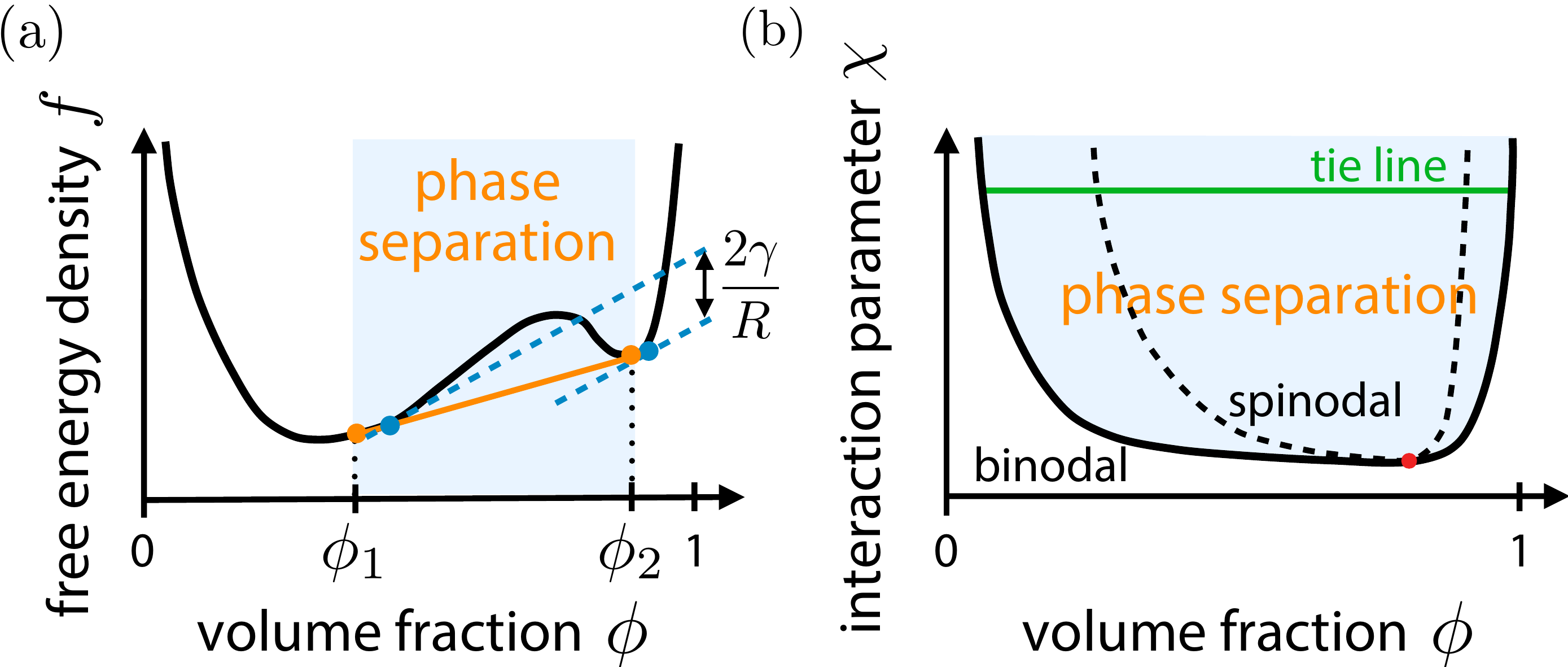} 
		\end{centering}
		\caption{(a) Sketch of an asymmetric free energy density $f(\phi)$ for an incompressible binary mixture as a function of volume fraction $\phi$, e.g.,
equation~\eqref{eq:Fmix_asym} for the case $n_B>n_A$ and $\chi>2$. In the presence of interactions, there can be a  range of volume fractions where the free energy density is concave with $\fe''(\phi)<0$. 
The Maxwell tangent construction modifies the free energy within the volume fraction range $[\phi_1, \phi_2]$ (orange line). As a result the free energy density becomes convex for all volume fractions. 
The equilibrium volume fraction of each phase are  $\phi_1$ and $\phi_2$. The impact of surface tension slightly increases the equilibrium volume fractions (blue dots, see section~\ref{sect:curvature_droplet}).
(b) Phase diagram as a function of the Flory-Huggins interaction parameter $\chi$ and volume fraction $\phi$. For large enough interaction parameters, phase separation can occur. The corresponding region in the phase diagram is bordered by the binodal line. The tie lines (green) connect the equilibrium volume fractions of two coexisting phases on the binodal line. The dashed line refers to the spinodal. Within the spinodal lines, the mixture can undergo a spontaneous partitioning into two phases, while between the spinodal and the binodal, only large enough phase separated domains can grow. This regime is referred to as nucleation \& growth.}
		\label{fig:ch2_1}
	\end{figure}

The inhomogeneous state is stable if it corresponds to a minimum of the free energy~\eqref{eq:freeen} consistent with the imposed constraint of particle conservation and in the absence of vacancies.
To find this minimum, we differentiate $F$ with respect to  $\phi_1$ and $V_1$, respectively, use the relationship  $\phi_2= (\phi V-\phi_1 V_1)/(V-V_1)$ and  set each expression to zero:
\begin{subequations}\label{eq:both_conditions_F_constr}
	\begin{align}
	\label{eq:equal_chem_pot}
	0 &= \fe'(\phi_1)- \fe'(\phi_2) \, ,
	\\
	\label{eq:equal_pressures}
	0 &= \fe(\phi_1)- \fe(\phi_2) +\left(\phi_2-\phi_1\right) \fe'(\phi_2)
	\, .
	\end{align}
\end{subequations}
The first equation is a balance of the exchange chemical potentials between phases, 
$\bar\mu_1=\bar\mu_2$,  with $\bar \mu_\alpha =  \bar{\mu} |_{\phi=\phi_\alpha}$ with $\alpha=1,2$; see \Eqref{eq:def_chem}).
The second equation corresponds to  the balance of the osmotic pressures between the two phases, $\Pi_1=\Pi_2$, with 
$\Pi_\alpha = \Pi |_{\phi=\phi_\alpha}$; see \Eqref{eq:def_p}.

Obviously, \Eqsref{eq:both_conditions_F_constr} are satisfied for homogeneous systems with $\phi_1 = \phi_2$.
To see that there also exist solutions with $\phi_1 \neq \phi_2$, the two conditions can be represented by a graphical tangent construction using the free energy density~$\fe(\phi)$, see \figref{fig:ch2_1}(a).
Here, condition~\eqref{eq:equal_chem_pot} implies that the slopes at the volume fractions $\phi_1$ and $\phi_2$ are the same and condition~\eqref{eq:equal_pressures} states that they are also equal to the slope of the line connecting the points $(\phi_1,\fe(\phi_1))$ and $(\phi_2, \fe(\phi_2))$.
Taken together, these conditions can only be satisfied by a common tangent to the two points.
This procedure of finding the equilibrium volume fractions is known as  {\it Maxwell's tangent construction} or {\it construction of the convex hull}.
The orange line in \figref{fig:ch2_1}(a) shows the result of such a construction.
In fact, inserting condition~\eqref{eq:equal_pressures} into equation~\eqref{eq:freeen} (and using conservation of $A$ particles) shows that 
this line corresponds to the volume weighted average of the free energy density of the two subsystems. 
Consequently, the corresponding demixed system is the thermodynamic equilibrium since it has a lower free energy than the mixed system described by the black line in \figref{fig:ch2_1}(a).
Accordingly, $\phi_{1}$ and $\phi_{2}$ are referred to as  
equilibrium volume fractions of the coexisting phases. 
Clearly, a separation into two phases with volume fractions $\phi_1$ and $\phi_2$ is only possible when the average volume fraction~$\phi$ obeys $\phi_1 < \phi < \phi_2$.
Outside this region, the homogeneous state is stable, since $f''(\phi) > 0$.

The parameter region where phase separation is possible can be determined from the solutions $\phi_1$ and $\phi_2$ as a function of the interaction parameter~$\chi$, see~\figref{fig:ch2_1}(b). 
The corresponding line is called the \textit{binodal line}.
In the simple case of the symmetric free energy of mixing (\Eqref{eq:Fmix}), 
$\fe_{\rm mix}(\phi_1)=\fe_{\rm mix}(\phi_2)$ and
$ \fe_{\rm mix}'(\phi_1)= \fe_{\rm mix}'(\phi_2)=0$. 
The binodal line is then given by $\chi_{\rm b}(\phi)=\ln(\phi/(1-\phi))/(2\phi-1)$ and phase separation occurs only for $\chi>\chi_{\rm b}$.
In particular, the minimal interaction parameter is $\chi_{\rm b}^{\rm min} = 2$, which is obtained at the critical point $\phi=\frac12$.
Near the critical point, the equilibrium volume fractions inside each phase obey
$\phi_{1}\simeq \frac12 - [\frac38(\chi-2)]^{1/2}$ and $\phi_{2}\simeq \frac12 + [\frac38(\chi-2)]^{1/2}$.
Note that the same results are obtained when \Eqref{eq:free_en} is used instead of $\fe_{\rm mix}$, since terms linear in $\phi$ do not alter the conditions given in \Eqsref{eq:both_conditions_F_constr}.

\subsubsection{Free energy of  inhomogeneous systems.}\label{sec:grad_free_energy}	
The discussion of the previous section neglected the contribution of the interfacial region on the equilibrium free energy.
This interfacial region is always present since the volume fraction is continuous in space and must thus interpolate between the values $\phi_1$ and $\phi_2$ in the two phases.
The additional free energy contribution associated with this spatial variation can be estimated within our lattice model.
For simplicity, we first consider a one-dimensional system with discrete lattice positions~$x_n$ for which the Hamiltonian given in \Eqref{eq:H} can be written as
\begin{align}	
H(\{\sigma\}) &=\sum_n 
\big[
	e_{AB} \bigl( \sigma_n (1-\sigma_{n+1}) + \sigma_{n+1} (1-\sigma_{n}) \bigr)
\notag\\ & \quad
	+ e_{AA}  \sigma_{n}  \sigma_{n+1} + e_{BB}  (1- \sigma_{n}) (1- \sigma_{n+1}) \big] \, .
\label{eq:spatvar}
\end{align}
We proceed analogously to \secref{sec:stat_mech} and perform a mean-field approximation after rewriting the coupling terms using $2 \sigma_n (1-\sigma_{n+1}) =   \left(\sigma_n-\sigma_{n+1}\right)^2  -\sigma_n^2-\sigma_{n+1}^2+2\sigma_n$.
Additionally, generalising to three dimensions and taking the continuum limit,
we replace $\sum_n \mapsto \nu^{-1}\int \dd^{3}r $, $\mean{\sigma_n} \mapsto \phi(\bbr)$, and $\mean{\sigma_{n+1} - \sigma_n} \mapsto \nu^{\frac{1}{3}} \nabla\phi(\vect r)$.
Hence,
\begin{align}
	E & \simeq
 	\frac{\kb T \chi}{\nu} \int \dd^{3}r \Bigl[  \phi(\bbr) \bigl(1-\phi(\bbr) \bigr)
		+ \frac{\nu^{\frac23}}{2} \left| \nabla \phi(\bbr) \right|^2  \Bigr]
	\notag
	\\
	& \quad +  \frac{z}{2\nu}\int \dd^{3}r \Big[e_{AA} \, \phi(\bbr) + e_{BB} \bigl(1-\phi(\bbr) \bigr)  \Big] \, ,
\end{align}
where 
$\chi$ is the Flory-Huggins parameter given in equation~\eqref{eq:chi}. 
The associated free energy~$F$ can be expressed as a functional of $\phi(\vect r)$,
\begin{align}
	F
	&=\frac{1}{\nu} \int \dd^{3}r \,  \Bigl[
	 \frac{\kappa}{2\nu} \left|\nabla \phi \right|^2
	 + \frac{z}{2}\left(e_{AA} \, \phi + e_{BB} \left(1-\phi \right) \right)
\notag\\ &\quad 
	+ \kb T \big[ \phi \ln (\phi) +(1 - \phi)\ln (1- \phi)  
	+ \chi \,  \phi (1-\phi) \big]
	\Bigr]
	\;,
	\label{eq:fmix_cont_bin}
\end{align}
where 
\begin{equation}\label{eq:kappa_lattice}
	\kappa = k_{\rm B}T \chi \, \nu^{\frac53}
\end{equation}	
characterises the change of free energy density due to concentration inhomogeneities.
Since the interaction parameter $\chi>2$ in the phase separating regime, we have $\kappa >0$ and the gradient term thus penalises spatial inhomogeneities.
We identify the integrand of the free energy given in  \Eqref{eq:fmix_cont_bin} as the total free energy density~$\fetot$.
Expressing it in terms of concentrations~$c  = \phi/\nu$,  we obtain
\begin{equation}
	\label{eq:freefuncttot}
	\fetot(c, \nabla c) = f_{\rm mix}(c)  + f_0(c) + \frac{\kappa}{2} \left|\nabla c\right|^2 \, ,
\end{equation}
where $f_{\rm mix}$ follows from \Eqref{eq:Fmix} and $f_0(c) =(z/2) \left( e_{AA} \, c + e_{BB} \left(\nu^{-1}-c \right) \right)$ is the free energy of the pure system before mixing.

\subsubsection{Ginzburg-Landau free energy.}\label{sec:GL}	     
In the previous sections we have derived the free energy density 
of an incompressible binary mixture using statistical mechanics.
We have seen that phase separation occurs when the free energy density exhibits a concave region enclosed by convex branches.
The simplest free energy for an incompressible mixture that has such a shape 
is  known as the {\it  Ginzburg-Landau free energy}~\cite{Bray_Review_1994}:
\begin{align}
	\label{eq:GLfree}
	F_{\rm GL}[c]=\int \dd^{3}r \left(
		\fe_{\rm GL}(c)+\frac{\kappa}{2} \left|\nabla c\right|^2
	\right)
	\; ,
\end{align}
with the corresponding free energy density given as 
\begin{align}
	\label{eq:GL_free_energy_density_asym}
	\fe_{\rm GL}(c) &= 
	\tilde{b} \left(c - \cc\right) -\frac{b}{2} \left(c - \cc\right)^2 \\
	\notag 
	& \quad + \frac{\tilde a}{3} \left(c-\cc\right)^3
	 + \frac{a}{4} \left(c-\cc\right)^4 
 \, .
\end{align}
This free energy density is parameterised by the phenomenological coefficients $a$, $\tilde a$, $b$, $\tilde b$, and $\cc$.
As discussed in section~\ref{sect:maxwell_contruct_homog}, phase separation equilibrium is not affected by the linear term, thus we choose $\tilde{b}=0$, and for reason of simplicity, we also consider the special case $\tilde{a}=0$, leading to the 
bi-quadratic form of the Ginzburg-Landau free energy density:
\begin{align}
	\label{eq:GL_free_energy_density}
	\fe_{\rm GL}(c) = 
	 -\frac{b}{2} \left(c - \cc\right)^2
	+ \frac{a}{4} \left(c-\cc\right)^4  \, ,
\end{align}
which is symmetric around the concentration $\cc$. 
This free energy density has a concave region if $b>0$ 
and the parameter $a>0$ characterises the convex branches of the energy density.
Using equation~\eqref{eq:equal_chem_pot} the equilibrium concentrations within each phase separated domain are
\begin{subequations}\label{eq:both_c1c2_GZ}
	\begin{align}
	c_1 & =\cc - \sqrt{ b/ a} \, ,
	\\
	c_2 & =\cc + \sqrt{ b/  a}
	\, .
	\end{align}
\end{subequations}
The Ginzburg-Landau free energy density $\fe_{\rm GL}(c)$ given in \Eqref{eq:GL_free_energy_density} can either be understood as a phenomenological free energy density that exhibits the qualitative feature necessary for phase separation, i.e.,  a concave domain enclosed by two convex domains, 
or as an expansion of the free energy of mixing
around a constant concentration $\cc$, typically the critical concentration~\cite{de2013non,Bray_Review_1994,Onuki_book,chaikin_b00}.
In particular,  expanding the symmetric free energy of mixing $f_{\rm mix}(c)$ (\Eqref{eq:Fmix})  around $\cc=1/(2\nu)$ up to the fourth order, we find
\begin{equation}
\label{eq_a_b}
	a=\frac{16}{3} \kb T \nu^3 \sep b= 2 (\chi-2) \kb T\nu
	\; ,
\end{equation}
which links the phenomenological parameters $a$ and $b$ with the molecular parameters 
of the lattice model.
Using the expressions above in equations~\eqref{eq:both_c1c2_GZ} we consistently obtain 
the equilibrium concentration found in section~\ref{sect:maxwell_contruct_homog}.

\subsection{Equilibrium states of a binary mixture}\label{sec:interface}	

In this section, we determine the equilibrium concentration profiles that minimise the free energy $F_{\rm GL}$.
Specifically, we calculate the extremal solutions of $F_{\rm GL}$ and discuss their stability in different regions of the phase diagram, see \figref{fig:ch2_1}(b).
An explicit expression of the interfacial profile will allow us to relate the free energy contribution characterised by $\kappa$ to the surface tension between phases.

\subsubsection{Stationary states.}
\label{sec:passive_stationary_states}
We start by determining the stationary states $c_*(\vect r)$ of the bi-quadratic Ginzburg-Landau free energy~$F_{\rm GL}$ given in \Eqref{eq:GLfree}.
Such states have an extremal free energy subjected to the constraint that the number of $A$ and $B$ molecules are conserved, \ie
\begin{equation}
	\label{eq:particle_conservation}
	\bar{c} = \frac1V \int_V \dd^{3}r \,  c(\bbr)
	\;, 
\end{equation}
where $\bar{c}=N_A/V$ denotes the mean concentration of $A$ molecules in the mixture of the finite volume $V$.
To enforce this constraint, we introduce a Lagrange multiplier $\lambda$ and vary the functional  $F_{\rm GL} -\lambda \int_V \dd^{3}r \, c(\vect r)$.
Here, the functional derivative of the free energy is the generalization of the exchange chemical potential generalized to inhomogeneous systems,
\begin{align}
	\bar{\mu} = \frac{\delta F}{\delta c}
	\label{def_inhomo_chem_pot}
	\;,
\end{align}
which reads $\bar{\mu}=a \left( c(\bbr)-\cc\right)^{3}- b \left(c(\bbr)-\cc \right)- \kappa \nabla^2 c(\bbr)$ for $ F_{\rm GL}$.
Consequently, the Euler-Lagrange equation for the stationary state is 
\begin{align}\label{eq:chem_pot_GL}
	\bar{\mu} = \lambda \, ,
\end{align}
where we dropped a boundary term proportional to $\kappa \nabla c$, assuming no-flux  boundary conditions at the system boundary. 
The stationary states thus correspond to a spatially uniform exchange chemical potential $\bar{\mu}$, which may be realised for both homogeneous and inhomogeneous concentration profiles $c_*(\vect r)$.

We start by considering the spatially homogeneous equilibrium states $c_*(\vect r) = c_0$.
Particle conservation implies $c_0 = \bar c$ (\Eqref{eq:particle_conservation}) and the Euler-Langrange equations read
$\lambda= a(\bar{c} - \cc)^3 - b(\bar{c} - \cc)$.
The homogeneous state $c(\vect r) = \bar c$ is thus an extremal state of the free energy $F_{\rm GL}$ and we will check whether it corresponds to a minimum in the next section.

We showed above that states with coexisting phases can be stable in some regions of the phase diagram shown in \figref{fig:ch2_1}(b).
Here, we determine the concentration profile that connects the two phases from extremising the free energy $F_{\rm GL}$.
In the following, we restrict ourselves to a flat interface oriented perpendicular to the $x$-axis at the position $x=0$ with a concentration $c(x=0) = \cc$.
For simplicity, we extend the system to infinity while keeping the position and concentration value of the interface fixed. For the symmetric free energy density (\Eqref{eq:GL_free_energy_density}), 
this implies $\bar \mu =0$, thus $\lambda=0$, at the interface, and in the case of an extremal inhomogeneous state, $\bar \mu =0$ at all positions.
Far away from the interface, the concentration inside the phases should be governed by the equilibrium concentrations (\Eqref{eq:both_c1c2_GZ}),
\begin{align}
	\lim_{x\rightarrow -\infty}  c(x) &= c_1\, ,
&
	\lim_{x\rightarrow \infty}  c(x)  &= c_2
	\;.
	\label{eq:interface_bcs}
\end{align}
With the boundary conditions above the unique solution is~\cite{krapivsky_b10}:
\beq \label{eq:tanh}
 c_{\rm I}(x) = \cc + \sqrt{\frac{b}{a}}\tanh\left(\sqrt{\frac{b}{2 \kappa}}x\right) \ .
\eeq
Since this interfacial profile varies substantially only within the region $|x| \lesssim \sqrt{2\kappa/b}$, we introduce the  {\it interfacial width}
\beq
\label{eq:Iwidth}
w=\sqrt{\frac{2\kappa}{b}}  \simeq \nu^{\frac13} \sqrt{\frac{\chi}{\chi-2}} \ .
\eeq
The right hand side relates $w$ to the lattice model using equations \eqref{eq:kappa_lattice} and \eqref{eq_a_b}.
In the limit of strong phase separation (large $\chi$) 
the interfacial width approaches the linear dimension $\nu^{1/3}$ of the molecules.

\subsubsection{Stability of stationary states.}
\label{sec:stability_stationary_states}
The homogeneous and inhomogeneous stationary states of the Ginzburg-Landau free energy, $c_*(x)=\bar c$ and $c_*(x)=c_{\rm I}(x)$, respectively, can be either stable or unstable.
They are stable if they correspond to a free energy minimum, \ie if all small concentration perturbations increase the free energy.
To test this, we consider concentration profiles $c=c_* +\epsilon$, where $\epsilon(\vect r)$ is a small, position dependent concentration perturbation.
To quadratic order, the change in the free energy due to this perturbation is 
\begin{align}
	\Delta F [c_*, \epsilon] &= F_{\rm GL} (c_* + \epsilon) - F_{\rm GL} (c_*) 
	\label{eq:triF}
\\
	&\simeq \int \diff^3r \Big[\frac{\epsilon^2}{2}\bigl( 3a (c_*-\cc)^2-b\bigr)
	\nonumber
	+\frac{\kappa}{2} \left(\nabla \epsilon \right)^2 \Big]
	\;.
\end{align}
The state $c_*(\vect r)$ is stable if all perturbations increase the free energy, i.e., if $\Delta F[c_*, \epsilon] > 0$ for all $\epsilon(\vect r)$.
In the case of the homogeneous state $c_*(\vect r) = \bar c$ both terms in the integrand are positive if $\abs{\bar c - \cc} > \sqrt{b/(3a)}$, which implies $\Delta F[c_*, \epsilon]>0$.
Conversely, $\Delta F[c_*, \epsilon]$ can be negative for $\abs{\bar c - \cc} < \sqrt{b/(3a)}$ in sufficiently large systems, \eg for the perturbation $\epsilon(x) \propto \tanh(x/w)$, which implies that the homogeneous state is unstable for these parameters.
Consequently, the stationary homogeneous state can be either stable or unstable to infinitesimal perturbations.
In contrast, the inhomogeneous state~$c_*(x)=c_{\rm I}(x)$ 
given by \Eqref{eq:tanh}
is always stable if it is a stationary state, \ie when $\abs{\bar c - \cc} < \sqrt{b/a}$, see \ref{sec:appendix_stability}.

Taken together, we can distinguish three different parameter regimes with different stable stationary states.
For mean concentrations~$\bar c$ far away from the symmetry point~$\cc$, \ie when $\abs{\bar c - \cc} > \sqrt{b/a}$, the homogeneous state is the only stable one.
Conversely, when $\abs{\bar c - \cc} < \sqrt{b/(3a)}$, only the inhomogeneous state is stable and phase separation will thus happen spontaneously.
This region is known as the \textit{spinodal decomposition} region~\cite{Cahn_61,Siggia_79,Tanaka_95,mecke1997morphology,hayward1987dynamic,wagner1998spinodal, Sappelt_1998} which is enclosed by the spinodal line (dashed line in \figref{fig:ch2_1}(b)).
Between the spinodal region and the homogeneous region, for $\sqrt{b/(3a)} < \abs{\bar c - \cc} < \sqrt{b/a}$, both states are stable to infinitesimal perturbations.
In this case, phases can only originate from the homogeneous state by large fluctuations, known as nucleation events.
Consequently, the respective region in the phase diagram is known as the \textit{nucleation and growth} regime.
All three phases are shown in the phase diagram in \figref{fig:ch2_1}(b).

\subsection{Surface tension of interfaces.}
The surface tension of the interface can be determined from the profile $c_{\rm I}(x)$.
To this end, we separate the energetic contributions of the bulk phases to the free energy from a contribution that is related to the interface.
For large volumes of the coexisting phases  $V_1$ and $V_2$,
the total free energy~$F$ can be written as
\begin{equation}
	\label{eq:freeenwithgamma}
	F = V_1 \fe(c_1)+V_2 \fe\left(c_2\right) + \gamma A
	\; ,
\end{equation}
where $c_1$ and $c_2$ are the corresponding equilibrium concentrations.
Here, $A$ is the area of the interface and $\gamma$ denotes the surface energy, which is also known as the \textit{surface tension}~\cite{Ip1994}.
Its value is obtained from the condition that the total free energy given in \Eqref{eq:freeenwithgamma} equals the free energy of the interfacial profile, $F=F[c_{\rm I}]$:
\begin{align}
	\gamma A &= \int_V \!\! \dd^{3}r \left[
			f(c_{\rm I}) + \frac{\kappa}{2} (\nabla c_{\rm I})^2
			 \right] 
		- V_1 \fe(c_1)- V_2 \fe\left(c_2\right)  \, . 
\end{align}
In the simple case where a flat interface is oriented perpendicular to the $x$-axis and the system is extended to infinity while keeping the position of the interface fixed at $x=0$, the surface tension reads
 \begin{align}
	\gamma  = \int_{-\infty}^{\infty} \text{d}x \left[
			f(c_{\rm I}) -\frac{1}{2} \Bigl[\fe(c_1) + \fe(c_2) \Bigr]+ \frac{\kappa}{2} (\nabla c_{\rm I})^2 \right]\, .
\end{align}
Considering the Ginzburg-Landau free energy and the corresponding interfacial profile $c_{\rm I}(x)$ (\Eqref{eq:tanh}), we find $\fe_{\rm GL}(c_1) = \fe_{\rm GL}(c_2) = -b^2/(4a)$ and thus
\begin{align}
	\gamma &= \int_{-\infty}^\infty \diff x \left[
			f_{\rm GL}(c_{\rm I}) + \frac{\kappa}{2} (\partial_x c_{\rm I})^2
			+ \frac{b^2}{4a}
		\right]
\notag\\[5pt]&=
	\frac{2\sqrt{2\kappa b^3}}{3 a}
	\simeq   \frac{\kb T \chi^{\frac12} (\chi -2)^{\frac32}}{2 \nu^{\frac23}} 
	\label{eqn:surfacetension}
	\;.
\end{align}
In the last approximation, we have used the expressions for the lattice model (equations  \eqref{eq:kappa_lattice} and \eqref{eq_a_b}), which shows that $\gamma$ scales like $\chi^2$ in the limit of strong phase separation (large $\chi$).

\subsection{Dynamical equations of phase separation}
\label{sec:dynamical_equation}
We next derive the dynamical equations describing how the binary mixture reaches its equilibrium state.
Considering an incompressible mixture, the volume fractions obey $\phi_A+\phi_B=1$, and thus $\partial_t \phi_A = - \partial_t \phi_B$. Using $\phi_i=c_i \nu_i$ ($i=A,B$) and considering for simplicity the case where molecular volumes of the two components are equal to $\nu_i=\nu$, the incompressibility condition leads to $\partial_t c_A = - \partial_t c_B$.
The particle conservation of $A$ and $B$ molecules can be expressed by the continuity equations
\begin{subequations}
\begin{align}
	\partial_t c_A &=- \nabla \cdot \vect{j}_A \, ,\\
	\partial_t c_B &=- \nabla \cdot \vect{j}_B  \, ,
\end{align}
\end{subequations}
where incompressibility and equal molecular volumes imply that the particle fluxes of component $A$ and $B$ read $\vect{j}_A=\vect{v} c_A +\vect{j}$ and $\vect{j}_B=\vect{v} c_B -\vect{j}$,  and that the volume flow velocity $\vect{v}$ obeys $\nabla \cdot \vect{v}=0$.
In the following sections, we restrict ourselves to a reference frame where $\vect{v}=\vect 0$. In this case the exchange current reads $\vect{j}=\left( \vect{j}_A -\vect{j}_B\right)/2$, which drives the time evolution of the concentration of components $A$, 
\begin{align}
	\partial_t c= -\nabla \cdot \bj \, ,
\end{align}
where we abbreviated $c=c_A$ for simplicity. 
In linear response, the exchange current is proportional to the thermodynamic force of the gradient of the exchange chemical potential  $-\nabla \bar{\mu}$, implying
 $\bj = - \Lambda(c) \nabla \bar{\mu}$;  see \ref{sec:appendix_entropy_production}.
 Here, $\Lambda(c)$ denotes a mobility coefficient,
 which is positive to ensure that the second law of thermodynamics is fulfilled, \ie that
 the corresponding entropy production, $-\int \diff^3 r \, \bj \cdot \nabla \bar{\mu}$, is positive~\cite{landau_lifschitz_6_hydro,balian2007microphysics2}. 
The resulting dynamical equation is
\begin{equation}
	\label{eq:dyneqn}
	\pp_t c = \nabla \cdot \bigl(\Lambda(c) \nabla  \bar{\mu}(c) \bigr)
	\, .
\end{equation}
This equation is also known as the deterministic version of the so-called model B~\cite{Hohenberg1977, chaikin_b00}.
Note, that by considering $\vect v=\vect 0$, we do not discuss the transport of momentum and the associated couplings to fluid flow.
The interested reader is referred to Refs.~\cite{Bray_Review_1994,de2013non, Onuki_book,cates_a12}. 

In the simple case of the Ginzburg-Landau free energy functional $F_{\rm GL}$ given in \Eqref{eq:GLfree}, the dynamical equation~\eqref{eq:dyneqn} reads
\begin{equation}
	\label{eq:dyneqnCH}
		\pp_t c =
		     \nabla \cdot \left[
		     	\Lambda(c) \, \nabla \bigl( a(c-\cc)^3 - b (c-\cc) - \kappa  \, \nabla^2 c \bigr) \right] \;,
\end{equation}
which is known as the \textit{Cahn-Hilliard equation}~\cite{Cahn_61}.
We can use this equation to scrutinize the stability of the homogeneous state, $c(\vect r)=\bar c$, by performing a linear stability analysis.
We denote the perturbed state as $c(\vect r, t) = \bar c + \epsilon \exp(\omega t + \ii \bk \cdot \bbr)$, where $\omega$ denotes the perturbation growth rate, $\bk$ the perturbation wave vector, and
 $\epsilon$ the associated small amplitude, $\abs{\epsilon} \ll \bar c$.
To  linear order, the growth rate  is
\begin{equation}
	\label{eq:linstab}
	\omega(\vect q) = 
		-\vect q^2 \Lambda(\bar c) \left[3a (\bar c - \cc)^2 -b +\kappa \vect q^2  \right] 
	\;.
\end{equation}
The homogeneous state is stable if all perturbations decay, \ie if $\omega(\vect q) < 0$ for all $\vect q$.
However, $\omega(\vect q)$ can become positive for small $\abs{\vect q}$ if $\abs{\bar c- \cc} <\sqrt{b/3a}$.
This parameter region corresponds to the spinodal decomposition that we found above (\figref{fig:ch2_1}).
The stability associated with the dynamical equations is therefore consistent with the one derived from the free energy discussed in \secref{sec:stability_stationary_states}.

Beyond the linear regime, \Eqref{eq:dyneqnCH} is difficult to solve as it is non-linear and involves fourth order spatial derivatives. 
However, inside the two coexisting phases and far away from the interface, concentration variations  are small, so we can ignore the fourth order derivative and linearize \Eqref{eq:dyneqn}
around the equilibrium concentrations $c_1$ and $c_2$  (\Eqref{eq:both_c1c2_GZ}). 
Hence, we arrive at two diffusion equations which are valid inside phase 1 and 2, respectively, 
\begin{equation}
	\label{eq:diff_equ_ch2}
	\pp_t  c \simeq D_\alpha \, \nabla^2  c
	\; ,
\end{equation}
where the collective diffusion coefficient inside phase $\alpha=1,2$ reads 
\begin{equation}
	\label{eq_diff_constant}
	D_\alpha = \mob(c_{\alpha}) \, f''(c_{\alpha})
	\; .
\end{equation}
Note that $D_{\alpha}$ is positive when phase separation occurs ($f''(c_\alpha) > 0$) and that an equivalent argument leads to positive diffusivity when phase separation is absent ($f''(\bar c) > 0$).
In the simple case of the symmetric Ginzburg-Landau free energy given in \Eqref{eq:GL_free_energy_density} and for a constant mobility $\mob$, the diffusion coefficients are identical in both phases and equal to $D=2 b \mob$.
Using the expressions corresponding to the lattice model (equation~\eqref{eq_a_b}), we obtain $D \simeq 4(\chi-2)  \nu \mob \kb T$ close to the critical point, which is positive as phase separation occurs only when $\chi > 2$.

\subsection{Dynamics of droplets}
In this section, we focus on droplets, which are small condensed phases coexisting with a large dilute phase.

\subsubsection{Impact of surface tension on the local equilibrium concentrations.}\label{sect:curvature_droplet}

One important difference between droplets and the condensed phases that we discussed so far is the curvature of the droplet interface, which is inevitable due to the finite size.
The surface tension~$\gamma$ of this curved interface affects the equilibrium concentrations inside and outside the droplet, which we denote by $\cEqIn$ and $\cEqOut$, respectively.
In the following, we consider the case where the surface tension $\gamma$ is constant and independent of the interface curvature, which is valid for droplets large compared to the
{\it Tolman length}~\cite{Tolman1949, Blokhuis2006}.
To derive how the equilibrium concentrations 
depend on the droplet curvature, 
we write \Eqref{eq:freeenwithgamma} for a spherical droplet of radius~$R$,
\begin{equation}
	F = \Vd \fe(\cEqIn)+(\Vsys - \Vd) \fe(\cEqOut) + 4\pi R^2 \gamma \, ,
\end{equation}
where $\Vd=\frac{4\pi}{3} R^3$ denotes the droplet volume and $\Vsys$ is the volume of the system.
Minimizing the free energy above analogously to section~\ref{sect:maxwell_contruct_homog}, we obtain the equilibrium conditions
\begin{subequations}\label{eq:both_conditions_F_constr_with_gamma}
\begin{align}
	\label{eq:shift1}
	0 &= \fe'(\cEqIn)- \fe'(\cEqOut) \, ,
\\[5pt]
	\label{eq:shift2}
	0 &= \fe(\cEqIn)- \fe(\cEqOut) 
\notag\\[-3pt] & \quad
	+\left(\cEqOut-\cEqIn\right) \fe'(\cEqOut)
	+\frac{2 \gamma}{R}
	\; .
\end{align}
\end{subequations}
Comparing these expressions to \Eqsref{eq:both_conditions_F_constr} in the thermodynamic limit, we find that the pressure balance \eqref{eq:shift2} contains an additional term~$2\gamma/R$, which is known as the \textit{Laplace pressure}.
Graphically, the Laplace pressure corresponds to the free energy difference of the tangents in the Maxwell construction, see Fig.~\ref{fig:ch2_1}(a).
The Laplace pressure is proportional to the interface curvature~$R^{-1}$ and thus disappears in the thermodynamic limit ($R \rightarrow \infty$).

The conditions~\eqref{eq:both_conditions_F_constr_with_gamma} determine the equilibrium concentrations~$\cEqIn$ and $\cEqOut$ inside and outside the droplet, respectively.
We derive approximate expressions by expanding $c^{\rm eq}_{\rm in/out} = c^{(0)}_{\rm in/out}+ \delta c_{\rm in/out}$ in  equations~\eqref{eq:both_conditions_F_constr_with_gamma} 
 to linear order in $\delta c_{\rm in/out}$.
Here, $c^{(0)}_{\rm in}$ and $c^{(0)}_{\rm out}$ denote the equilibrium concentration in the thermodynamic limit in the condensed and dilute phase, respectively, so $\delta c_{\rm in/out}$ captures the effects of Laplace pressure.
We find
\begin{subequations}
\begin{align}
	\label{eq:phiout}
	\delta c_{\rm out} &\simeq 
		\frac{2 \gamma}{\bigl(\cBaseIn - \cBaseOut\bigr) \fe''\bigl(\cBaseOut\bigr) R}
		\, ,
	\\[5pt]
		\label{eq:phiin}
	\delta c_{\rm in} &\simeq 
		\frac{ \fe''\bigl(c^{(0)}_{\rm out}\bigr)}{ \fe''\bigl(c^{(0)}_{\rm in}\bigr)} 	\delta c_{\rm out}
	\;,
\end{align}
\end{subequations}
which are known as the {\it Gibbs-Thomson relations}.
Since both expressions are positive, the Laplace pressure elevates the concentrations both inside and outside the droplet.
This effect is stronger for smaller droplets,
which becomes explicit when writing the equilibrium concentrations as 	
\begin{subequations}\label{eq:final_GT_relations}
\begin{align}
	c^{\rm eq}_{\rm out}&=c^{\rm (0)}_{\rm out} \left(1+\frac{\ell_{\gamma,{\rm out}}}{R}\right)
	\; ,
\\
	\label{eq:cineq}
	c^{\rm eq}_{\rm in}&=c^{\rm (0)}_{\rm in} \left(1+\frac{\ell_{\gamma,{\rm in}}}{R} \right)
	\; ,
\end{align}
where we defined for both phases the {\it capillary lengths}
\begin{align}
	\ell_{\gamma,{\rm out}} &=
		\frac{2 \gamma}{\bigl(\cBaseIn - \cBaseOut\bigr) \fe''\bigl(\cBaseOut\bigr)\cBaseOut}
\, ,
\\[5pt]
\ell_{\gamma,{\rm in}} &= 
 \frac{ \fe''(\cBaseOut)\cBaseOut }{ \fe''(\cBaseIn)\cBaseIn} \, \ell_{\gamma,{\rm out}}
\;.
\end{align}
\end{subequations}
In the following, we are interested in the limit of strong phase separation with  $c^{(0)}_{\rm in} \gg c^{(0)}_{\rm out}$ and thus  $\ell_{\gamma,{\rm in}} \ll \ell_{\gamma,{\rm out}}$.
In this case, the impact of the Laplace pressure on the equilibrium concentration inside the droplet can be neglected, \ie $c^{\rm eq}_{\rm in} \simeq c^{\rm (0)}_{\rm in}$.
Since this leaves us with a single capillary length, which we define $\ell_\gamma = \ell_{\gamma,{\rm out}}$.
When the  phase outside is dilute, \ie the chemical potential can be written as  $\bar{\mu}(c) \simeq k_{\rm B}T \ln (\nu c)$~\cite{pitaevskii_b81},
 we find
\begin{equation}
	\ell_{\gamma}  \simeq
	\frac{2 \gamma}{c^{(0)}_{\rm in} \kb T} 
	\;.
	\label{eq:lc_app}
\end{equation}
If we additionally assume that the condensed phase is highly packed such that $\cBaseIn \simeq \nu^{-1}$, \Eqref{eq:lc_app} provides a useful estimate of the capillary length when the surface tension~$\gamma$ is known~\cite{pitaevskii_b81}. 
Using equation~\eqref{eqn:surfacetension} from our lattice model, the capillary length can also be expressed as $\ell_{\gamma}  \simeq \chi^{1/2} \left( \chi -2\right)^{3/2}\nu^{1/3}$.
This expression demonstrates that for interaction parameters not too far away from the 
critical value, $\chi_{\rm b}^{\rm min} = 2$,
it is typically on the order of the molecular length scale~$\nu^{1/3}$.
Consequently, we have $\ell_\gamma \ll R$ and the increase of the  equilibrium concentrations predicted by the Gibbs Thomson relations~\eqref{eq:final_GT_relations} is actually small, supporting the validity of the linear approximation.

\subsubsection{Growth of a single droplet in a supersaturated environment.}\label{sect:intro_growth_single_droplet}

The dynamics of the droplet size and its shape are linked to the movement of its interface, which we assume to be thin compared to the droplet size in the following.
To describe the dynamics of the interface, we consider a spherical coordinate system~$(r, \varphi, \theta)$ centered on the droplet.
Assuming the interface does not deviate strongly from a spherical shape, we parameterize its shape $\vect{R}(\varphi, \theta; t)= \mathcal{R}(\varphi, \theta; t) \vect{e}_r$ by the radial distance $\mathcal{R}(\varphi, \theta; t)$ as a function of the polar angle~$\varphi$ and the azimuthal angle~$\theta$.
The movement of the interface is most naturally described in the local coordinate system spanned by the two tangential directions $\vect{e}_1=\partial \vect{R}/\partial \varphi$ and $\vect{e}_2=\partial \vect{R}/\partial \theta$ and the outward normal vector $\vect{n}=\frac{\vect{e}_1 \times \vect{e}_2}{|\vect{e}_1 \times \vect{e}_2|}$. 
Note that the droplet shape is only affected by the normal component~$v_n$ of the interfacial velocity, while the tangential components transport material along the interface.
Material conservation implies that this normal component is proportional to the net material flux toward the interface,
\begin{equation}
	\label{eq:interface_vel}
		v_n =\frac{\vect{j}_{\rm in}-\vect{j}_{\rm out}}{\cEqIn-\cEqOut} \cdot \vect{n} 
	\; ,
\end{equation}
where $\vect{j}_{\rm in}=\lim_{\epsilon\to 0} \vect{j} (\vect{R} - \epsilon\vect{n} )$
and
$\vect{j}_{\rm out}=\lim_{\epsilon\to 0} \vect{j} (\vect{R} + \epsilon\vect{n} )$
 are the local material fluxes right inside and outside of the interface, respectively.
Expressing the time evolution of the interface in 
the spherical coordinate system gives
$\partial_t \vect{R}= \partial_t \mathcal{R} \, \vect{e}_r $,
while in the local coordinate system of the interface,
\begin{align}
	\partial_t \vect{R}
	= v_n \vect{n} + v_{\text{t},1} \vect{e}_1 + v_{\text{t},2} \vect{e}_2
\, .
\end{align}
We can use the connection between the local and the global coordinate system to identify
 conditions $\partial_t \vect{R} \cdot \vect{e}_\theta = 0$ 
and $\partial_t \vect{R} \cdot \vect{e}_\varphi = 0$, which can be used to obtain the in plane velocity components $v_{\text{t},1}$ and $v_{\text{t},2}$.
The radial interface velocity then reads
\begin{align}
	\label{eq_interface_evolution}
	\partial_t \mathcal{R} &= 
	v_n \left[
		1 + \left( \frac{ \partial_\theta \vect{R}}{R}\right)^2 + 
		\left(\frac{\partial_\varphi \vect{R}}{R \sin( \theta) } \right)^2
	\right]^{\frac12}
	\; .
\end{align}
In the case where the dynamics within the phases are described by the diffusion equation~\eqref{eq:diff_equ_ch2}, the material flux is given by $\vect j=-D \nabla c$ and \Eqref{eq_interface_evolution} directly determines the time evolution of the interface.

\begin{figure*}[tb]
	\begin{center}
		\includegraphics[width=1.0\textwidth]{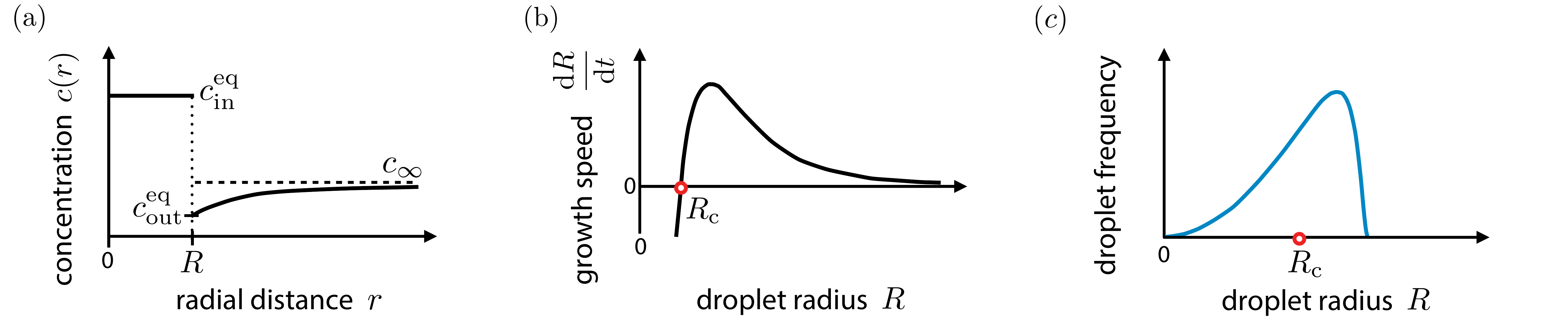}
	\end{center}
	\caption{
(a) Illustration of the concentration field inside and outside of the droplet; see equations~\eqref{eq:csol_inside_outside}. These concentration fields can be obtained from solving the diffusion equations~\eqref{eq:diff_equ_ch2} using quasi-static approximation and considering the case of an infinitely thin interface. 
(b) The droplet growth speed $\dd R /\dd t$ is shown as a function of droplet radius $R$. There is a critical radius $R_{\rm c}(t)=\ell_\gamma/\varepsilon(t)$ above/below which a droplet grows/shrinks. As the supersaturation $\varepsilon(t)$ decreases with time, the critical radius increases. 
(c) Frequency of droplets as a function of droplet radius $R$. As the critical radius increases with time, the distribution broadens. Rescaling the radius by $R_{\rm c}(t) \propto t^{1/3}$, leads to a collapse of all droplet radius distributions. }
	\label{fig:ch2_2}
\end{figure*}

Before we consider shape perturbations in the subsequent chapters, we here focus on a spherical droplet, $\mathcal R(\varphi, \theta; t) = R(t)$, in a spherically symmetric system.
To derive its growth dynamics, we employ the \textit{quasi-static approximation}, which assumes that
 the droplet radius varies slowly such that 
transients in the diffusion equation~\eqref{eq:diff_equ_ch2} can be neglected.  
Within this approximation, the diffusion equation~\eqref{eq:diff_equ_ch2} reduces to
a Laplace equation inside and outside the droplet,
\begin{equation}
	\label{eq:laplace}
	0 \simeq \nabla^2 c(r)=\frac{1}{r^2}\frac{\pp}{\pp r}\left(r^2 \frac{\pp c}{\pp r} \right) \, ,
\end{equation}
where we have written the Laplace operator in spherical coordinates considering that there is no polar and azimuthal dependence of the concentration field.
The associated boundary conditions are given by the Gibbs-Thomson relations~\eqref{eq:final_GT_relations} at the droplet interface and no flux conditions at the droplet centre.
Moreover, 
we consider the case where the droplet is embedded in a large system and the concentration far away is fixed to $c_\infty$:
\begin{subequations}\label{eq:boundary_conditions}
\begin{align}
	\pp_r c (r) & = 0 \,  \qquad \text{at} \qquad r=0
	\; ,
\\
	\lim_{r\rightarrow \infty} c (r) & = c_\infty \; .
\end{align}
\end{subequations}
Using these boundary conditions the solutions inside and outside the droplet read
\begin{subequations}\label{eq:csol_inside_outside}
\begin{align}
	\label{eq:diff}
	c(r) & = c_\infty + \left(c_{\rm out}^{\rm eq} -c_\infty\right)\frac{R}{r} \; ,
	& r&> R \; ,
	\\
	c(r) & = c^{\rm eq}_{\rm in} \; , & r&<R \, ,
\end{align}
\end{subequations}
which are illustrated in \figref{fig:ch2_2}(a).
%
%
These solutions imply that the fluxes $\vect j_{\rm in} = \vect 0$ and $\vect j_{\rm out} = DR^{-1}(\cEqOut - c_\infty) \vect e_r$ inside and outside of the interface, respectively.
The  growth rate of the droplet then follows from \Eqref{eq:interface_vel},
\begin{equation}
	\label{eq:dRdt}
	\frac{\dd R}{ \dd t} 
	= 	\frac{D \, \cBaseOut}{R \, \cBaseIn}
		\left(\supSat - \frac{\ell_\gamma}{R} \right)
	\ ,
\end{equation}
where we consider the case of strong phase separation ($\cBaseIn \gg \cBaseOut$). We have defined the \textit{supersaturation}
\begin{equation}
	\supSat  = \frac{c_\infty}{\cBaseOut} - 1 
	\;,
\end{equation}
which measures the excess concentration relative to the equilibrium concentration $\cBaseOut$ in the dilute phase.
Equation~\eqref{eq:dRdt} shows that the droplet only grows in sufficiently supersaturated environments where $\varepsilon > \ell_\gamma R^{-1}$.
The droplet dynamics are thus directly linked to the concentration of droplet material in its environment.
In particular, we can define the  \textit{critical radius} $\Rcrit=\ell_\gamma\supSat^{-1}$, below which the droplet shrinks, see \figref{fig:ch2_2}(b).
Although this deterministic description cannot account for the spontaneous emergence of droplets, the critical radius is key to estimate the frequency of such nucleation events~\cite{Binder_Staufer_76, binder1984nucleation}.
In essence, nucleation relies on large fluctuations that spontaneously enrich droplet material in a region of radius~$\Rcrit$.
In this case, the resulting droplet starts growing spontaneously according to \Eqref{eq:dRdt}.

\subsubsection{Droplet coarsening by Ostwald ripening.}\label{sect:Ostwald_intro}
So far, we focused on a single droplet, but most phase separated systems  contain many droplets.
In such emulsions, large droplets typically grow at the expense of smaller droplets, which vanish eventually.
This phenomena is referred to as {\it Ostwald ripening}~\cite{ostwald1897studien}.

In the following we consider the interactions of many droplets that are far apart from each other in a dilute system with small supersaturation.
In this case, nucleation events are rare and the surrounding of droplets can be considered to be spherically symmetric with a common concentration~$c_\infty$  far away from each droplet.
This implies a common supersaturation~$\supSat$, which depends on time and mediates the interactions between the droplets.
As the supersaturation  $\supSat$ can be determined from the total amount of material, the state of the system is fully specified by the radii~$R_i$ of the $N$ droplets in the system.
Their dynamics follows from \Eqref{eq:dRdt} and reads \cite{Lifshitz_Slyozov_61}: 
 \begin{subequations} 	
\label{eq:multidrop}
\begin{align}
	\difffrac{R_i(t)}{t} &=
		\frac{D \, \cBaseOut}{R_i(t) \, \cBaseIn}
			\left(\frac{\cInfty(t)}{\cBaseOut}-1 -\frac{ \ell_\gamma}{R_i(t)} \right) \, ,
\\
	\label{eq:multidrop_solvent}
	  \bar c \Vsys &=
	 	\cBaseIn \sum_{i=1}^N \frac{4\pi}3 R_i(t)^3
		 \\ \notag & \quad 
		 + c_\infty(t) \left[ V- \sum_{i=1}^N \frac{4\pi}3 R_i(t)^3 \right]  
	\;. 
\end{align}
\end{subequations}
Equation~\eqref{eq:multidrop_solvent} states that the material is shared between the droplets of radius $R_i$ and the dilute phase of concentration $\cInfty(t)$. 
In order to neglect the spatial correlations between the droplets~\cite{yao1992ostwald,yao1993theory},
we assumed that the system volume $\Vsys$ is large compared to all droplets, $\Vsys \gg \sum_i V_i$ with $V_i = \frac{4\pi}3 R_i^3$.
In this limit,  \Eqref{eq:multidrop_solvent} can also be approximated as
$\bar c \Vsys \simeq \cBaseIn \sum_{i=1}^N V_i(t) + c_\infty(t) V$.

In the limit of many droplets, the system can be described by a continuous droplet size distribution.
If additionally the supersaturation is small, Lifshitz and Slyozov~\cite{Lifshitz_Slyozov_61} demonstrated that this size distribution converges to a universal form 
\begin{align}
P(\tilde R) &= \frac{4}{9} \, 
 \tilde{R}^2 \left(1+\tilde{R}/3 \right)^{-7/3} \\
&\quad\times \left(1-2 \tilde{R}/3 \right)^{-11/3}
\exp{\left(-\frac{1}{1-2 \tilde{R}/3} \right)} \, ,
\notag
\end{align}
when normalised by the critical radius~$\Rcrit$, i.e., $\tilde R=R/\Rcrit$,
irrespective of the initial size distribution, see \figref{fig:ch2_2}(c).
In such a coarsening system where droplets grow and shrink, 
the critical radius scales with the average droplet radius, $\Rcrit(t) \simeq \mean{R(t)}$.
Moreover, a droplet radius $R_i$ is typically in the order of 
 the critical radius $\Rcrit(t)=\ell_\gamma/ \left(\frac{\cInfty(t)}{\cBaseOut}-1\right)$.
 Thus, equation~\eqref{eq:multidrop} gives $\difffrac{\Rcrit(t)}{t} \simeq \frac{D \ell_\gamma \cBaseOut} { \cBaseIn \Rcrit^2 }$ leading to the scaling 
 \begin{equation}\label{eq_R_t_Ostwald}
	 \mean{R(t)} \simeq \Rcrit(t) \propto \left(\frac{D\ell_\gamma \cBaseOut}{\cBaseIn} \, t \right)^{\frac13} \, ,
 \end{equation}
which  is the {\it Lifshitz-Slyozov scaling law}. 
In summary, the increasing mean droplet radius and critical radius reflect coarsening dynamics  where large droplets grow at the expense of smaller ones.

\subsubsection{Droplet coarsening by coalescence.}
\label{sect:coalescence}
Another coarsening mechanism in emulsions, besides Ostwald ripening, is the coalescence of droplets driven by their Brownian motion~\cite{siggia_pra79,cates_a12}.  
Brownian coalescence is not included in the theory presented here, since we neglected momentum transport and thermal fluctuations.
However, the evolution of the mean droplet size due to droplet coalescence can be determined by estimating the change in radius $\Delta \mean{R}$ for a typical fusion event and the frequency $\Delta t^{-1}$ of inter-droplet encounters assuming that most encounters lead to coalescence.
Since the droplet volume is conserved during fusion, two equally sized droplet of  size $\mean{R}$ lead to a change of the mean radius  of $\Delta \mean{R} \simeq  \left( 2^{\frac{1}{3}} - 1 \right) \mean{R}$.
The frequency of inter-droplet encounters can be estimated by
the diffusion time leading to $\Delta t^{-1} \simeq \lambda D_R/\ell_\text{p}^2$, where 
$D_R=\kb T/(6 \pi \etaR \mean{R})$ is the Stokes-Einstein diffusion constant of a spherical droplet with
$\etaR$ denoting the viscosity of the surrounding fluid experienced by the droplet of average size $\mean{R}$.
Moreover, 
$\ell_\text{p}=V/\left( \pi N \mean{R}^2 \right)$
 is the mean free path between the droplets, where the droplet number can be estimated as $N \simeq V_\text{tot}/(\frac{4\pi}{3} \mean{R}^3)$ and the volume occupied by droplets $V_\text{tot}$ is determined by particle conservation, i.e.,
$V_\text{tot}/V = (\bar c - \cEqOut)/(\cEqIn-\cEqOut)$. 
Not every inter-droplet encounter leads to a coalescence event in particular in the presence of surfactants~\cite{van2004coalescence, dai2008mechanism}.
To account for the stochastic initiation of a coalescence event we have introduced the parameter $\lambda \in [0,1]$ characterising the average fraction of encounters that lead to coalescence. 
By writing $\mean{R}^{-1} \text{d}\mean{R}/\text{d}t \simeq \mean{R}^{-1} \Delta\mean{R}/\Delta t \simeq \lambda \kb T V_\text{tot}^2/(V^2 \etaR \mean{R}^3)$, skipping the numerical prefactors, 
 we obtain a differential equation for the mean radius $\mean{R}$, where integration gives the scaling for the mean radius arising from fusion of droplets:
\begin{equation}\label{eq_R_t_Coalescence}
	\mean{R(t)} \propto  \left( \lambda \left(\frac{\bar c - \cEqOut}{\cEqIn-\cEqOut}\right)^2 \frac{\kb T}{ \etaR}  \, t \right)^{\frac13} \, .
\end{equation}
Remarkably, the coarsening due to droplet coalescence has the same scaling with time as the growth of droplets by Ostwald ripening described by equation~\eqref{eq_R_t_Ostwald}.

\subsubsection{Comparison between coarsening via Ostwald-ripening and coalescence.}
\label{sect:comparision_Ostwald_coalescence}
We can use our lattice model to determine the relative contributions of the two coarsening mechanisms to the growth of droplets.
In the case of Ostwald ripening, we have to estimate the molecular diffusion constant $D$, the capillary length $\ell_\gamma$ and the relative dilution of the minority phase, $c^{(0)}_{\rm out}/ c^{(0)}_{\rm in}$.
We use the Stokes-Einstein relationship to express the diffusivity as $D \simeq \kb T/ \left(6\pi \eta_{\rm m} \nu^{1/3} \right)$, where  
$\eta_{\rm m}$ denotes the fluid viscosity felt by the molecules of volume $\nu$.
Moreover, from equations~\eqref{eqn:surfacetension}  and \eqref{eq:lc_app}, the capillary length $\ell_\gamma \simeq \frac12 \nu^{1/3} \chi^{1/2}(\chi-2)^{3/2}$.	
Finally,  the binodal line corresponding to our lattice model 
with equal-size molecules $A$ and $B$ (see end of section~\ref{sect:maxwell_contruct_homog})
can be used to estimate the relative dilution of the minority phase. 
It turns out that the fraction between the equilibrium concentrations   $c^{(0)}_{\rm out}/c^{(0)}_{\rm in} \propto \exp(-\chi)$, i.e., it  decreases exponentially to zero  as the interaction strength $\chi$ becomes large (limit of strong phase separation), while $c^{(0)}_{\rm out}/c^{(0)}_{\rm in} \simeq 1-\sqrt{\left(6 (\chi-2) \right)}$   changes only weakly close to the critical point (weak phase separation). 
By comparing equation~\eqref{eq_R_t_Ostwald} to~\eqref{eq_R_t_Coalescence}, we find that Ostwald ripening dominates coarsening if 
 \begin{equation}
 	 \chi^{\frac12}(\chi-2)^{\frac32} \frac{c^{(0)}_{\rm out} } {c^{(0)}_{\rm in} } 	\left( \frac{\cEqIn-\cEqOut}{\bar c - \cEqOut} \right)^2 \frac{\etaR}{\etaM} \lambda^{-1} \gg 1 \, ,
	\label{eqn:coarsening_ratio}
\end{equation}
where we dropped all numerical prefactors.
In the simple case of a constant size-independent viscosity, $\etaM = \etaR$, our estimates from the simple binary lattice model indicate that coalescence  typically dominates Oswald ripening for most interaction parameters $\chi$.
In particular, the left hand side of \Eqref{eqn:coarsening_ratio} goes to zero close to the critical point ($\chi = 2$) and in the limit of strong phase separation ($\chi \rightarrow \infty$).
However, for intermediate $\chi$-values, the dominant coarsening mechanism could still be Ostwald ripening because coalescence events may be suppressed by surfactants  ($\lambda \ll 1$) or when the ratio of the viscosities often satisfies $\etaR/\etaM \gg 1$.
Such different viscosities are particularly relevant for condensed phases in polymer or protein solutions, or droplet-like compartments in living cells.
These complex, phase separated liquids can even show visco-elastic effects leading to a dramatic slow down of the movements of large droplet-like phases~\cite{Tanaka_viscoelastic_Review_2000,Takeaki_viscoelastic_3d_2001,Tanaka_2006}. In particular, inside cells, diffusion of  very large compartments is strongly suppressed by the cytoskeleton~\cite{weiss_biophysj04}, while the diffusion of small molecules may experience less hinderance.
We therefore expect that Ostwald ripening is the dominant mechanism of droplet coarsening inside cells since it relies on evaporation and condensation of small diffusing molecules.


\section{Positioning of condensed phases}\label{sect:phase_sep_in_gradients}

In this chapter, we discuss the positioning of condensed phases (\eg droplets) by external fields and non-equilibrium concentration gradients.
We focus on the case where two components phase separate while a third  component, referred to as regulator, influence the phase separation. 
Here we discuss two scenarios of how to affect the position of condensed phases: 
(i) The position of a condensed phase can be influenced by an external field such as gravitation, electric or magnetic fields~\cite{julicher2009generic}. These fields position the phase of higher mass density, larger charge or larger magnetic moment toward regions of lower potential energy. The resulting stationary states correspond to a minimum of the total free energy of the system and are inhomogeneous thermodynamic states. 
 (ii) Positioning of a condensed phase can also be affected by a regulator  gradient that is driven and maintained by boundary conditions. The presence of a regulator flux may let the system settle in  (stationary)  non-equilibrium states. 
Such a concentration gradient could be generated for example  by  concentration boundary conditions, or  sources and sinks~\cite{saha2016polar}, or via position-dependent reaction kinetics with broken detailed balance~\cite{tenlen08, griffin11}. 
 
In  section~\ref{sect:stat_states}, we discuss a simple 
system of two phase separating components and illustrate how  an external field can affect the average position of the phase separated concentration profiles. The corresponding stationary states are inhomogeneous thermodynamic states and can thus be accessed through a minimisation of the free energy.
Section~\ref{sect:droplet_ripening} is then devoted to discuss how a concentration gradient of a regulator can affect the dynamics of droplet position.

\subsection{Positioning of condensed phases by external fields}\label{sect:stat_states}

External fields can influence the position of components in a mixture and thereby also the position of condensed phases. In this section, we investigate how external fields affect a mixture which undergoes phase separation. 
To this end, we briefly review the thermodynamics with external fields.

\subsubsection{Thermodynamics of binary mixtures in external fields.}

Here we discuss the thermodynamics of binary mixtures in the presence of external fields such as gravitation with a gravitational acceleration $g$, and electric or magnetic, position-dependent  potentials denoted as $U(x)$.
The presence of such an inhomogeneous external potential can influence the shape and the mean position of the concentration profiles $\A (x)$ and $\B (x)$, which can be defined as 
\begin{equation}
	x_{i}= 	 \frac{1}{L} \int_0^{L} \text{d}x \, x \, \frac{c_i(x)}{ \bar c_i } \, ,
\end{equation}
where $\bar c_i = L^{-1} \int_0^{L} \text{d}x \, c_i(x)$ denotes the mean concentration and $L$ is the size of the system.

For a compressible system the binary mixture is described by two concentration fields $\A$ and $\B$. Considering the case where the external fields vary along the $x$-coordinate, the total free energy density reads
\begin{align}	\notag
	\fetot&= \fe\left( \A, \B,  \nabla \A, \nabla \B \right)\\
	\label{eq:free_energy_density}
	& \quad + \rho g x + U_A(x) \A + U_B(x) \B 
	\, .
\end{align} 
The interactions between the components are governed by the free energy density $f$,
 $\rho=m_A \A+ m_B \B$ is the mass density,  and $m_A$ and $m_B$ denote the molecular mass of each component.
The  contributions of the external potentials can be combined
to  $\tilde{U}_A(x) \A  + \tilde{U}_B(x) \B $, where $\tilde{U}_i(x)= U_i(x)+ m_i g x$, $i=A,B$. 

Thermodynamic equilibrium for systems with external fields can be defined at each position. 
The position-dependent equilibrium profile $c_i(x)$ is then determined by a spatially constant generalised  chemical potential,
${\mu}_{{\rm tot},i}=\mu_i(x) + \tilde{U}_i(x)$, where  $\mu_{i}(x)=\delta F/\delta c_i$ with $F=\int \text{d}^3 x f $.
As an example of a system with such a position-dependent equilibrium profile 
we consider 
an incompressible system  ($\nu_A \A+ \nu_B \B=1$) with gravitation as the only external potential
and where 
the $A$-molecules are dilute, $\nu_A \A \ll 1$~\cite{julicher2009generic}.
 The generalised exchange chemical potential then reads
$\bar{\mu}_{\rm tot}  \simeq k_{\rm B}T \ln (\nu \A) + \nu_A \Delta \rho g x$ with the density difference $\Delta \rho = m_A/\nu_A-m_B/\nu_B$.
At thermodynamic equilibrium, this gives the barometric height formula, $\A(x)\propto \exp({ -\Delta \rho \nu_A g x / ( k_{\rm B}T) })$. 
Thus gravitation always positions the molecules of highest mass density toward the lower gravitational potential.
A similar result occurs if the binary mixture phase separates. In this case the condensed phase of larger mass density is  positioned toward the region of lower gravitational potential.
However, for fixed difference in mass density and gravitational potential, there is no way to switch the concentration profiles or the position of the condensed phases with respect to gravitation. 
In order to create a possibility to switch the position of condensed phases, we ask what happens if 
 an additional component that affects phase separation is added to the system and subject to an external potential?
Specifically, we wonder what are the stationary concentration profiles and which physical parameters control their positions?

\subsubsection{Positioning of condensed phases by a regulator potential.}\label{sect:lift_of_deg}

\begin{figure*}[tbh]
\centering
\includegraphics[width=170mm]{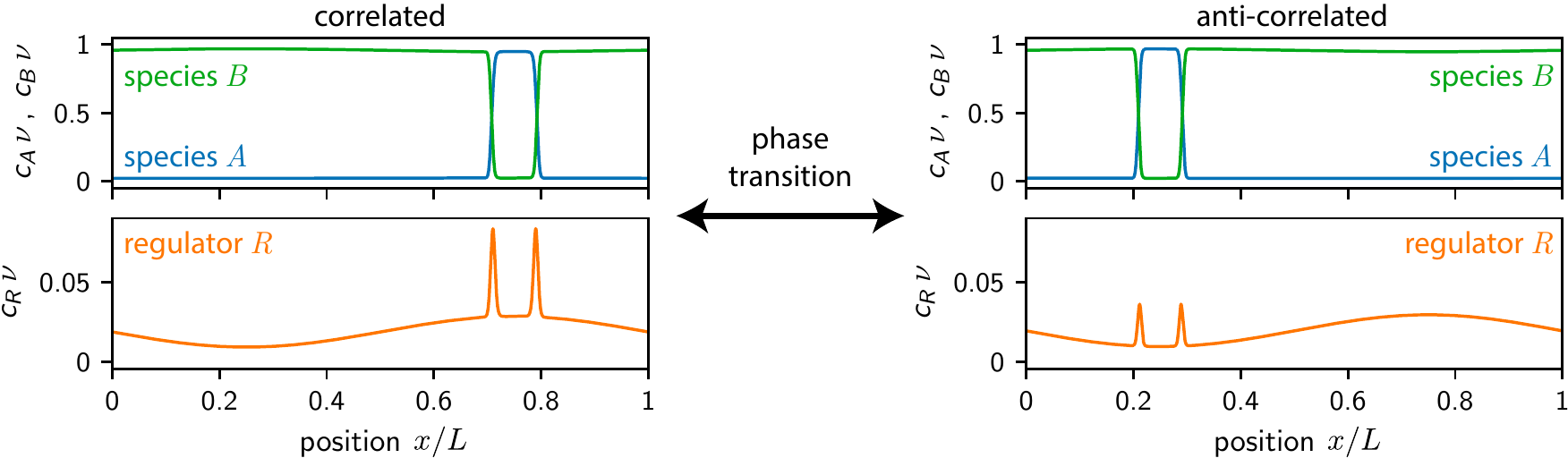} 
\caption{\label{fig:Fig_3_profiles} 
Spatial regulation of phase separation in an external potential by a discontinuous phase transition. The regulator forms a
spatially inhomogeneous profile due to an external potential. As the interactions
with the regulator are changed, the spatial distribution of component
$A$  switches from a spatially correlated (left) to an anti-correlated (right) distribution with respect to the regulator. 
The switch corresponds to a discontinuous phase transition.
}
\end{figure*}

To explore the propensity to switch the position of condensed phases using such an additional component we propose a simple ternary model~\cite{kruger2017switching}.
This model accounts for the demixing of two components, $A$ and $B$, and a regulator component $R$ that interacts with the other components and thereby affects phase separation between $A$ and $B$. 
The regulator component $R$ is influenced by an external potential $U(x)$.
Interactions between the components $i$ and $j$ are captured by the mean-field interaction parameters $\chi_{ij}$. 
The scenario of regulation of phase separation
can be described by the following free energy density 
\begin{align}\nonumber
	 \fetot  &=\kb T \bigg[ \sum_{i={\rm{A,B,R}}} c_i \ln (\nu c_i) +  \chi_{AB} \, \nu\,\A \B \\
	 	\nonumber
	 &\quad + \R (\chi_{BR} \,  \B+\chi_{AR} \A) \, \nu  \bigg] + U(x) \R \\
	  \label{eq:free_energy_ternary_functional_theory_hom}
	 & \quad + \frac{\kappa_R}{2}\left| \nabla \R \right|^2+\frac{\kappa_A}{2}\left| \nabla \A \right|^2 \, ,
\end{align}
which is a Flory-Huggins free energy density~\cite{flory1942thermodynamics,huggins42} for three components.
Analogously to the case of the binary system discussed in section~\ref{sect:bin_phasesep}, the ternary free energy given above can be derived from a partition sum using a mean-field approximation (see \cite{lajzerowicz1975spin,sivardiere1975spin} and Appendix of Ref.~\cite{kruger2017switching}).
In equation~\eqref{eq:free_energy_ternary_functional_theory_hom}  we consider an incompressible system where the molecular volumes are constant 
and equal to $\nu$ for all components, and the concentrations thus obey $\B=\nu^{-1}-\R-\A$. 
The logarithmic terms in equation~\eqref{eq:free_energy_ternary_functional_theory_hom} correspond to entropic contributions  related to the number of possible configurations.
The remaining contributions characterise the interactions between the three components with the (dimensionless) mean field interaction parameter $\chi_{ij}$, also referred to as Flory-Huggins interaction parameter. The interaction parameter between $A$ and $B$, $ \chi_{\rm{AB}}$, determines the tendency of  $A$ and $B$ to phase separate. 
The two terms in equation~\eqref{eq:free_energy_ternary_functional_theory_hom} proportional to the regulator concentration $\R$ describe the interactions between the regulator $R$ and the demixing  components $A$ and $B$. To ensure that $R$ acts as a regulator we choose these interaction parameters such that the regulator $R$ does not demix from $A$ or $B$. 
The  terms in equation~\eqref{eq:free_energy_ternary_functional_theory_hom} with spatial derivatives represent contributions to the free energy associated with spatial inhomogeneities~\footnote{We neglected a mixed term proportional to $\nabla \A \cdot \nabla \R$ since it has only little quantitative impact on the spatial profiles of the phase separated profiles~\cite{kruger2017switching}}.

In the following we consider
a periodic system with a periodic potential $U(x)$ that varies solely along the $x$-coordinate.
We choose
a potential that affects the distribution of the regulator component  of the form
\begin{equation}\label{eq:potential} 
U(x)= - \kb T \ln \left( 1 - Q \sin{\left(2 \pi x/L \right)}\right) \, ,
 \end{equation} 
 where $0<Q<1$ characterises the strength of the potential and $L$ denotes the size of the system along the $x$-direction. 

 In the case where  the components $A$ and $R$ are dilute,  $\nu \Ahom \ll 1$ and $\nu \Rhom \ll 1$, and for weak  external potentials ($\kappa_R (Q/L)^2 \ll 1$) such that the the gradient terms in free energy can be neglected, the profile of the regulator component is solely given by the external potential $U(x)$ with the regulator profile $\R(x)$  assuming the shape of negative sine function, $-\sin{\left(2 \pi x/L \right)}$. 
Thus the regulator profile has  one minimum and one maximum in the periodic domain.
 The interaction of such a regulator profile with the components $A$ and $B$
 causes a positional dependence of their concentration profiles.
 We would like to understand how these interactions affect the system if $A$ and $B$ phase separate.

For simplicity, we also consider a  one dimensional system of size $L$  in the absence of  boundaries.
In this one dimensional system the periodic boundary conditions are $c_i (0)= c_i (L)$ and $c_i' (0)= c_i' (L)$, where the primes denote spatial derivatives. For the considered case of an external potential $U(x)$ varying only along the $x$-coordinate,
the restriction to a one dimensional, phase separating system 
represents a valid approximation for large system sizes,  where the interface between the condensed phases becomes flat.

\subsubsection{Minimisation of free energy.}\label{sect:min_free_energy}

To find the equilibrium states in a phase separating system in the presence of a regulator gradient induced by the external potential $U(x)$, we determine the concentration profiles $c_i(x)$ of all components $i=A,B,R$ by minimising the total free energy (equation~\eqref{eq:free_energy_ternary_functional_theory_hom}; see Ref.~\cite{kruger2017switching}).
Due to particle number conservation, there are two constraints for the minimisation imposing  $\bar c_i = L^{-1} \int_0^L \text{d}x \, c_i(x)$ for $i=A,R$, 
where  $\bar c_i $ are the average concentrations and $\bar c_B =\nu^{-1}-\Ahom-\Rhom$. 
Variation of the total free energy with the constraints of particle number conservation implies ($i=A,R$): 
\begin{equation}\label{eq:minimization_2}
	0=  \int_{0}^{L}\text{d}x \,  \left( \frac{\partial \fetot}{\partial c_i} -
	 \frac{\text{d}}{\text{d}x} \frac{\partial{\fetot}}{\partial c_i'}
 + \lambda_i \right) \delta c_i  + \kappa_i \delta c_i  c_i' \bigg{|}_0^L\, ,
\end{equation}
where $\lambda_R$ and $\lambda_A$ are  Lagrange multipliers and the $\delta c_i$ is the variation of the concentration corresponding to component $i$.
The boundary term in equation~\eqref{eq:minimization_2} is zero in case of periodic boundary conditions. 
Using the explicit form of the free energy density (equation~\eqref{eq:free_energy_ternary_functional_theory_hom}), 
a set of  Euler-Lagrange equations can be derived from equation~\eqref{eq:minimization_2} (see Ref.~\cite{kruger2017switching}).
These equations  can be solved numerically using a finite difference solver but also approximately investigated analytically (see section~\ref{sect:analytic_argument}). 
As control parameters we consider the three interaction parameters 
$\chi_{AR}$, $\chi_{AB}$ and $\chi_{BR}$, the strength of the external potential $Q$ and the mean concentration of $A$-material, $\Ahom$.
The mean concentration of the regulator material is fixed to a small volume fraction $\nu \Rhom =0.02$ in all presented studies to avoid phase separation of the regulator.
Moreover, we focus on the limit of strong phase segregation, where the interfacial width determined by $\kappa_A$ are small compared to the system size.
We verified that our results depend only weakly
on the specific values of $\kappa_A$ and $\kappa_R$.

\begin{figure*}[tbh]
\centering
\includegraphics[width=170mm]{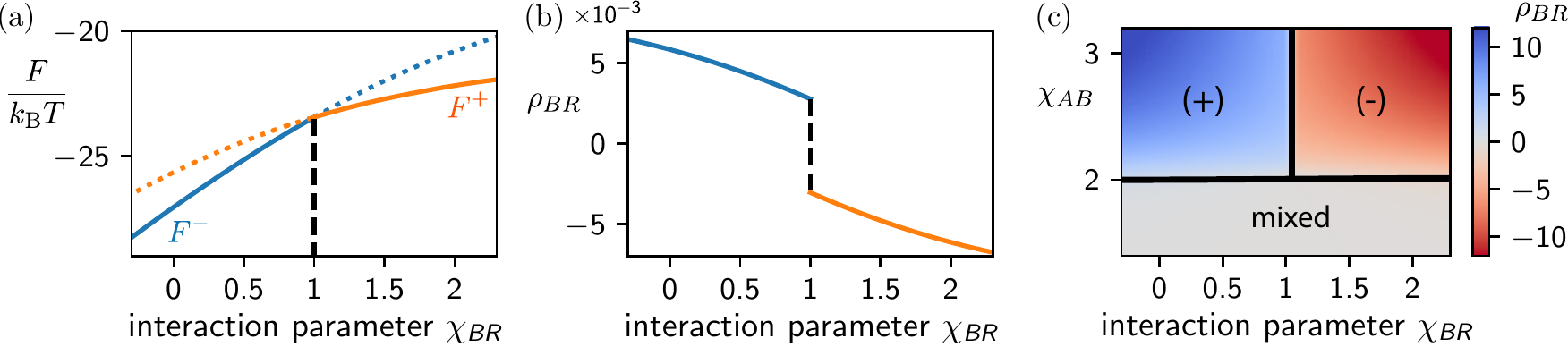} 
\caption{\label{fig:Fig_3_phasetransition} 
Discontinuous phase transition of a ternary phase separating systems in a periodic potential and with periodic boundary conditions.
(a) Free energy $F$ as a function of the $B$-$R$ interaction parameter $\chi_{BR}$.
$\Fl$ and $\Fr$ are the free energies of the correlated and anti-correlated 
stationary solution with respect to the regulator gradient, respectively. 
Lines are dashed when solutions are metastable. 
At $\chi_{BR}^*$, $\Fl$ and $\Fr$ intersect and the solution of minimal free energy exhibits a kink. 
This shows that the transition between  correlation and anti-correlation
 is a discontinuous phase transition.
(b) The order parameter $\rho_{BR}$ corresponding to the solution of minmimal free energy
jumps at $\chi_{BR}^*$.
(c) Phase diagrams of our ternary model for spatial regulation in a periodic potential and periodic boundary conditions ($\nu \Ahom=0.1$). 
The color code depicts  the order parameter $\rho_{BR}$.
Component $B$ is spatially correlated ($+$) with the regulator profile if 
$\rho_{BR}>0$, and anti-correlated ($-$) otherwise. When the system is mixed,  $\rho_{BR} \approx 0$, and spatial profiles of all components are only weakly inhomogeneous (no phase separation).
The black lines corresponds to the transition  where the free energy has a kink.
Parameters for (a-c):  $\chi_{AR}=1$, $\nu \Ahom=0.5$, $\nu \Rhom=0.02$, $\kappa_R/(\kb T \nu L^2)=7.63 \cdot 10^{-5}$, $\kappa_A/(\kb T \nu L^2)=6.10 \cdot 10^{-5}$,
$Q=0.5$. For (a) and (b), $\chi_{AB}=4$.
For plotting, $\nu=L/256$ was chosen.
}
\end{figure*}

\subsubsection{Discontinuous switching of average position of phase separated concentration profiles in external fields.}

Solving the Euler-Lagrange equations, we find two 
extremal profiles of the phase separating component $A$,  $\Al(x)$ and  $\Ar(x)$,
and  two corresponding profiles of the regulator component $R$, 
denoted as $\Rl(x)$ and $\Rr(x)$ 
(the profile of $B$ follows from number conservation). 
 The phase separating material $A$  can be accumulated at larger regulator concentration  and correlates ($+$) with the concentration of the regulator material (figure~\ref{fig:Fig_3_profiles}(a)). The corresponding solutions are $\Ar(x)$ and $\Rr(x)$.
 Alternatively,  the $A$-material accumulates at  smaller regulator concentrations ($\Al(x)$ and $\Rl(x)$)  corresponding to an anti-correlation ($-$) with respect to the regulator profile (figure~\ref{fig:Fig_3_profiles}(b)).
The free energies of the correlated and the anti-correlated states, $\Fr=F[\Ar, \Rr]$ and $\Fl=F[\Al, \Rl]$,  are different for most interaction parameters. 
 The free energies only intersect at one point $\chi_{BR}=\chi_{BR}^*$ (figure~\ref{fig:Fig_3_phasetransition}(a)).
At this point the minimal free energy exhibits a kink.
This means that the system undergoes a {\it discontinuous phase transition} when switching between a spatially anti-correlated ($-$) and a spatially correlated ($+$) solution with respect to the regulator.

To study this phase transition
 the appropriate set of  order parameters can be defined from the changes of the free energy upon varying the interaction parameters (see figure~\ref{fig:Fig_3_phasetransition}):
\begin{align}
\label{eq:order_parameter2}
	\rho_{ij} &= \left( \kb T  \mathcal{N}_{ij} \nu L \right)^{-1} 
	\frac{\text{d}}{\text{d}\chi_{ij}} 
	\Delta F(c_i(x),c_j(x))  \, ,
\end{align} 
where 
$\Delta F(c_i(x),c_j(x))=F(c_i(x),c_j(x)) - F(\bar c_i, \bar c_j)$.
The  normalisation
$\mathcal{N}_{ij}$ is chosen such that  $-1<\rho_{ij}<1$.
When
inserting  equations \eqref{eq:free_energy_ternary_functional_theory_hom}, the order parameter $\rho_{ij}$ 
becomes the covariance,
\begin{align}
\label{eq:order_parameter}
	\rho_{ij} &=\left(\mathcal{N}_{ij} L\right)^{-1} \int_0^L \text{d}x \,  \left( c_i(x) c_j(x) - \bar c_i \bar c_j\right) \, ,
\end{align} 
which characterises the spatial correlation between the concentration profiles $c_i(x)$ and $c_j(x)$. 
If the fields are spatially correlated ($+$), $\rho_{ij}>0$, and if they are anti-correlated ($-$), $\rho_{ij}<0$,
 and $\rho_{ij}=\pm1$ if the  concentration profiles of component $i$ and $j$  follow spatially correlated or anti-correlated step functions. 
 If the regulator is homogeneous, $c_R(x)=\bar c_R$, the order parameter is zero, $\rho_{iR}=0$ for $i=A,B$.

Varying the interaction parameter $\chi_{BR}$  (figure~\ref{fig:Fig_3_phasetransition}(b)), the order parameters $\rho_{BR}$ and  $\rho_{AR}$ jump at the threshold value $\chi_{BR}^*$.
 The jump of both order parameters in the presence of a regulator gradient indicates that the spatial correlation of $A$ and $B$ with respect to $R$ changes abruptly, which is expected in case of a first order phase transition.

 By means of the order parameter 
$\rho_{BR}$ (equation~\eqref{eq:order_parameter}) we can now discuss the phase diagrams as a function of the interaction parameters.
In the case of a spatial correlation ($+$), we have $\rho_{ij}>0$, while for an anti-correlation ($-$), $\rho_{ij}<0$.  We thus find three regions (figure~\ref{fig:Fig_3_phasetransition}(c)): 
A mixed region, where concentration profiles are only weakly inhomogeneous and no phase separation occurs, and two additional regions, where components $A$ and $B$ phase separate and $A$ is spatially correlated or anti-correlated with the regulator $R$, respectively.  There exists a triple point where all three states have the same free energy.

In summary, the presence of a concentration gradient in phase separating systems leads to  equilibrium states of different spatial correlation with the regulator profile. 
The regulator gradient creates a bias in the position of the phase separated concentration profiles for almost all parameters in the phase diagram. 
If the external potential acting on the regulator has exactly one minimum and one maximum, 
there are two stationary states with different mean positions of the phase separating material. 
One of these stationary states corresponds to a global minimum of the free energy while the other state may only be locally stable.
The parameters characterising the interactions between the regulator and the phase separating material determine which of these states  corresponds to  equilibrium.
There is a  {\it discontinuous phase transition} between both states upon changing these interaction parameters.  

For simplicity we have discussed a  phase separating system in the presence of an external potential restricting  to  inhomogeneities in one dimension. 
However, preliminary Monte-Carlo studies in three dimensions with phase separated  droplets in a regulator gradient suggest that 
the position of droplets can be switched in a discontinuous manner~\cite{in_prep}.

\subsubsection{Analytic argument of the occurrence of a discontinuous phase transition}
\label{sect:analytic_argument}

In the previous section we have considered the numerical minimisation of
 a set of non-linear Euler-Lagrange equations derived from the free energy density (equation~\eqref{eq:free_energy_ternary_functional_theory_hom}). 
 Here we give some approximate analytic arguments to understand the minimal ingredients for the occurrence of the   
discontinuous phase transition. To this end, we would like to simplify the system further 
and consider 
the dilute limit of the regulator, i.e.,
$ \nu \Rhom \ll 1$ and thus approximate $\bar c_B \simeq 1 - \nu^{-1} \Ahom$ in equation~\eqref{eq:free_energy_ternary_functional_theory_hom}.
For such dilute conditions, the equilibrium concentrations (for $A$ component) in each phase, $\cBaseIn$ and $\cBaseOut$, are then well given by 
the binary $A$-$B$ system. In addition, for strong enough external potential $U(x)$,
the regulator profile follows well, $\R(x)= \tilde A - \tilde  B \sin{\left(2 \pi x/L \right)}$, apart from the peaks (see figure~\ref{fig:Fig_3_profiles}), where $\tilde A>0$ is some concentration offset and $\tilde B>0$ characterises the strength of the spatial modulations of the regulator profile. For a discussion of the relevance of the regulator peaks at the interface see Ref.~\cite{kruger2017switching}; here we simply neglect these peaks for simplicity. 
We expect that the extrema of the regulator profile  $\R(x)$ at $x=L/4, 3L/4,$ determine the position of the $A$-rich condensed phase. As obtained from the numerical analysis presented in the last section,
there are two solutions, either spatially correlated ($+$) or anti-correlated ($-$) with the regulator. These solutions can be approximated in the dilute limit of the regulator as:
 \begin{subequations} 	
\begin{align}
\notag
	\A^+(x) &\simeq \cBaseOut + \left( \cBaseIn-  \cBaseOut \right) \Theta \left( x- 3L/4 + x_0/2 \right) \\
	& \quad \times \Theta \left(  3L/4 + x_0/2  -x  \right) \, ,\\
	\notag
	\A^-(x) &\simeq \cBaseOut + \left( \cBaseIn-  \cBaseOut \right) \Theta \left( x- L/4 + x_0/2 \right) 
	\\
	& \quad \times 
	\Theta \left(  L/4 + x_0/2  -x  \right) \, ,
\end{align} 
 \end{subequations} 	
 where $\Theta(\cdot)$ denotes the Heaviside step function. These solutions describe the $A$-rich domains either localised around the maximal ($+$) or minimal ($-$) amount of regulator.
The domain size of the $A$-rich phase denoted as $x_0$ is determined by conservation of particles, i.e., $\Ahom L= \int \text{d} x \A^+(x) = \int \text{d} x \A^-(x)$, leading to $x_0=L (\Ahom-\cBaseOut)/(\cBaseIn-\cBaseOut)$.
Using the approximate solutions above we can calculate the difference in free energy 
corresponding to correlated and anti-correlated states,
\begin{equation}\label{eq_deltaF_analytic}
\Fr - \Fl \simeq W \,{\kb T \, L} \left( \chi_{AR} - \chi_{BR} \right) \tilde B   \, ,
\end{equation}
where $W= 2 \pi^{-1} 
\sin \left(  \pi \frac{x_0}{L}\right)>0$
is a positive constant. Please note that all contributions apart from the 
$A$-$R$ and $B$-$R$ interaction vanish in the dilute limit and due to conservation of $A$ and $B$ material.
As $B$ characterises the concentration modulations of the regulator profile,  
the free energy difference $(\Fr - \Fl)$ consistently vanishes for zero $\tilde B$ (equation~\eqref{eq_deltaF_analytic}). 
In the case of non-zero $\tilde B$, the free energy difference is determined by the difference in the interactions of $A$ and $B$ with respect to the regulator $R$, $\Delta \chi =\chi_{AR} - \chi_{BR}$. 
Most importantly, at $\Delta \chi=0$, the two solutions switch their thermodynamic stability: for $\Delta \chi>0$, the anti-correlated state is favoured, while for $\Delta \chi<0$, the correlated state is preferred; $\Delta \chi=0$ indicates the transition point.
In addition, the slopes of correlated and anti-correlated free energy, $\Fr$ and $\Fl$,
with  respect to $\Delta \chi$ at the transition point are not equal. 
The difference in slopes implies that the minimal free energy exhibits a kink at the transition point $\Delta \chi=0$, which means that the system undergoes a discontinuous phase transition switching from a correlated to an anti-correlated state for increasing $\Delta \chi$. 
The discontinuous phase transition occurs at $\Delta \chi=0$, which agrees with the numerical predictions  shown in figure~\ref{fig:Fig_3_phasetransition}(c) (note that the corresponding volume fraction of the regulator is rather dilute). 
Our approximate analytic treatment indicates that the occurrence of a discontinuous phase transition solely relies on the existence of the position-dependent profile of the regulator and the interactions of the regulator molecules with the liquid condensed phases. 
It seems that the peaks of regulator material at the interface of the condensed phase, which we neglected in this approximate analytic argument,  are not necessary to observe the discontinuous transition. Maybe the discontinuous nature is also preserved if the regulator gradient is driven by boundary conditions? We leave this question to future research.

\subsection{Dynamics and coarsening of droplets in concentration gradients}\label{sect:droplet_ripening}

In this section we discuss the dynamics of multiple droplets in a one-dimensional gradient of a regulator component that affects the phase separation of droplets. 
For simplicity, we consider the case where  
the regulator profile is not affected by the phase separating components. 
Given a  regulator gradient $\R(x)$ we introduce a set of  physical quantities such as the position dependent supersaturation, which determine the inhomogeneous ripening dynamics. 
These quantities depend on position and will be used to develop a generic theory of droplet ripening in concentration gradients. This theory extends the classical laws of droplet growth derived by Lifschitz \& Slyozov~\cite{Lifshitz_Slyozov_61} and  
can explain the positioning of droplets in concentration gradients by droplet drift and spatially dependent growth.

\subsubsection{Spatially varying supersaturation.}

Here we discuss the ripening dynamics of droplets in a  regulator gradient that varies only along the $x$-coordinate.
To this end, we modify the concept of a common far field concentration introduced in section~\ref{sect:Ostwald_intro} to a concentration field $\cinf(x) $ that changes on the length scale of the system size $L$.

In the absence of a regulator gradient, the concentration outside  approaches the ``far field'' of the droplet, $\cinf$, 
as the distance to the droplet interface  increases. 
The far field is created by the surrounding droplets and is well reached 
if the length scale corresponding to the mean distance between droplets~$\ell$ exceeds the droplet radius $R$, \ie $\ell \gg R$. 

In the presence of a regulator gradient varying along the
$x$-coordinate, the far field seen by the droplet $\cinf(x)$ is now also position dependent
  and can be approximately written as: 
\begin{equation}
	\cinf(x) \simeq \frac{1}{L_y L_z  }  
	 \int_{0}^{L_y} \text{d}y \int_{0}^{L_z} \text{d}z \,  \cOut_\text{out}(x,y,z) \, ,
\end{equation}
where $\cOut_\text{out}(x,y,z)$ is the concentration field outside the droplets 
and $L_y$ and $L_z$ denote the system size in the $y$ and $z$-direction, respectively. 
The expression above is an approximation because we have neglected 
weak concentration perturbations close to droplet interfaces described by the Gibbs Thomson relationship~\eqref{eq:final_GT_relations}. 
However, for the typical case of droplet radii exceeding the capillary length ($R \gg \ell_\gamma$) and the mean inter-droplet distance being larger than the droplet size ($\ell \gg R$), these  concentration perturbations are very small (see section~\ref{sect:curvature_droplet}).

Finally, to make sure that the spatial variations of the position dependent far field are large on the system size but comparably small on the droplet scale, 
 we consider the case where all these length scales separate, $\ell_\gamma \ll R \ll \ell \ll L$. 
 This separation of length scales will allow us to investigate weak perturbations of the droplet shape parallel to the concentration gradient 
  and also to construct the equilibrium concentration at each position $x$ along the gradient.

\begin{figure}[tb]
\centering
\begin{centering}
\includegraphics[width=1.0\columnwidth]{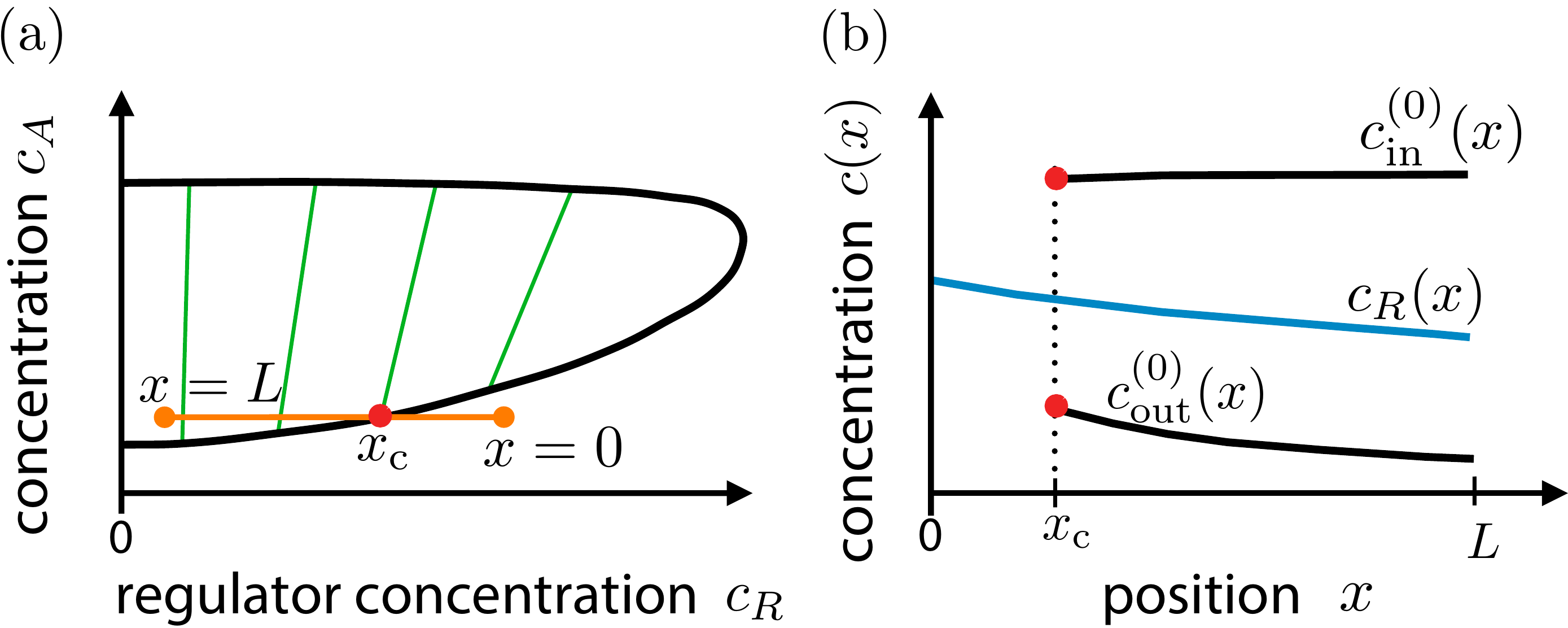} 
\end{centering}
\caption{\label{fig:Fig_3_sketch_phasediagram_and_equil_conc} 
(a) Sketch of a ternary phase diagram as a function of the homogenous regulator concentration $c_R$ and the droplet material $c_A$. The tie lines (green) connect the equilibrium concentration values of the coexisting phases. 
Each position $x$ of a system subject to a different regulator gradient (e.g.\ blue line in (b)) may correspond to a point in the phase diagram along the orange line.  
 If the position is inside the phase separation region, droplets can form, while outside phase separation is absent. 
The position $\xd$, referred to as dissolution boundary,  marks the location below which there is no phase separation and vice versa. 
 (b) Sketch of a representative regulator gradient $c_R(x)$ (blue) and 
 the equilibrium concentration  
 $\ceqbaseIn(x)$ and $\ceqbase(x)$ obtained from the phase diagram (a).
}
\end{figure}

The separation of length scales suggests to divide the system into independent slices of a size corresponding to the intermediate length scale $\ell$. 
The  phase separation dynamics can then be discussed locally
for each position $x$ corresponding to a slice element of linear length $\ell$.
For this discussion, we consider a simplified model with a free energy density given by \Eqref{eq:free_energy_ternary_functional_theory_hom} and calculate the corresponding phase diagram (figure~\ref{fig:Fig_3_sketch_phasediagram_and_equil_conc}(a)).
If the droplet material  is roughly constant each position~$x$ maps on a single point in the phase diagram because the regulator profile is fixed.
The corresponding curve in the phase diagram due to the spatial dependence of the regulator  
defines local values of the equilibrium concentrations inside and outside of the droplet (figure~\ref{fig:Fig_3_sketch_phasediagram_and_equil_conc}(b)). 
In other words, droplets in the slice corresponding to the position $x$ feel the local equilibrium concentrations, $\ceqbaseIn(x)$ and $\ceqbase(x)$.
For simplicity, we restrict ourselves to a special case where the equilibrium concentration inside is position independent, $\ceqbaseIn(x)\simeq \ceqbaseIn$, thus droplet growth is solely determined by the conditions outside of the droplet. 
Moreover, as concentration inhomogeneities of droplet material are in general weak,
we do not consider weak transients of $\ceqbase(x)$ due to the space and time varying $\cinf(x)$.
The actual concentration of droplet material outside, $\cinf(x)$,
together with the equilibrium concentration outside, $\ceqbase(x)$, 
determine a spatially dependent  supersaturation
\begin{equation}\label{eq:supersaturation}
	\varepsilon(x)=\frac{\cinf(x)}{\ceqbase(x)}-1 \, .
\end{equation}
There is a dissolution boundary located at the position $x=\xd$ where the supersaturation  
 $\varepsilon(\xd)=\ell_\gamma/R$.
 For the  considered case of $\ell_\gamma \ll R$, this boundary approximately  corresponds to a vanishing supersaturation $\varepsilon(\xd)\simeq 0$.
 In the illustration in figure~\ref{fig:Fig_3_sketch_phasediagram_and_equil_conc}(a), the fluid is mixed for $x<\xd$, while droplets can form ($\varepsilon>0$) for $x>\xd$. 
In the absence of droplets, the concentration field $\cinf(x)$ evolves in time satisfying a diffusion equation, which we will derive in the next section. 
When droplets are nucleated, their local dynamics of growth or shrinkage is guided by the local supersaturation $\varepsilon(x)$ as well as $\ceqbase(x)$ (see section~\ref{sect:Ostwald_intro}).
This local droplet dynamics then in turn also influences the concentration field $\cinf(x)$.
As time proceeds, diffusion of droplet material occurs on length scales larger than the intermediate length scale $\ell$. For this regime, we will derive a dynamical theory and extend the Lifschitz \& Slyozov theory to concentration gradients.

\subsubsection{Dynamics of a single droplet in a concentration gradient.}

A regulator concentration gradient generates 
 a position-dependent 
 equilibrium concentration $\ceqbase(x)$ and a position-dependent supersaturation $\varepsilon(x)$ (equation~(\ref{eq:supersaturation})). 
This  supersaturation will drive the droplet dynamics and lead to a position-dependent growth,  drift of droplets and even deformations of their shape.
 In the following we  discuss  the dynamics of growth of a single droplet where 
 the equilibrium concentration, $\ceqbase(x)$, and the concentration of droplet material, $\cinf(x)$, are position dependent. The case of multiple droplets is studied in the next section.

\begin{figure}[tb]
\centering
\includegraphics[width=1.0\columnwidth]{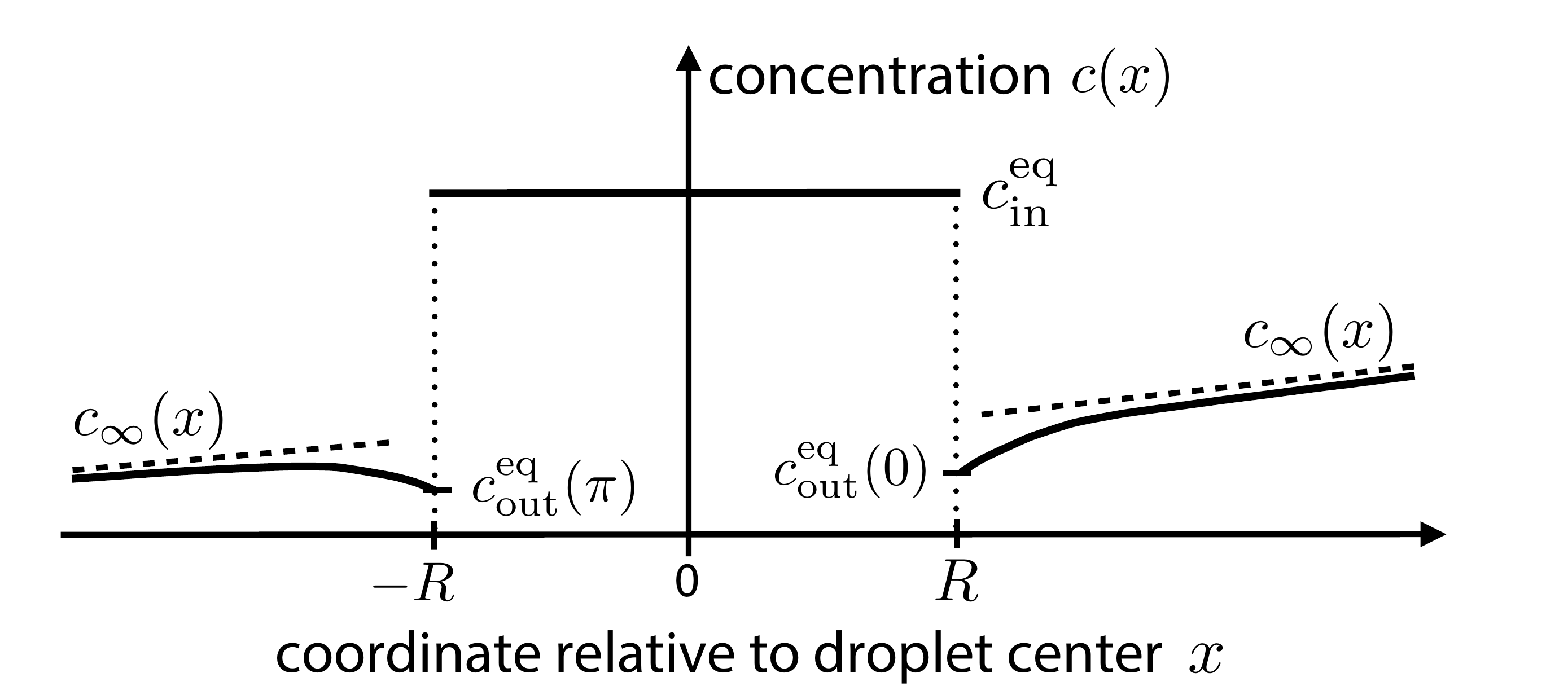} 
\caption{\label{fig:Fig_3_sketch_length_scales} 
Sketch of the concentration field inside and outside of a droplet in a position-dependent supersaturation field. 
The droplet center is located at $x=0$. The equilibrium concentration inside is $\cEqIn$. Right outside the droplet  at $\pm R$  the concentration is given by the Gibbs-Thomson relation $\cEqOut$ (equation~\eqref{eq:bc1}). 
Far away from the droplet center, the
concentration approaches 
$\cinf(x)$ (equation~\eqref{eq:bc2}).
}
\end{figure}

The  concentration inside the droplet 
can be approximated by the equilibrium concentration
$\cEqIn \simeq \ceqbaseIn$ in the limit of strong phase separation ($c^{(0)}_{\rm in} \gg c^{(0)}_{\rm out}$; see section~\ref{sect:curvature_droplet}). 
This allows us to restrict the analysis to the concentration field~$\cOut(r,\theta,\varphi)$ outside of a  droplet.
Here, we use spherical coordinates centred at the droplet position $x_0$,
with  $r$ denoting the radial distance from the centre, and $\theta$ and $\varphi$ 
are the azimuthal and polar angles relative to the $x$-axis.
Within the quasi-stationary approximation (see section~\ref{sect:intro_growth_single_droplet}) the concentration outside but near the droplet  obeys the steady state
of a diffusion equation~\eqref{eq:laplace}.
For large $r$
the concentration field approaches the ``far field''  
which in the presence of a linear gradient reads
(figure~\ref{fig:Fig_3_sketch_length_scales}): 
\begin{equation}\label{eq:bc1}
\cinf = \lim_{r\to\infty} \cOut(r, \theta, \varphi) \simeq \al + \be  \, r \cos \theta  \, .
\end{equation}
This inhomogeneous far field concentration is locally (with respect to inter-droplet distance $\ell$) characterised by the concentration  $\al=\cinf(x_0)$ and the gradient $\be=\partial_x\cinf(x)|_{x_0}$ at the  position of the droplet $x_0$. 
  At the surface of the spherical droplet, $r=R$,
  the boundary condition is 
  \begin{align}\label{eq:bc2}
  	\cOut(R,\theta, \varphi) &=\cEqOut(\theta)\\
	\nonumber
	&\simeq\left(\ceqbase + R \cos(\theta)\partial_x \ceqbase(x)|_{x_0} \right) \left(1+\ell_\gamma/R \right) \, .
\end{align} 
 Here, $\ell_\gamma$ is the capillary length as introduced in section~\ref{sect:curvature_droplet}.
 Equation~\eqref{eq:bc2} corresponds to the Gibbs-Thomson relation (see section~\ref{sect:intro_growth_single_droplet}), which
 describes the increase of the local concentration at the droplet interface relative to the equilibrium concentration due to the surface tension of the droplet interface.
The presence of  spatial inhomogeneities on the scale of the droplet $R$ lead to an additional term in the Gibbs-Thomson relation.
To linear order, this contribution to $\cEqOut$  reads $R \cos(\theta)\partial_x \ceqbase(x)|_{x_0}$.
The values of $\be$ and $\al$ characterising the far field, $\cinf(x_0)$, together with the local equilibrium concentration at the droplet surface, $\cEqOut(\theta)$,  determine the local rates of growth or shrinkage of the drop at $x=x_0$ in a spatially inhomogeneous regulator gradient.

  The solution to the Laplace equation with cylindrical symmetry 
  is  of the form $ \cOut (r,\theta) = \sum_{i=0}^\infty \big(A_i r^i +B_i r^{-i-1}\big) P_i (\cos \theta)$,
where $P_i (\cos \theta)$ are the Legendre polynomials. Using the boundary conditions~(\ref{eq:bc1}) and (\ref{eq:bc2}), we find
  \begin{eqnarray}\label{eq:sol_conc_grad}
 \label{eq:phi_quasi_static}
\cOut(r, \theta)&= \al \left( 1- \frac{R}{r}\right)  + \be \cos \theta \left( r - \frac{R^3}{r^2} \right)
\\& \quad + \left(\ceqbase + R \cos(\theta)\partial_x \ceqbase(x)|_{x_0} \right) \left( 1+\frac{\ell_\gamma}{R}\right) \frac{R}{r} \, .
\nonumber
\end{eqnarray}
The droplet could grow, drift or deform due to normal fluxes of droplet material  at the interface leading to a movement of the interface. The speed normal to the interface reads $v_\mathrm{n}=\vect n \cdot \vect{v}_\mathrm{n}$, where
$\vect n$ denotes the normal vector to the interface.
In case of a spherical droplet, $\vect n = \vect e_r$, where 
$\vect e_r$ is the radial unit vector in spherical coordinates. 
In the limit of strong phase separation, i.e., $c^{(0)}_{\rm in} \gg c^{(0)}_{\rm out}$,
 $ \cEqIn \simeq \ceqbaseIn$, and
the velocity normal to the interface, $v_\mathrm{n}$ (equation~\eqref{eq:interface_vel}), can be 
expressed by 
 $v_\mathrm{n}\simeq \vect n \cdot (\vect j_\mathrm{in}-\vect j_\mathrm{out})/\ceqbaseIn$.
 For weak spatial variations inside and outside of the droplet, 
 the local flux is defined as $\vect j=-D \nabla c$.
The concentration inside the droplet is approximately constant and 
for simplicity we consider it to be independent of the droplet position. 
Thus the flux inside the droplet vanishes, $\vect j_\mathrm{in}=0$,
while the flux outside  reads  $\vect j_\mathrm{out}=-D \nabla \cOut|_R$.

Now we discuss how the normal speed $v_{\rm n}$ can be used to calculate 
 the droplet growth speed $v_0$, 
the droplet drift velocity $v_1$, and the rate of deformations from the spherical shape, $v_2$.
To this end, we parametrise the surface of the droplet 
in terms of Legendre polynomials as there is no dependence on the polar angle,
which gives
$\mathcal{R}(\theta, t)=\sum_i d_i(t) P_i \left(\cos\theta\right)$, where
 $d_i(t)$ are the expansion coefficients characterising the shape of the interface. 
 The corresponding interface velocity  of an approximately spherical droplet is $v_{\rm n} \simeq\partial_t \mathcal{R}(\theta, t)=\sum_i v_i(t) P_i \left(\cos\theta\right)$, 
where the 
speeds for droplet growth, drift and deformations along the regulator gradient read  $v_i(t)=\partial_t d_i(t)$. 
 We can now identify the  radius $R$ as
$d_0=   \langle \mathcal{R}, P_0 \rangle/\langle P_0, P_0 \rangle$,
 the position of the droplet center $x_0$ as $d_1=\langle \mathcal{R}, P_1 \rangle/\langle P_1, P_1 \rangle$, and
 the deformations are characterised by
  $d_2=\langle \mathcal{R}, P_2 \rangle/\langle P_2, P_2 \rangle$
Here, the brackets indicate the scalar product  $\langle h , g \rangle =  \int_0^\pi  \mathrm{d}\theta \sin \theta  \, h \, g $ between the functions $g$ and $h$.
Most importantly,  we can identify $v_0= \mathrm{d}R/\mathrm{d}t$ as the rate of change of the radius and $v_1= \mathrm{d}x_0/\mathrm{d}t$ as the drift velocity of the droplet center.

Using equation~\eqref{eq:sol_conc_grad}, we find for the droplet growth speed
\begin{equation}\label{eq:LS_inhomog}
	\frac{ \mathrm{d}R}{\mathrm{d}t }=
	\frac{D}{\ceqbaseIn  R}\; \left[ \al -\ceqbase(x_0) \left(1 + \frac{\ell_\gamma}{R} \right)\right] \, . 
\end{equation}
In the presence of concentration gradients there also exists
 a net droplet drift speed
\begin{equation}\label{eq:drift}
	\frac{\mathrm{d}x_0}{\mathrm{d}t} = \frac{D}{\ceqbaseIn }  \left[ 3\be - \partial_x \ceqbase (x)|_{x_0}  \left(1 + \frac{\ell_\gamma}{R} \right) \right] \, ,
\end{equation}
where we found in contrast to Ref.~\cite{NJP_Weber_ChiuFan_Juelicher} an additional factor of $3$ in front of the coefficient $\beta$.
Note that both the growth rate and the drift speed are proportional to the molecular diffusion constant $D$ of droplet material. 

If the far field $\cinf(x)$ is well parametrised by a linear gradient (equation~\eqref{eq:bc1}),   there are no deformations from the spherical shape, $v_2=0$. Deformations of the spherical shape can only occur if higher order polynomials $P_n (\cos \theta)$ with $n \ge 2$ are necessary to describe the far field. 
If we include the quadratic order in the parametrisation of the far field (equation~\eqref{eq:bc1}),  $(r \cos \theta)^2 \partial^2_x \cinf(x)|_{x_0}$, the deformation speed reads  $v_2=(10/3) (R D/\ceqbaseIn) \partial^2_x \cinf(x)|_{x_0} $. This quadratic contribution does not affect 
the droplet drift $v_1$ but the growth law $v_0$ is changed. The quadratic term gives an extra contribution inside the brackets of equation~\eqref{eq:LS_inhomog} of the form $(5/3) R^2 \partial^2_x \cinf(x)|_{x_0}$.  
Thus, shape deformations and their impact on the growth law in the case of a non-linear far field gradient are negligible  if
\begin{equation}
	 \frac{5 \partial^2_x \cinf(x)|_{x_0}}{3\cinf(x)|_{x_0}} R^2 \ll 1 \, .
\end{equation}
For the system under consideration where length scales separate, i.e., $R \ll \ell \ll L$, deformations  from the spherical shape are weak because gradients of the far field $\cinf(x)$ occurs on the length scale of the system size $L$.
In recent numerical studies considering a continuous phase separating Flory-Huggins model in a regulator gradient maintained by sink and source terms, droplet deformations are indeed visible when droplets approach the order of the system size~\cite{saha2016polar}.

\subsubsection{Dynamical equation of multiple droplets in a concentration gradient.}

 {We can now describe the  dynamics of many droplets, $i=1,\dots, N$,
with positions $x_i$ and  radii 
$R_i$. If droplets are far apart from each other, the rate of growth of  droplet $i$ reads}
\begin{subequations}\label{eq:all_ripening_grad}
\begin{equation}\label{eq:LS_inhomogi}
	\frac{\mathrm{d}R_i }{\mathrm{d}t} =
	\frac{D}{R_i}\; \frac{\ceqbase(x_i)}{\ceqbaseIn }  \left[ \varepsilon(x_i) - \frac{\ell_\gamma}{R_i}\right] \, .
\end{equation}
The  drift velocity of droplet $i$, ${v}_1(x_i)=\mathrm{d} x_i/\mathrm{d}t$, is  
\begin{equation}\label{eq:drifti}
	\frac{\mathrm{d} x_i}{\mathrm{d}t}= \frac{D}{\ceqbaseIn}  \left[ 3\partial_x  \cinf(x)\vert_{x_i}  - \partial_x \ceqbase(x)\vert_{x_i}   \left(1 + \frac{\ell_\gamma}{R_i} \right) \right]  \, .
\end{equation}
If the distance between droplets is large relative to their size,
droplets only interact  via the far field concentration field  $\cinf(x,t)$. 
It is governed by a diffusion equation including gain and loss terms associated with growth or shrinkage of drops:
\begin{align}\label{eq:beta_eq}
	 \partial_t \cinf(x,t)  &= D \frac{
	 	\partial^2}{
	 	\partial  x^2}  \cinf(x,t) \\
		\nonumber
		& \quad - H(t) \sum^{N}_{i=1}\delta(x_i-x) \frac{4\pi}{3}\frac{\mathrm{d} }{\mathrm{d} t} R_i(t)^3
	  \, ,
\end{align}
\end{subequations}
where the time-dependent function
\begin{equation}
H(t)= \frac{\ceqbaseIn - \cinf(t)  }{V- \sum^{N}_{i=1}\delta(x_i-x) \frac{4\pi}{3} R_i(t)^3}
 \end{equation}
 is approximately constant, $H\simeq \ceqbaseIn/V$, in the limit of strong phase separation
 $\ceqbaseIn \gg \cinf(t)$ and for very large inter-droplet distances corresponding to 
 $V \gg \sum^{N}_{i=1}\delta(x_i-x) \frac{4\pi}{3} R_i(t)^3$ with 
$V$.

Equation~(\ref{eq:beta_eq})
 describes the effects of large scale spatial inhomogeneities on the ripening dynamics for the case of a regulator gradient varying along the $x$-axis.
 Since large scale variations of $\cinf(x,t)$ only build up along the $x$-directions, derivatives of $\cinf$ along the $y$ and $z$ directions do not contribute as $\cinf$ and $\ceqbase$ are constant along these directions.

In the absence of a regulator gradient, $\ceqbase$ and  $\cinf$ are constant in space implying a position-independent and common supersaturation level $\varepsilon$ for all droplets (equation~(\ref{eq:supersaturation})).
In this case equation~(\ref{eq:LS_inhomogi})  gives the  classical 
law of droplet ripening derived by Lifschitz-Slyozov ~\cite{Lifshitz_Slyozov_61,wagner61} (also referred to as Ostwald ripening),  and the net drift vanishes  (equation~(\ref{eq:drifti})).
In the case of Ostwald ripening, droplets larger than the critical radius  $R_{\mathrm{c}}= \ell_{\gamma}/\varepsilon$ grow at the expense of smaller shrinking drops which then disappear.
This  competition between smaller and larger drops causes an increase of the average droplet size and a broadening of the droplet size distribution with time. 
 Ostwald ripening is characterised by a 
 supersaturation that decreases with time, leading to an increase of the  critical droplet radius  $R_\mathrm{c}=\ell_\gamma/\varepsilon(t)\propto t^{1/3}$. 
 The droplet size distribution $P(R)$
 exhibits an universal shape and is nonzero only in the interval $[0, 3R_\mathrm{c}/2]$ (figure~\ref{fig:Fig_4_dotR_droplet_size_distribution}(b), blue graph).
 In other words, there are no droplets larger than $3R_\mathrm{c}/2$ and thus also no droplets exist beyond the maximum of 
 $\mathrm{d}R/\mathrm{d}t$  at $R=2R_\mathrm{c}$ 
 where larger droplet could grow slower. 
Therefore, in homogeneous systems
  the broadening of $P(R)$ follows from larger droplets growing at a larger rate $\mathrm{d}R/\mathrm{d}t$ than smaller droplets and because all droplets feel the same supersaturation level droplet positions remain homogeneously distributed in the system.
In the presence of a regulator gradient  the droplet dynamics exhibits a different behaviour.

 \subsubsection{Droplet positioning via position dependent growth and drift.}

There are two novel possibilities of how droplet material is transported: 
There is exchange of material between droplets at different positions along the concentration gradient due to a {\it position dependent droplet growth}, and
 {\it droplets can drift} along the concentration gradient. 

 Droplets grow or shrink with rates that vary along the gradient because the  local equilibrium concentration $\ceqbase(x)$ and the far field concentration $\cinf(x)$ are position dependent (equation~(\ref{eq:LS_inhomogi})). 
 For a supersaturation $\varepsilon(x)=\left(\cinf(x)/\ceqbase(x)-1\right) > \ell_\gamma/R$, a droplet located at position $x$ grows, and shrinks in the opposite case. 
In other words, the critical droplet radius depends on position, and droplets with radii  below or above $R_{\mathrm{c}}(x)= \ell_\gamma/\varepsilon(x)$   shrink or grow.
 This position dependence implies a movement of the dissolution boundary $x_{\rm c}(t)$ which is defined where the supersaturation $\varepsilon(x_{\rm c}(t))=\ell_\gamma/R$ (equation~\ref{eq:supersaturation}).
 This definition can be simplified for the case where the capillary length, which is typically in the order of the molecular size, is small relative to the droplet radii, i.e.,
  $\ell_\gamma \ll R$, 
 leading to $\cinf(x_{\rm c}(t)) \simeq \ceqbase(x_{\rm c}(t))$.
 Taking the derivative in time gives the speed of the dissolution boundary, $v_{\rm c} (t)= \frac{\mathrm{d} x_{\rm c}(t)}{\mathrm{d}t} $:
 \begin{equation}\label{eq_diss_boundary}
 	v_{\rm c} (t) = {\frac{\mathrm{d} \cinf (x)}{\mathrm{d}t}}\bigg{|}_{x=x_{\rm c}(t)} \bigg{/}{\frac{\mathrm{d} \ceqbase(x)}{\mathrm{d}x}}\bigg{|}_{x=x_{\rm c}(t)} \, .
 \end{equation} 
We can now discuss the movement direction of the dissolution boundary. 
If droplets can grow in the system, the far field concentration should decay, $\frac{\mathrm{d} \cinf}{\mathrm{d}t}<0$. Thus the dissolution boundary always moves  toward positions corresponding to  smaller values of $\ceqbase$. When the dissolution boundary moves through the system it dissolves all droplets on its way.
The dissolved droplet material will diffuse and feed the growth of the remaining droplets. 
Thus the moving dissolution boundary positions the phase separated material toward one boundary of the system corresponding to the lower values of the position-dependent equilibrium concentration $\ceqbase(x)$.

 Another mechanism of positioning is via droplet drift (equation~(\ref{eq:drifti})).
 The drift of a droplet results from an asymmetry of material flux at the interface parallel to the regulator gradient.
 To be more specific, 
 we have to distinguish between the scenario of many droplets and the case of a single droplet. 
 In the case of many droplets to the right of the dissolution boundary,
 the presence of these droplets keep the position-dependent supersaturation 
 small, thus
 $\frac{\mathrm{d} \cinf}{\mathrm{d}x} \simeq \frac{\mathrm{d} \ceqbase}{\mathrm{d}x}$.  
 Using the derived equation for droplet drift~\ref{eq:drift}, the drift of droplet `d' in a system with many droplets roughly follows
  \begin{equation}\label{eq:drift_many}
	\frac{\mathrm{d} x_{\rm d}}{\mathrm{d}t} \simeq 2 \frac{D}{\ceqbaseIn}  \frac{\mathrm{d} \ceqbase}{\mathrm{d}x} \,.
 \end{equation}
 This equation implies that droplets drift into the opposite direction of the dissolution boundary. 
 Thus droplets are pushed toward dissolution making it impossible for them to escape  their dissolution via drift. 
 This picture can be different for a single droplet. 
 When diffusion is fast, the far field concentration is expected for be roughly homogeneous, $\frac{\mathrm{d} \cinf}{\mathrm{d}x} \simeq 0$.  
 As a consequence the drift points parallel to the direction of the dissolution boundary in the case of single droplet:
   \begin{equation}
  \frac{\mathrm{d} x_0}{\mathrm{d}t}\simeq - \frac{D}{\ceqbaseIn}  \frac{\mathrm{d} \ceqbase}{\mathrm{d}x} \, .
   \end{equation}
   The droplet may thereby escape its dissolution by drift.
   The drift of a single droplet in a roughly homogeneous far-field concentration  should be driven by the efflux of material at the back, which diffuses to the front of the droplet (droplet front is faced into the direction of movement of the dissolution boundary).
   Droplet movement then arises from asymmetric droplet growth between the back and the front of the droplet. The associated time-scale of this process is roughly given by the time to diffuse the droplet radii, i.e.,  $R^2/D$. If this time-scale is smaller than the time-scale $R/v_{\rm c}$ necessary for the dissolution boundary to move the distance $R$, a single droplet may escape the dissolution boundary.
   Using equation~\ref{eq_diss_boundary} 
   the condition for a single droplet to drift can thus be written as 
 \begin{equation}
  D \frac{\mathrm{d} \ceqbase}{\mathrm{d}x} < R \frac{\mathrm{d} \cinf}{\mathrm{d}t} \, .
 \end{equation}
 Otherwise, the droplet would dissolve and recondense via nucleation typically 
 in domains corresponding to lower values of the position-dependent equilibrium concentration $\ceqbase(x)$.

\subsubsection{Narrowing of the droplet size distribution.}

\begin{figure}[tb]
\centering
\begin{centering}
\includegraphics[width=1.0\columnwidth]{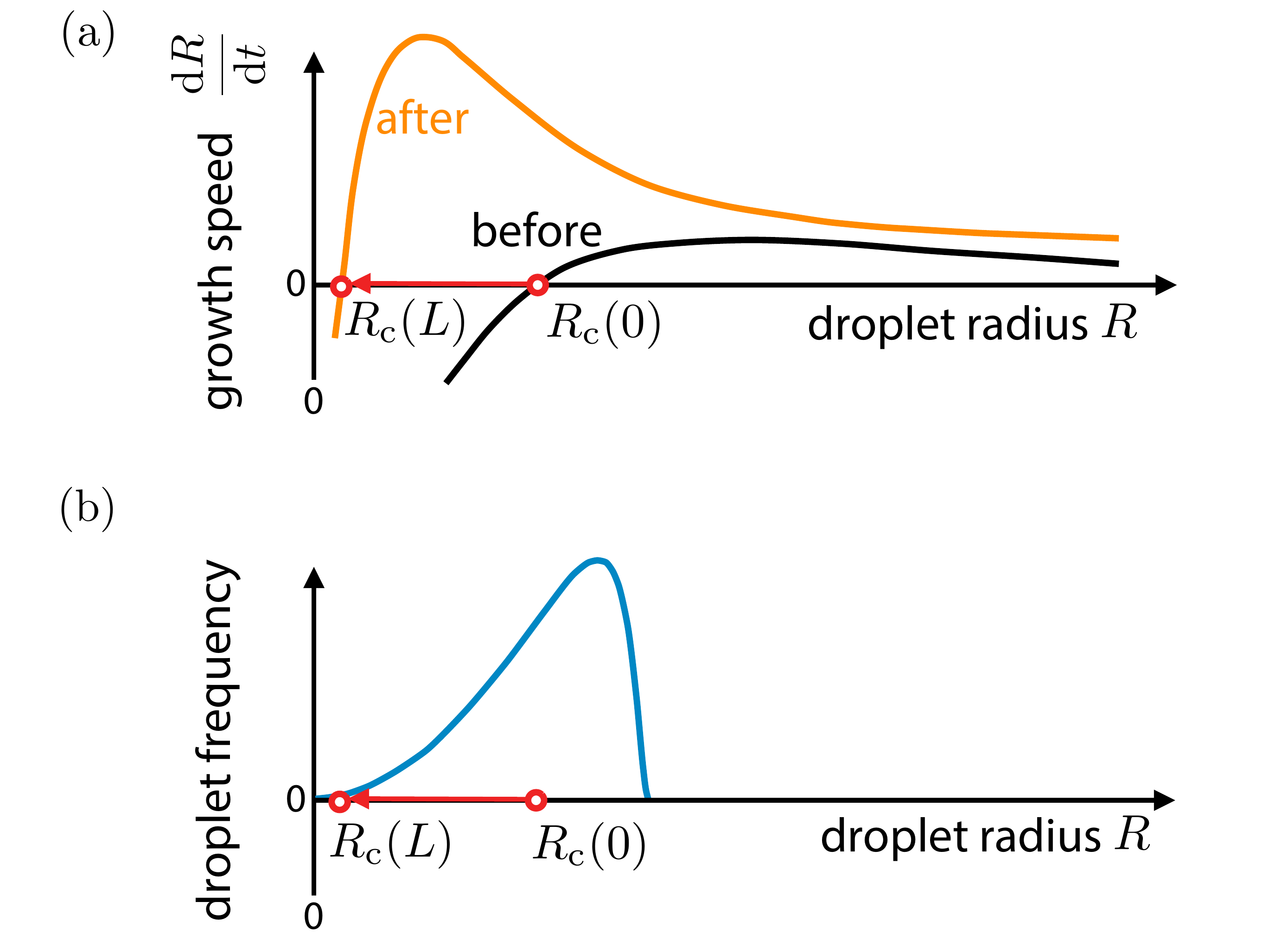} 
\end{centering}
\caption{\label{fig:Fig_4_dotR_droplet_size_distribution} 
Sketch depicting the mechanism of the narrowing of the droplet size distribution due to the presence of concentration gradients. 
(a) The black curve depicts the droplet growth speed $\text{d} R/\text{d}t$ before the spatial quench, where the system undergoes Ostwald ripening with a homogenous far field $\cinf$ and a homogenous equilibrium concentration $\ceqbase(0)$. The corresponding droplet size distribution is shown in (b). 
The spatial quench
$\ceqbase(x)= \ceqbase(0) \,  (1- m \, x)$ reduces the equilibrium concentration at $x= L$. Therefore, the local supersaturation increases which amounts to a decrease of the critical radius at $x= 0$ from $R_\mathrm{c}(0)$ to $R_\mathrm{c}(L)$ (indicated by red arrow). 
This change in supersaturation changes the droplet growth velocity $\text{d} R/\text{d}t$ (orange).
Subsequent to such a spatial quench mostly all droplets at $x=L$ grow. 
However, larger droplets typically grow less than smaller drops. Consequently, the size distribution will narrow.
The narrowing is most pronounced close to the rightmost boundary at $x=L$ since the moving dissolution boundary dissolves all droplets at $x<L$.
In addition, the dissolution of these droplets will keep the far field concentration $\cinf(x)$ at $x=L$ at increased level which maintains a small value of critical radius until the dissolution boundary has reached the boundary at $x=L$.
}
\end{figure}

Numerically solving equations~\eqref{eq:all_ripening_grad}  for a large number of droplets we find that the droplet size distribution narrows during the positioning of droplets toward one boundary of the system. For details on the numerics, please refer to reference~\cite{NJP_Weber_ChiuFan_Juelicher}.
This narrowing  of the droplet size distribution in a concentration gradient 
is fundamentally 
different from the  broadening of the droplet size distributions during
classical Ostwald ripening~\cite{Lifshitz_Slyozov_61,wagner61};
see figure~\ref{fig:Fig_4_dotR_droplet_size_distribution} for an illustration of the mechanism underlying the narrowing.
Imagine we spatially quench  the system by imposing a  spatially varying equilibrium concentration $\ceqbase(x)= \ceqbase(0) \,  (1- m \, x)$,
where $ \ceqbase(0)$ denotes the equilibrium concentration before the quench and $m$ is the slope of the  ``spatial quench''.
Such a spatial quench reduces the critical radius at the right boundary at $x= L$ from $R_\text{c}(0)$ (critical radius before the quench) to 
 $R_\mathrm{c}(x= L)=\ell_\gamma/\epsilon(x= L)$ 
(equation~(\ref{eq:supersaturation})). 
This quench also shifts the maximum of $\mathrm{d}R/\mathrm{d}t$ for droplets
 at $x= L$ to smaller radii since  the  radius corresponding to the maximum occurs at $R=2 R_\mathrm{c}$. 
After the spatial quench there are many droplets  with  radii  $R>2 R_\mathrm{c}(x= L)$.
According to $\text{d} R/\text{d}t$ (figure~\ref{fig:Fig_4_dotR_droplet_size_distribution}(a))
these droplets grow more slowly than those around $R=2R_\mathrm{c}$ which
leads to a {\it narrowing of the droplet size distribution} $P(R)$ at $x= L$. 
 The critical radius $R_\mathrm{c} (x = L)$ remains small because dissolution of droplets at $x<L$ leads to a diffusive flux toward $x= L$ and thus keeps the concentration $\cinf(L)$ at increased levels.
 These conditions hold for a longer time if the spatial quench has a steeper slope. As a result the distribution  narrows more for steeper quenches. For weak enough slopes of the quench, 
the narrowing of the droplet size distribution vanishes, however, droplet positioning still occurs as long as this slope is not zero.

 When the dissolution boundary reaches the rightmost boundary close to $x=L$ the critical radius catches up with the mean droplet size. 
Concomitantly,  narrowing of the droplet size distribution stops.
Because all droplets are approximately of equal size the exchange of material between droplets via Ostwald ripening is slowed down dramatically.
This slowing down of inter-droplet diffusion via Ostwald ripening leads to a long phase where droplet number and size is  almost constant.  Close to the end of this arrest phase, the droplet distribution begins to broaden slowly and the dynamics approaches classical Ostwald ripening.

In summary, a concentration gradient of a regulator component  that affects phase separation can significantly change the dynamics of droplet coarsening. 
The regulator gradient causes an inhomogeneity 
of the equilibrium concentration and the concentration field far away from the droplet. 
Both induce a position-dependent ripening process where droplets can drift along the gradient and dissolve everywhere besides to a region close to one boundary of the system. 
During this positioning process of droplets to one boundary
 the droplet size distribution can dramatically narrow for steep enough quenches which causes a transient arrest of droplet growth.  
 After this arrest phase the positioned droplets are subject to a locally homogenous concentration environment and the system recovers the dynamics of  classical Ostwald ripening.



\section{Droplets driven by chemical turnover}\label{sect:phase_sep_with_turn_over}

Droplets can also be controlled by chemical reactions that directly affect the concentrations of the segregating species. 
For instance the building blocks~$B$ that form droplets could emerge from precursors~$P$ by a chemical reaction.
While simple conversion reactions typically suppress phase separation, chemical reactions that are driven by an external energy input allow to control the droplet size as well as the droplet count and can even lead to spontaneous droplet division.
To describe such phenomena, we start by deriving the dynamical equations from a thermodynamic consistent description of phase separation in the presence of chemical reactions.

\subsection{Thermodynamics of chemical reactions}
\label{sec:thermodynamics_reactions}

Before we consider the coupling of phase separation and chemical reactions,
we review the thermodynamics of chemical reactions in homogeneous systems.
To highlight the core concepts, we here focus on very simple chemical reactions.

\subsubsection{Chemical reactions in homogeneous systems}
We start by considering an isolated system where the two chemical species, the building block $B$ and the precursor $P$, are converted into each other by the reaction
\begin{equation}
	\tag{R1}
	P \rightleftharpoons B
	\label{eqn:reaction1}
	\;.
\end{equation}
At constant temperature~$T$ and volume~$V$, the system is described by a free energy $F(N_P,N_B)$, where $N_i$ are the particle numbers of type $i=P,B$.
The thermodynamic equilibrium of the system corresponds to the minimum of $F$.
The necessary condition for this minimum reads $\mu_P \diff N_P + \mu_B \diff N_B = 0$, where the chemical potentials are $\mu_P = \takenat{\partial F/\partial N_P}{N_B}$ and $\mu_B = \takenat{\partial F/\partial N_B}{N_P}$.
Note that in contrast to phase separation without reactions, species can now be converted into each other and only the total number of particles, $M=N_P + N_B$, is conserved.
This implies $\diff N_P = -\diff N_B$, such that the equilibrium condition requires $\mu_P - \mu_B=0$.
Consequently, at equilibrium, the chemical reaction equalises the chemical potentials of the two species.

The difference between the chemical potentials, $\mu_P - \mu_B$,  also affects the relaxation rate toward equilibrium.
This rate is quantified by the total reaction flux $s=-\diff c_P/\diff t = \diff c_B/\diff t$, where $c_i = N_i/V$ denotes the concentrations in the homogeneous system for $i=P,B$.
Since the reaction can proceed in both directions, the total reaction flux~$s= \sF - \sB$ is given by the difference of the forward reaction flux~$\sF$ associated with the conversion of $P$ to $B$ and the reverse flux~$\sB$.
As a consequence of detailed balance, the ratio of the two reaction fluxes obey (see \ref{sec:appendix_detailled_balance})
\begin{align}
	\frac{\sF}{\sB} &= \exp\left( -\frac{\mu_B - \mu_P}{\kb T} \right)
	\label{eqn:foward_backward_ratio}
	\;,
\end{align}
which we call \textit{detailed balance of the rates}~\cite{Julicher1997}.
The relation shows that the net direction in which the reaction proceeds depends on the sign of the chemical potential difference $\mu_B - \mu_P$.
Moreover, the net reaction flux~$s$ vanishes at chemical equilibrium ($\mu_P = \mu_B$) since $\sF = \sB$.
Close to chemical equilibrium, \Eqref{eqn:foward_backward_ratio} can be linearized and the reaction flux $s= \sF - \sB$ can be expressed as
\begin{align}
	s &= -\mobR(c_P,c_B) \, (\mu_B - \mu_P)
	\label{eqn:reaction_rate}
	\;,
\end{align}
 where the function $\mobR(c_P, c_B)$
determines the reaction rate. 
$\mobR(c_P, c_B)$ is an Onsager coefficient, which must be positive to ensure a positive entropy production rate~\cite{landau_lifschitz_6_hydro,balian2007microphysics2};
see~\ref{sec:appendix_entropy_production}.

\subsubsection{Chemical reactions in inhomogeneous systems and stability of homogeneous states}
\label{sec:active_passive_conversion}

To describe chemical reactions in inhomogeneous systems, we assume local thermodynamic equilibrium, \ie there exist local volume elements where thermodynamic quantities such as concentrations and temperature can be defined.
This is possible when the local volumes equilibrate quickly compared to the rates of the processes that we want to describe.
In particular, the exchange with neighbouring volumes, associated with diffusive transport, and the conversion of particles into different species, associated with chemical reactions, should take place on timescales longer than the equilibration timescale of the volumes.
If this is the case, a system of reacting and diffusing particles can be described by concentration fields $c_i(\vect r)$ for all species~$i$.

To study the interplay of the chemical reaction \eqref{eqn:reaction1} with phase separation, we first consider a binary, incompressible system described by the concentration of the droplet component, $c_B(\vect r)=c(\vect r)$, while $c_P(\vect r) = \nu^{-1} - c(\vect r)$ with $\nu$ denoting the molecular volume of $P$ and $B$.
The behavior of the system is governed by the free energy~$F[c]$, which is now a functional of the concentration field~$c(\vect r)$.
For simplicity, we here consider the form 
\begin{align}
	F[c] &= \int \diff^3  r \left( f(c) + \frac{\kappa}{2} \abs{\nabla c}^2 \right)
	\;,
\end{align}
which combines a local contribution of the free energy density~$f(c)$ with a term that accounts for the free energy costs of spatial inhomogeneities proportional to $\kappa$, analogous to  \secref{sec:GL}. 
The exchange chemical potential $\bar\mu = \mu_B - \mu_P$ is thus given by $\bar\mu = \delta F[c]/\delta c$.
The resulting equilibrium condition of the chemical reaction is $\bar\mu(\vect r)=0$, which includes the equilibrium condition for phase separation, $\bar\mu(\vect r) = \text{const.}$, see \secref{sec:passive_stationary_states}.

The dynamical equation of the system follows from the conservation law
\begin{align}
	\partial_t c + \nabla \cdot \vect j = s
	\label{eqn:conservation_law}
	\;,
\end{align}
where $\vect j$ is the diffusive flux and $s$ is the net flux of the production of species $B$ by the reaction~\eqref{eqn:reaction1}.
These two thermodynamic fluxes are driven by their respective conjugated forces $\nabla \bar\mu$ and $\bar\mu$; see~\ref{sec:appendix_entropy_production}.
Using linear response theory, $\vect j = -\mob \nabla \bar\mu$ and  $s=-\mobR\bar\mu$, we arrive at the dynamical equation
\begin{align}
	\partial_t c &= \nabla \cdot \bigl[ \mob(c) \, \nabla \bar\mu(c) \bigr]
		- \mobR(c)  \, \bar\mu(c) 
	\label{eqn:active_pde_equilibrium}
	\;,
\end{align}
which describes a binary system that exhibits phase separation and chemical reactions.
Note that we recover the Cahn-Hilliard equation if chemical reactions are absent ($\mobR=0$). 
Conversely, in the limit where $\mobR$ is constant and diffusive fluxes vanish ($\mob=0$), we 
obtain the Allen-Cahn model~\cite{Allen1979}, which is the deterministic version of model A~\cite{Hohenberg1977}.

\begin{figure}[tb]
	\begin{center}
		\includegraphics[width=1.0\columnwidth]{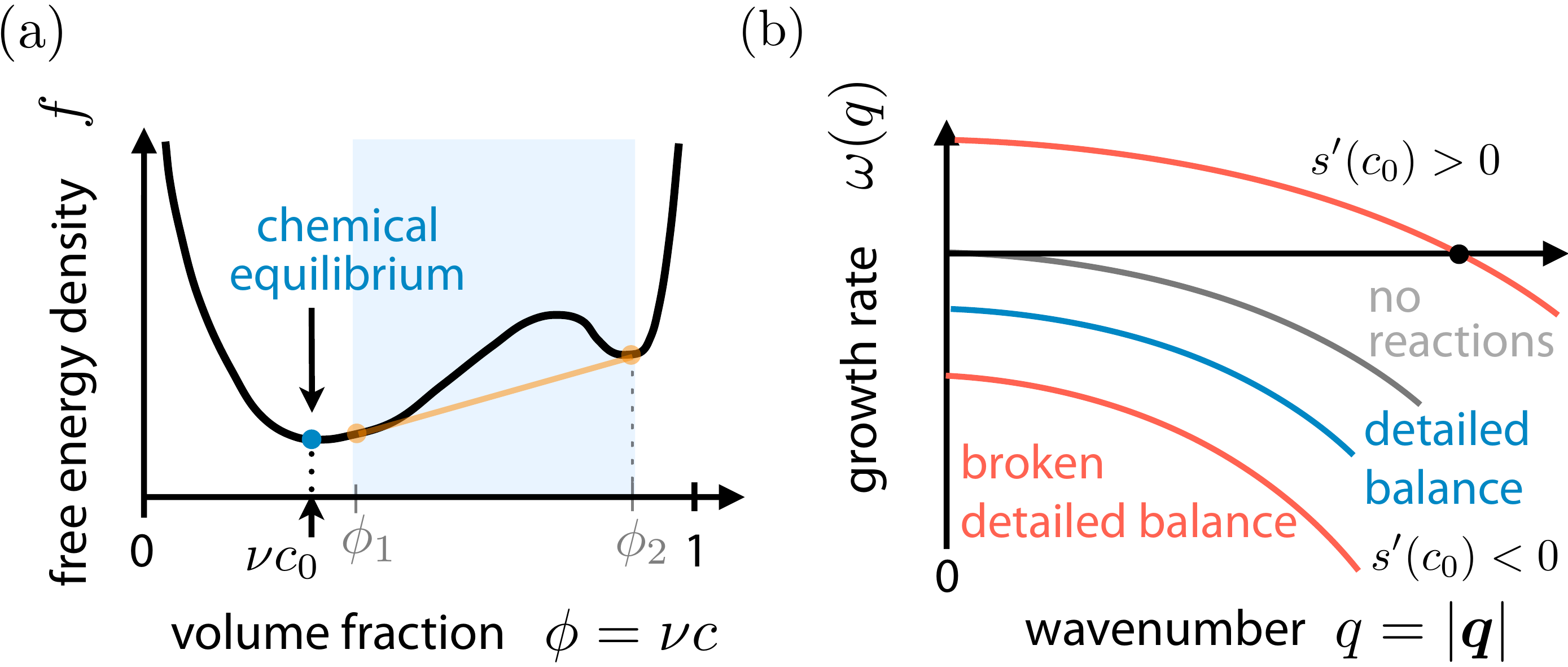}
	\end{center}
	\caption{
	Schematic representation of the impact of chemical reactions on phase separation.
	(a) In chemical equilibrium, a system that is able to phase separate settles in the minimum of the free energy density corresponding to the homogeneous concentration $c_0$. Phase separated states with concentrations $c_1=\phi_1/\nu$  and $c_2=\phi_2/\nu$ are no more stable.
	(b)
	Growth rate as a function of the wavenumber $q=|\vect{q}|$ ($\vect{q}$: wave vector) for phase separation in the absence of chemical reactions (grey), 
phase separation in the presence of a chemical reaction tending toward chemical equilibrium satisfying detailed balance of the rates (blue), and phase separation combined with non-equilibrium chemical reactions which break detailed balance of the rates (red).   The sign of $s'(c_0)$ can be adjusted by the chemical potential difference between fuel and waste, $\bar\mu_2$, for example. The homogeneous  concentration $c_0$ is defined as the concentration at which $s(c_0)=0$. 
	}
	\label{fig:4_free_energy_dispersions}
\end{figure}

We study the effects of chemical reactions by first  analyzing homogeneous equilibrium states $c(\vect r)=c_0$, which are governed by the equilibrium condition $\bar\mu(\vect r)=0$.
This condition implies vanishing chemical reaction flux, see \Eqref{eqn:reaction_rate}, and  $f'(c_0) =0$, so that $c_0$ is a (local) extremum of the free energy density $f(c)$. 
To assess the stability of these states, we consider harmonic perturbations with wave vector~$\vect q$, as described in \secref{sec:dynamical_equation}.
In the linear regime, these perturbations grow with a rate
\begin{align}
	\omega(\vect q) 
	&= -\bigl[\mob(c_0) \vect q^2 + \mobR(c_0)\bigr]
		 \bigl[f''(c_0) + \kappa \vect q^2\bigr]
	\;.
	\label{eqn:pert_growth_rate_passive}
\end{align}
The system is stable if all perturbations decay, \ie if $\omega(\vect q) < 0$ for all wave vectors~$\vect q$.
Since $\mob,\mobR\ge0$, the stability is governed by the sign of the second bracket in \Eqref{eqn:pert_growth_rate_passive} and the homogeneous state becomes unstable when $f''(c_0) < 0$ \cite{Gitterman1978}.
This condition is identical to the condition for the spinodal instability in the case without chemical reactions ($\mobR=0$).
However, in the presence of chemical reactions, only the homogeneous states with $f'(c_0)=0$ are stationary states because particles numbers of $P$ and $B$ are not conserved.
In contrast, in the absence of chemical reactions, all homogeneous states are stationary with the homogeneous concentrations of $P$ and $B$ being conserved. 
Particle conservation also implies $\omega(0)=0$, while  the $q=0$ mode is unstable when chemical reactions are present, see figure~\ref{fig:4_free_energy_dispersions}(b).

Taken together, we showed that if the system settles in a homogeneous state it will attain minimal free energy~\cite{Gitterman1978, Lamorgese2016}, see figure~\ref{fig:4_free_energy_dispersions}(a).
The major difference to the case without chemical reactions is that the species are not conserved individually and the system can thus relax by altering the composition locally.
In the next section, we will show that this local conversion destabilises all inhomogeneous states including the ones corresponding to coexisting phases.

\subsubsection{Dissolution of droplets by chemical reactions}
\label{sec:droplet_dissolution}

We now investigate the stability of inhomogeneous states to see how chemical reactions that can relax to equilibrium ($\bar\mu=0$) affect droplets.
As an example, we first discuss the Ginzburg-Landau free energy presented in \Eqref{eq:GLfree}.
When chemical reactions are present, this free energy permits two stable homogeneous solutions, corresponding to the two minima of the free energy density.
Without chemical reactions, we have shown in section~\ref{sec:interface} that there is also a stable inhomogeneous stationary state, which consists of two bulk phases separated by an interfacial region.
In the simple case of a one-dimensional system, the interfacial profile~$\cInt(x)$ is given by \Eqref{eq:tanh}.
This interfacial profile is also a stationary state in the case with chemical reactions, since the symmetric free energy implies vanishing chemical potentials in the two bulk phases and thus satisfies the equilibrium condition $\mu(\vect r)=0$.

To scrutinise the stability of the interfacial profile in the presence of chemical reactions, we determine whether a small change $\delta c$ in the concentration profile could possibly decrease the free energy.
Mathematically, this corresponds to calculating the second variation $\Delta F [\cInt,\delta c]$ of the free energy, which is given by \Eqref{eq:triF}. 
The stationary state is stable if $\Delta F [\cInt,\delta c]$ is positive for every nonzero variation~$\delta c$.
We show in \ref{sec:appendix_stability} that indeed almost all perturbations~$\delta c$ increase the free energy.
The only exception is $\delta c = \partial_x \cInt$, for which $\Delta F = 0$, implying that this perturbation does not decay in time and the state is marginally stable.
This perturbation corresponds to an infinitesimal translation, indicating that interfaces can move without changing the total free energy when chemical reactions are present.
Note that this perturbation does not conserve the mass of the individual species and is thus forbidden in the case without chemical reactions discussed in \secref{sec:stability_stationary_states}.
Taken together, we showed that the sigmoidal interface profile given by \Eqref{eq:tanh} is neither stable nor unstable in the special case of the Ginzburg-Landau free energy, where both minima have the same energy.

In the general case where the minima of the free energy density are at different energies (see figure~\ref{fig:ch2_1}(a)), the global free energy decreases when the interface moves such that the phase with the smaller free energy density expands~\cite{Lamorgese2016}.
Consequently, coexisting phases in asymmetric free energies are unstable due to presence of chemical reactions. 
Similarly, curved interfaces are unstable (also in the case of a symmetric free energy density), since the associated Laplace pressure implies elevated concentrations on both sides of the interface, see \Eqsref{eq:final_GT_relations}.
The Laplace pressure, and thus the concentrations and the free energy, decrease when the interface moves towards its concave side, implying that droplets shrink~\cite{Lamorgese2016}.
Taken together, these arguments show that inhomogeneous states are generally unstable when chemical reactions are present and the system attains the global free energy minimum everywhere, see figure~\ref{fig:4_free_energy_dispersions}(a).

\begin{figure*}
	\raggedleft
	\includegraphics[width=\textwidth]{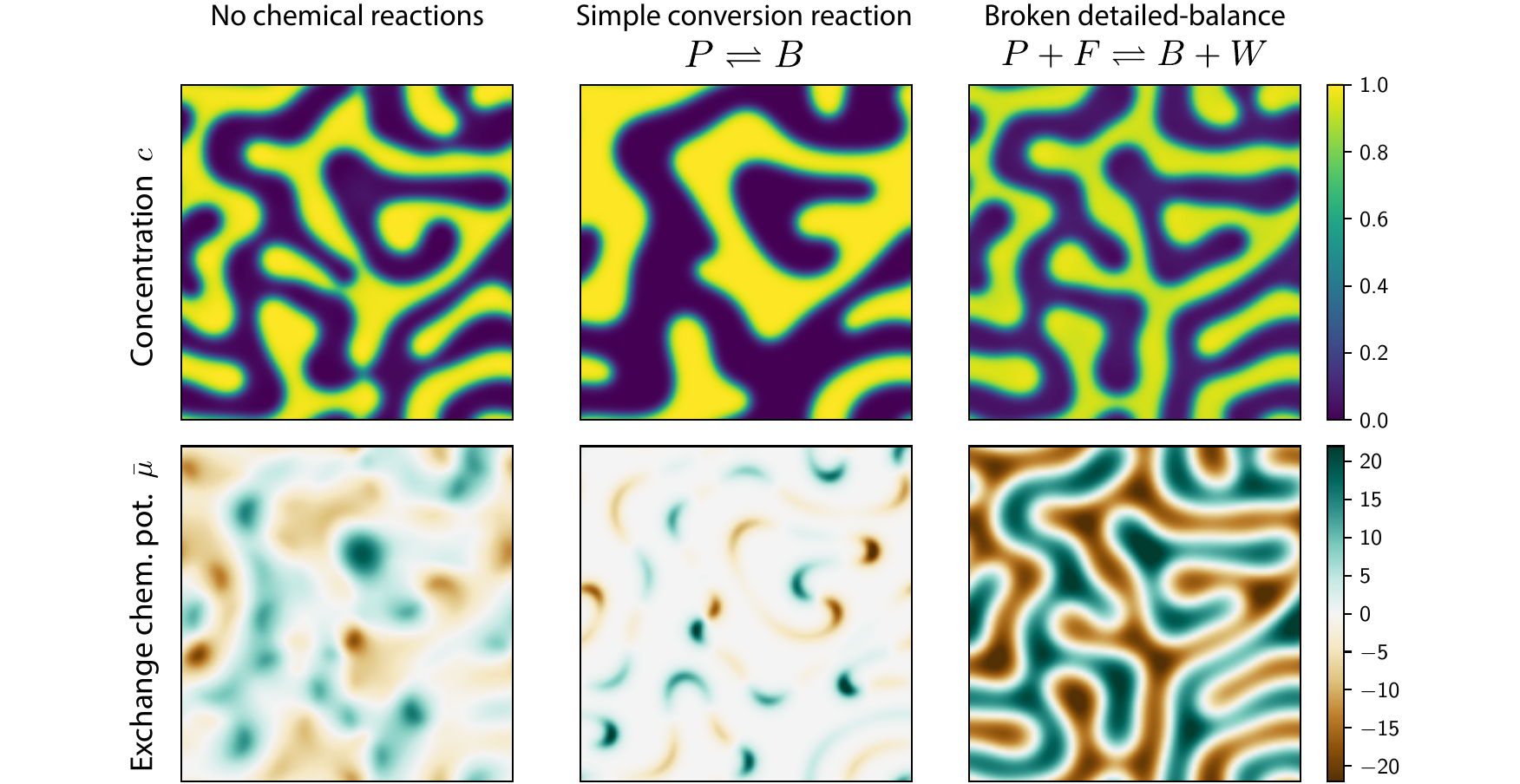}
	\caption{
	   Influence of chemical reactions on phase separation.
	Shown are concentration profiles (upper row) and associated chemical potentials (lower row) 
	of a binary system with equal amounts of $P$ and $B$ in different situations:
	Without chemical reactions the system undergoes coarsening by diffusive transport (left column).
	With an additional conversion reaction between the two species, the bulk phases relax quickly and the interface moves due to localised reactions (middle column).
	If detailed balance is broken, the system is driven away from equilibrium everywhere and more complex dynamics can occur (right column).
	}
	\label{fig:4_phasefield}
\end{figure*}

A hint of these dynamics is visible in the simulation results shown in the middle column in  \figref{fig:4_phasefield}, where the chemical potential difference~$\bar\mu$ is very close to zero in the bulk phases but finite at the interfaces.
In fact, the chemical potential deviates from zero more strongly at interfacial regions of larger curvature.
One consequence of this observations is that the local entropy production $\Lambda_r\bar\mu^2$ by the chemical reactions
is also largest at the interfaces.


Taken together, we showed that droplets are destabilised by the simple conversion reaction \eqref{eqn:reaction1} obeying detailed balance of the rates.
In particular, the system always settles in a homogeneous state, even if droplets appear initially due to a spinodal instability or nucleation.
The reason for the initial appearance of droplets is that 
diffusion is fast on small length scales and the formation of droplets locally reduces the free energy density. On longer time scales, the chemical reaction drives the system into a homogeneous equilibrium state.
 To compensate this destabilisation of droplets due to the reaction \eqref{eqn:reaction1}
we will introduce an additional chemical reaction in the bulk phases.
 This chemical reaction is driven by fuel, and thereby breaks detailed balance of the rates and can prevent the chemical reaction driving the system to a homogeneous equilibrium state.
 This additional chemical reaction allows us to control the behaviour of droplets.
 In particular, we will show that systems where droplet material is produced in one phase while it is destroyed in the other can lead to interesting phenomena
 such as mono-disperse emulsions or dividing  droplets.

\subsection{Phase separation with broken detailed balance of the rates}

\begin{figure}[tb]
	\begin{center}
		\includegraphics[width=1.0\columnwidth]{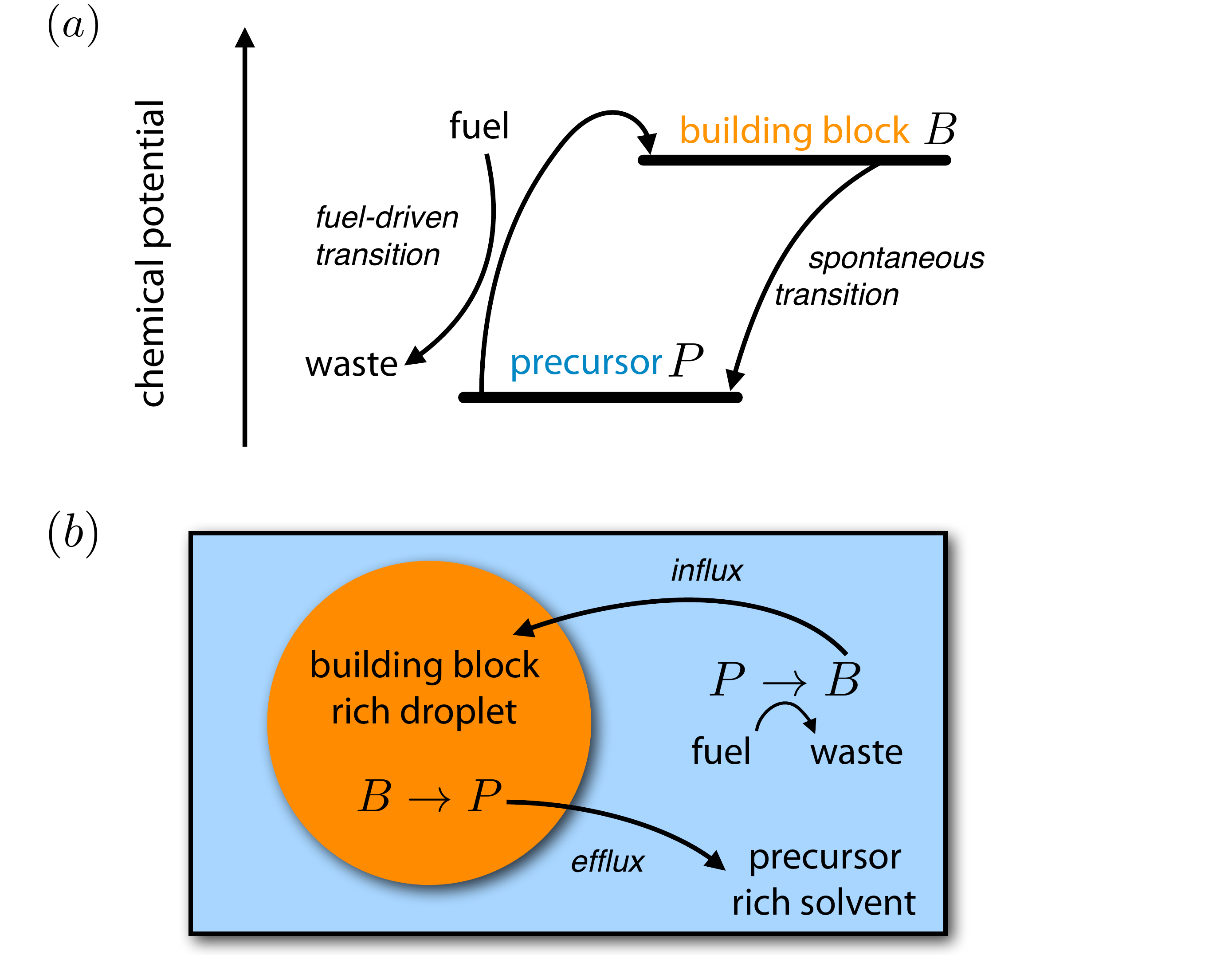}
	\end{center}
	\caption{
	Illustration of the fuel-driven chemical reactions and the diffusive fluxes relevant for an active emulsion. 
	(a) The key feature required for an active emulsion is that precursor molecules~$P$ are turned into building blocks~$B$ by a chemical reaction that consumes energy.
	This energy could be supplied by fuel that turns into waste during this transition.
	Energy is dissipated when the building block of higher chemical potential spontaneously transitions to the precursor of lower chemical potential.
	The reaction rates, and thus also the lifetime of the building block, depend strongly on the chemical composition, \ie whether the reaction takes place in a droplet or not~\cite{boekhoven2015transient, tena2017non}. 
	(b) 	This schematic depicts the processes occurring inside and outside of a building block rich droplet in an active emulsion considering the example of an externally maintained droplet. 
Building blocks inside such a droplet are turned into precursors.
 The resulting precursors diffusive out of the droplet where they may get transformed into building blocks again by the consumption of fuel.  These building blocks may in turn diffusive into the droplet and replenish the spontaneously degraded building blocks inside the droplet.  
	}
	\label{fig:4_illustration_active_emulsions}
\end{figure}

The detailed balance of the conversion reaction between the precursor $P$ and the building block $B$ can be broken effectively by coupling the system to an external energy supply.
This can for instance be achieved by adding a second reaction that converts $P$ into $B$ using  additional energy supplied by fuel and waste components, which we respectively denote by $F$ and $W$; see \figref{fig:4_illustration_active_emulsions}(a).
The associated chemical reactions,
\begin{align}
	\tag{R1}
	P &\rightleftharpoons B \;,
\\
	\tag{R2}
	P + F &\rightleftharpoons B + W
	\;,
	\label{eqn:reaction2}
\end{align}
will allow us to drive the system out of equilibrium by controlling the concentrations of $F$ and $W$ externally. 
To study the interplay of these reactions with phase separation, we describe the system using a free energy density $f_4(c_P, c_B, c_F, c_W)$.
A simple choice is
\begin{align}
	f_4 &=\kb T \biggl[
		\sum_i c_i \log(\nu c_i) + \chi \nu c_P c_B
	\biggr]
	\;,
\end{align}
where the sum captures the entropic contributions of all species $i=P,B,F,W$ and the last term accounts for the interaction between $P$ and $B$.
For simplicity, we consider the case where the additional components~$F$ and $W$ are dilute and do not interact with $P$ and $B$.
In this case, the additional components do not affect the phase separation of $P$ and $B$.
Without chemical reactions, this system  phase separates similar to the binary system discussed in \secref{sect:passive_phase_sep}.
In particular, the chemical potentials~$\mu_i =  \partial f_4/\partial c_i$ for $i=P,B,F,W$ are homogeneous throughout the system and the chemical potential difference $\bar\mu = \mu_B - \mu_P$ between $P$ and $B$ is generally non-zero and equal to the Lagrange multiplier fixing the conservation of components; see equation~\eqref{eq:chem_pot_GL}.

When chemical reactions are present, they are driven by the chemical potential differences of their products and reactants.
If the system is isolated, it will typically evolve toward a homogeneous thermodynamic equilibrium, as described in \secref{sec:droplet_dissolution}.
In contrast, this thermodynamic equilibrium may not be reached in open systems, for instance when the concentrations of the fuel and waste components are controlled at the boundary by coupling the system to a reservoir.
In particular, the chemical potential difference $\bar\mu_2 = \mu_F - \mu_W$  between fuel and waste can be directly controlled at the boundary.
To show that imposing $\bar\mu_2 \neq 0$ breaks detailed balance of the conversion rates between the precursor $P$ and the building block $B$, we next investigate the forward and backward fluxes of the two reaction pathways~\eqref{eqn:reaction1} and \eqref{eqn:reaction2}.
These fluxes must obey conditions analogous to \Eqref{eqn:foward_backward_ratio},
\begin{subequations}
\label{eqn:detailed_balance_2reactions}
\begin{align}
	\frac{\sF[1]}{\sB[1]}
	&= \exp\left(-\frac{\bar\mu}{\kb T}\right)
\\
	\frac{\sF[2]}{\sB[2]} &= \exp\left(\frac{\bar\mu_2 - \bar\mu}{\kb T}\right)
	\;.
\end{align}
\end{subequations}
These equations show that detailed balance of the individual reaction pathways can only be obtained for vanishing exchange chemical potentials, $\bar\mu = \bar\mu_2=0$.
Moreover, the total forward flux $\sF=\sF[1]+\sF[2]$ and total reverse flux $\sB=\sB[1]+\sB[2]$ of the conversion between $P$ and $B$ obey
\begin{align}
	\frac\sF\sB = e^{-\frac{\bar\mu}{\kb T}} \left[
			1 + \frac{1}{1 + \frac{\sB[1]}{\sB[2]}}\left(e^{\frac{\bar\mu_2}{\kb T}} - 1\right)
		\right]
	\;,
	\label{eqn:breaking_detailed_balance_rates}
\end{align}
which is only compatible with the detailed balance of the rates given by  \Eqref{eqn:foward_backward_ratio} in the special case $\bar\mu_2 = 0$.
Consequently, detailed balance of the rates is broken for $\bar\mu_2\neq0$.

The detailed balance conditions given by equations~\eqref{eqn:detailed_balance_2reactions}  only constrain the ratios of the forward and backward fluxes of the two chemical reactions~\eqref{eqn:reaction1} and \eqref{eqn:reaction2}.
Analogous to \Eqref{eqn:reaction_rate}, the total reaction fluxes $\s[i]=\sF[i]-\sB[i]$ for each pathway can be expressed in the linearized regime as
\begin{subequations}
\label{eqn:active_reaction_flux}
\begin{align}
\label{eqn:active_reaction_flux_a}
	\s[1] &= -\mob[1](\vect c) \, \bar\mu \;,
\\
\label{eqn:active_reaction_flux_b}
	\s[2] &= -\mob[2](\vect c)(\bar\mu - \bar\mu_2) 
	\;,
\end{align}
\end{subequations}
where $\mob[1]$ and $\mob[2]$ are Onsager coefficients 
that set the reaction rates.
These coefficients can depend on the concentrations $\vect c=\set{c_P, c_B, c_F, c_W}$ but  must be positive to ensure that the entropy productions $\mob[1]\bar\mu^2$ and $\mob[2] (\bar\mu - \bar\mu_2)^2$ are positive; see~\ref{sec:appendix_entropy_production}.
For simplicity, we here consider the case where the fuel $F$ and the waste $W$ are dilute and diffuse fast, so the local composition can be described by a single concentration $c=c_B \simeq \nu^{-1} - c_P$, where we consider the case of equal molecular volume~$\nu$ for precursors $P$ and building blocks $B$.
In particular, $\mob[1]$ and $\mob[2]$ mainly depend on $c$ in this case and we do not need to describe the dynamics of the additional components~$F$ and $W$ explicitly.
However, their chemical energy $\bar\mu_2$ affects the conversion between $P$ and $B$.
In particular, detailed balance of the conversion between $P$ and $B$ is broken in the reduced system where only these two components are described.

The dynamical equation of the system with broken detailed balance of the rates is given by the conservative diffusion fluxes driven by $\nabla\bar\mu$ together with the non-conservative reaction flux $\sTot = \s[1] + \s[2]$ given in \Eqsref{eqn:active_reaction_flux}.
Using a conservation law analogous to \Eqref{eqn:conservation_law}, we obtain
\begin{align}
	\partial_t c &=
		\nabla \cdot \bigl( \mob(c) \, \nabla \bar\mu(c) \bigr)
		+ \kappa \mobTot(c) \, \nabla^2 c
		+ \sLoc(c)
	\;,
	\label{eqn:pde_broken_db}
\end{align}
where $\mobTot=\mob[1] + \mob[2]$ and we split the total reaction flux $\sTot$ into a contribution akin to a diffusion term related to surface tension and a local contribution $\sLoc(c) = \mob[2](c) \bar\mu_2 - \mobTot(c) f'(c)$.
We will show below that the term $s(c)$ can affect the dynamics significantly, while the additional diffusion term has only a minor influence since it only increases the effective diffusion constant.
Note that if detailed balance is obeyed ($\bar\mu_2=0$), \Eqref{eqn:pde_broken_db} reduces to \Eqref{eqn:active_pde_equilibrium} when $\mobTot$ is replaced by $\mobR$.
In this case, the system approaches the homogeneous equilibrium states that we discussed in \secref{sec:thermodynamics_reactions}.
Conversely, more complex behavior can be expected when detailed balanced of the rates is broken by maintaining a non-zero chemical potential difference $\bar\mu_2$. 

Equation~\eqref{eqn:pde_broken_db} without the additional diffusion term has been proposed before as a simple combination of phase separation with chemical reactions~\cite{Glotzer1994, Christensen1996}, albeit without an explicit breaking of detailed balance~\cite{Lefever1995}.
Instead, simple reaction rate laws~$s(c)$ have been analyzed~\cite{Tran-Cong1996, Christensen1996, Glotzer1994, Toxvaerd1996, Zhu2003}.
For instance, it has been shown that first-order rate laws are equivalent to systems with long-ranged interactions of the Coulomb type~\cite{Liu1989, Christensen1996} and that such interactions affect pattern formation~\cite{Motoyama1996, Motoyama1997, Muratov2002, Sagui1995, Sagui1995b}, \eg in block copolymers~\cite{Leibler1980, Ohta1986, Liu1989, Li2015, Glasner2015}.

\label{sec:active_homogeneous_state}
The effect of the chemical reactions with broken detailed balance of the rates can be highlighted by considering the stationary homogeneous states $c(\vect r) = c_0$.
Equation \eqref{eqn:pde_broken_db} implies that $c_0$ must satisfy $\sLoc(c_0) = 0$, \ie chemical reactions are balanced.
When detailed balance is obeyed ($\bar\mu_2=0$), this condition is equivalent to the equilibrium condition $f'(c_0)=0$ that we encountered before.
Conversely, the homogeneous stationary states are altered when detailed balance of the rates is broken ($\bar\mu_2 \neq 0$).

We examine the stability of the homogeneous states using a linear stability analysis, as described in \secref{sec:dynamical_equation}.
We find that perturbations described by wave vectors~$\vect q$  grow in the linear regime with a rate 
\begin{align}
	\omega(\vect q) &= 
	s'(c_0)
	- \vect q^2 \zeta(c_0)
	- \vect q^4 \mob(c_0) \kappa
	\;,
	\label{eqn:perturbation_growth_rate}
\end{align}
where $\zeta(c_0) = \mob(c_0) f''(c_0) + \kappa \mobTot(c_0)$ and 
 primes denote derivatives with respect to $c$.
The associated stationary state is stable only if $\omega(\vect q)$ is negative for all $\vect q$, which is the case if the maximum
\begin{equation}
	\max_{\vect q}\bigl(\omega(\vect q)\bigr)
	= \begin{cases}
		\sLoc'(c_0)
			& \zeta(c_0) \ge 0
		\\[4pt]
	s'(c_0) + \dfrac{\zeta^2(c_0)}{4\kappa  \Lambda (c_0)}
			& \zeta(c_0) < 0
	\end{cases}
	\label{eqn:perturbation_growth_rate_max}
\end{equation}
is negative.
In the simple case without reactions ($\mobTot(c)=0$, implying $\zeta=\mob f''(c_0)$ 
and $s'(c_0)=0$), we obtain the spinodal instability for $f''(c_0) < 0$, which is discussed in  \secref{sec:dynamical_equation} and shown by the grey line in figure~\ref{fig:4_free_energy_dispersions}(b).
If the reactions obey detailed balance of the rates, implying $s(c) = -\mobR f'(c)$ and thus $s'(c_0)=-\mobR(c_0) f''(c_0)$, the homogeneous states corresponding to minima of the free energy density are stable; see \secref{sec:active_passive_conversion} and the blue line in figure~\ref{fig:4_free_energy_dispersions}(b).



Chemical reactions that break detailed balance of the rates ($\bar\mu_2\neq0$) can modify the stability of stationary states; 
see red lines in figure~\ref{fig:4_free_energy_dispersions}(b).
Equation~\eqref{eqn:perturbation_growth_rate_max} implies that a homogeneous state can become unstable when \mbox{$\sLoc'(c_0) > 0$}~\cite{Carati1997,Bazant2017}.
This case corresponds to an auto-catalytic reaction, since it implies that the production of building blocks~$B$ accelerates with larger concentration of $B$. 
Conversely, homogeneous states are generally stabilized by auto-inhibitory reactions, where $\sLoc'(c_0) < 0$. 
In particular, these effects can be observed for homogeneous states close to the equilibrium states that we discussed before, where $f''(c_0) > 0$ and thus $\zeta>0$.
In this regime, our stability analysis suggests that droplets form and grow easily in auto-catalytic systems, while spontaneous formation might be suppressed by auto-inhibitory reactions.
In particular, the behavior can be regulated independent of the free energy density using the externally controlled chemical potential difference~$\bar\mu_2$.

Although it is not directly relevant to active droplets, the dynamics of homogeneous stationary states with $\zeta<0$ is also interesting.
It implies $f''(c_0) < 0$ and the corresponding homogeneous stationary states are thus closer to maxima of the free energy density.
In this case, there exists a regime with weakly auto-inhibitory reactions ($-\zeta^2/(4\kappa\mob) < \sLoc'(c_0) < 0$) where a range of finite wave vectors~$q$ becomes unstable and pattern formation is expected.
This regime has been studied extensively in the literature~\cite{Glotzer1994, Glotzer1994a, Christensen1996} and typically leads to stripe-like patterns.
Since we here focus on active emulsions, we exclusively consider the case $\zeta > 0$ below.

In this section, we discussed a model system where additional fuel and waste components break the detailed balance of the rates of the conversion of the main components $P$ and $B$.
We showed that this combination of phase separation with non-equilibrium chemical reactions  can suppress the instability associated with spinodal decomposition and we will now analyse the impact on the dynamics of droplets.

\subsubsection{Coarse-grained description of active droplets.}
\label{sec:coarse_grained_active_droplets}
We next discuss the dynamics of active droplets in the case of strong phase separation, where the interfacial width~$\width$ is small compared to the droplet radius~$R$.
In this case, the volume occupied by the interface and thus the chemical reactions inside the interfacial region are negligible.
Conversely, we will show that the chemical reactions producing and destroying droplet material in the bulk phases influence the droplet dynamics significantly.
We here consider a coarse-grained description, where a thin interface separates the inside of the droplet with a high concentration of droplet material~$B$ from the dilute phase outside, analogous to \secref{sect:intro_growth_single_droplet}.
The dynamics in both phases is described by \Eqref{eqn:pde_broken_db}, but since the concentration variations are small within the phases, it can be approximated by a reaction-diffusion equation~\cite{Zwicker2015},
\begin{equation}
	\partial_t c \simeq D_\alpha \nabla^2 c + \sLoc(c)
	\label{eqn:4_reaction_diffusion}
	\;,
\end{equation}
where $D_\alpha = \mob(\cBase_\alpha) f''(\cBase_\alpha) + \kappa\mobTot(\cBase_\alpha)$ denotes the diffusivity in the two phases $\alpha=\text{in}, \text{out}$.
The first term in the expression for~$D_\alpha$ stems from the conservative fluxes and is thus equivalent to \Eqref{eq_diff_constant} while the second term captures the apparent diffusion due to chemical conversion.
Note that the diffusivity~$\DIn$ inside the droplet is generally different than the diffusivity~$\DOut$ outside.
The same approximation that led to \Eqref{eqn:4_reaction_diffusion} can also be used to linearize the reaction flux in the two phases,
\begin{equation}
	\sLoc(c) \simeq \begin{cases}
		\sBaseIn - \kIn (c - \cBaseIn) & \text{inside} \\
		\sBaseOut - \kOut (c - \cBaseOut) & \text{outside}
	\end{cases}
	\label{eqn:reaction_flux_linear}
	\;,
\end{equation}
where $\sBaseIn = \sLoc(\cBaseIn)$ and $\sBaseOut = \sLoc(\cBaseOut)$ are the reaction fluxes when the concentrations are at their equilibrium values~$\cBaseIn$ and $\cBaseOut$ in the two phases, respectively.
Deviations from these values are accounted for by $\kIn = -\sLoc'(\cBaseIn)$ and $\kOut = -\sLoc'(\cBaseOut)$, which can be interpreted as elasticity coefficients of the chemical reactions~\cite{Kacser1995}.

The basal fluxes~$\sBaseIn$ and $\sBaseOut$ need to have opposite sign for droplets to exist.
If they had the same sign, droplet material would either be destroyed ($\sBaseIn, \sBaseOut < 0$) or produced everywhere ($\sBaseIn, \sBaseOut > 0$), which both implies a homogeneous stationary state.
To see under which conditions the basal fluxes have opposite sign, we express them as
\begin{subequations}
\label{eqn:active_base_reaction}
\begin{align}
	\sBaseIn &=
		-\mob[1](\cBaseIn) \, \bar\mu
		-\mob[2](\cBaseIn)(\bar\mu - \bar\mu_2)
	\;,
\\
	\sBaseOut &=
		-\mob[1](\cBaseOut) \, \bar\mu
		-\mob[2](\cBaseOut)(\bar\mu - \bar\mu_2)
	\;,
\end{align}
\end{subequations}
using \Eqsref{eqn:active_reaction_flux}.
To give a concrete example, we here consider weak chemical reactions, so that the diffusive fluxes almost equilibrate the exchange chemical potential~$\bar\mu$, which is therefore approximately homogeneous.
If the building block~$B$ is of higher energy than the precursor~$P$ ($\bar\mu > 0$), equations \eqref{eqn:active_base_reaction} imply that droplet material is produced outside the droplet ($\sBaseOut>0$) and destroyed within ($\sBaseIn < 0$) if the fuel supplies sufficient energy ($\bar\mu_2 > \bar\mu > 0$) and the Onsager coefficients obey
\begin{align}
		\frac{\mob[1](\cBaseOut)}{\mob[2](\cBaseOut)}
		&<  \frac{\bar\mu_2}{\bar\mu} - 1
		< 	\frac{\mob[1](\cBaseIn)}{\mob[2](\cBaseIn)}
	\label{eqn:reaction_onsager_condition}
	\;.
\end{align}
For instance, if $\bar\mu$ is equal in the two phases and the external energy input is given by $\bar\mu_2 = 2\bar\mu$, reaction~\eqref{eqn:reaction1} must be faster than reaction~\eqref{eqn:reaction2} inside the droplet, $\mob[1](\cBaseIn)>\mob[2](\cBaseIn)$, while the opposite is true outside, $\mob[1](\cBaseOut) < \mob[2](\cBaseOut)$.
In this case, the conversion reaction~\eqref{eqn:reaction1} proceeds from the building block $B$ to the precursor $P$ spontaneously ($\s[1]<0$), while reaction~\eqref{eqn:reaction2} converts fuel to waste to produce the high-energy building block $B$ from $P$ ($\s[2]>0$); see figure~\ref{fig:4_illustration_active_emulsions}.
If the coefficients~$\mob[1]$ and $\mob[2]$ obey \Eqsref{eqn:reaction_onsager_condition} the total reaction flux~$\sTot = \s[1] + \s[2]$ is then positive outside the droplet, while it is negative inside.
A concrete implementation of such a system is discussed in the Supporting Information of \refcite{Zwicker2017}.

We showed that droplet material can be produced in one phase while it is destroyed in the other when the basal fluxes $\sBaseIn$ and $\sBaseOut$ have opposite sign.
Do similar conditions apply to the elasticity coefficients~$\kIn$ and $\kOut$?
We showed in \secref{sec:active_homogeneous_state} that negative $\sLoc'(c)$ has a stabilising effect and we would thus expect that droplets can persist if $k>0$; see also figure~\ref{fig:4_free_energy_dispersions}(b).
Conversely, since positive $\sLoc'(c)$ can destabilise homogeneous phases, the active droplets we discuss here might  be unstable if $k<0$.
However, there is a smallest length scale~$q_{\rm max}^{-1}$ below which the instability cannot develop.
This length scale can be determined from \Eqref{eqn:perturbation_growth_rate} and the condition $\omega(q_{\rm max})=0$, yielding
\begin{align}
	q_{\rm max}^{-1}
&\simeq
	\left[\frac{\zeta(\cBaseIn)}{ s'(\cBaseIn)} \right]^{\frac12}
	\;,
\end{align}
for weak reactions, $s'(\cBaseIn) \ll \zeta(\cBaseIn)$. 
In the case where $q_{\rm max}^{-1}$ is large compared to the droplet radius, the instability is effectively suppressed and the droplet could be stable even if $\kIn < 0$.
Conversely, the dilute phase will typically be large compared to $q_{\rm max}^{-1}$, such that an instability would develop there if $\kOut < 0$.
In the following, we thus consider all values of $\kIn$, but restrict our discussion to $\kOut > 0$.

\begin{figure}[tb]
	\begin{center}
		\includegraphics[width=1.0\columnwidth]{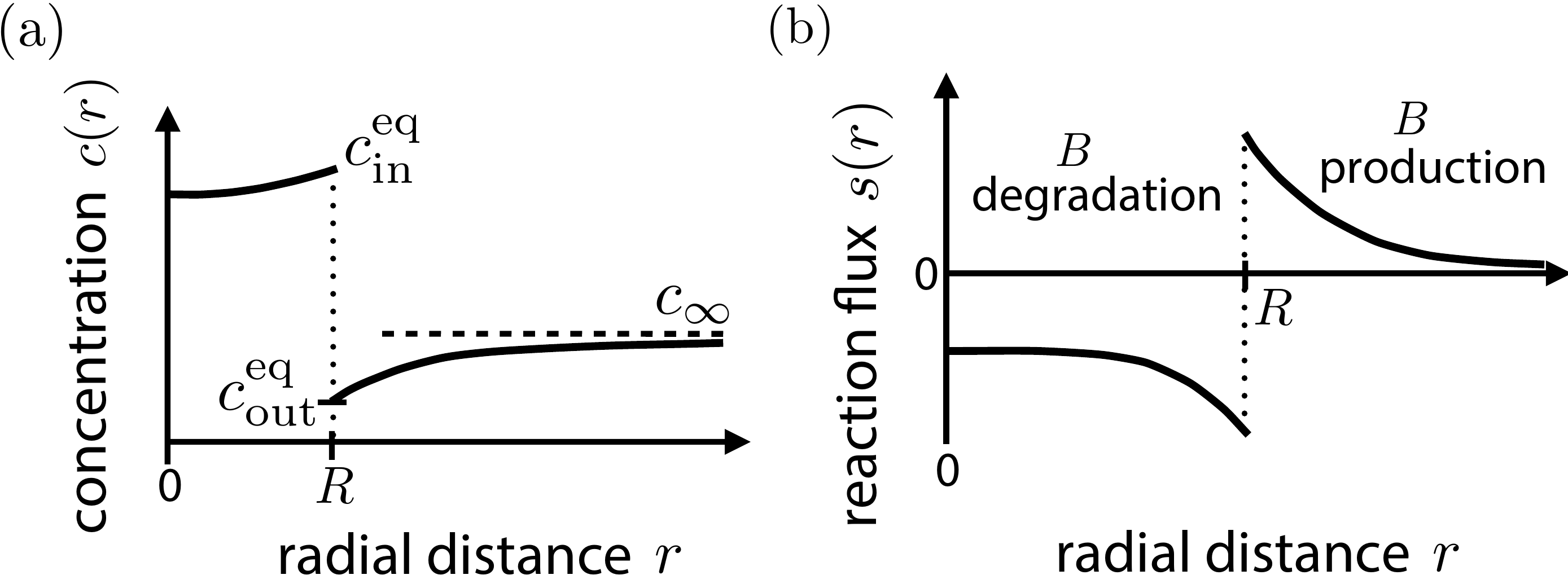}
	\end{center}
	\caption{
	Schematic picture of an externally maintained active droplet, where droplet material is produced in the solvent phase.
	(a) Concentration $c$ of droplet material as a function of the radial distance~$r$.
	The chemical reactions modify the profiles compared to the passive droplet shown in Figure~\ref{fig:ch2_2}(a).
	(b) Reaction flux $\sLoc$ as a function of $r$.
	Droplet material is produced outside the droplet while it is degraded inside.
	The droplet dynamics can strongly deviate from the passive case.
	In particular, Ostwald ripening can be suppressed and droplets even divide spontaneously (see section \ref{sect:Ostwald_supression} and \ref{sect:spon_division}).
		}
	\label{fig:4_concentration_profile}
\end{figure}

We distinguish different classes of active droplets based on where droplet material is produced.
If the reaction fluxes $\sBaseIn$ and $\sBaseOut$ have the same sign, droplet material is produced or destroyed in the entire system and the only stable stationary state are homogeneous. 
Consequently, stable droplets can only exist if $\sBaseIn$ and $\sBaseOut$ have opposite sign:
We denote the case where droplet material is produced outside the droplets ($\sBaseOut > 0$, $\sBaseIn < 0$) as \textit{externally maintained droplets}, while the opposite case where droplets produce their own material ($\sBaseIn > 0$, $\sBaseOut < 0$) is called \textit{internally maintained droplets}.

\subsubsection{Droplet growth equation.}
\label{sec:active_single}

We begin by studying a single, isolated active droplet surrounded by a large dilute phase.
The droplet grows if there is a net flux of droplet material toward its surface, see \Eqref{eq:interface_vel}.
In the simple case of a spherical droplet of given radius~$R$ and volume $\Vd= \frac{4\pi}{3} R^3$, the growth rate reads
\begin{align}
	\difffrac{\Vd}{t} \simeq \frac{\FluxIn - \FluxOut}{\cBaseIn - \cBaseOut}
	\label{eqn:droplet_growth_sym}
	\;,
\end{align}
where we neglected surface tension effects in the denominator and defined the integrated surface fluxes $J_\mathrm{in/out}=4\pi R^2 \, \vect{n} \cdot \vect{j}_\mathrm{in/out}$.
These fluxes can be determined from the stationary solutions $\steady{c}(r)$ that follow from solving \Eqref{eqn:4_reaction_diffusion} with the boundary conditions at the interface given in \Eqref{eq:final_GT_relations}, see \figref{fig:4_concentration_profile}(a).

The flux~$\FluxIn$ inside the droplet interface can be obtained in the quasi-static limit, where it equals the reaction flux $\sFluxIn = \int \diff^3 r \, \sLoc(\steady{c}(\vect r))$ inside the droplet volume.
In the typical case where the radius $R$ is small compared to the length $\DiffLenIn= (\DIn/\abs{\kIn})^{\frac12}$ generated by the reaction-diffusion system, the concentration inside the droplet is $c_*(\vect r) \simeq \cBaseIn$, implying
\begin{align}
	\FluxIn \simeq \sBaseIn \Vd
	\;.
	\label{eqn:active_flux_inside}
\end{align}
Droplet material is thus transported toward the interface ($\FluxIn>0$) only in internally maintained droplets ($\sBaseIn > 0$).

The integrated flux $\FluxOut$ outside the droplet interface can also be obtained in a quasi-static approximation.
In contrast to the passive case discussed in \secref{sect:intro_growth_single_droplet}, the supersaturation $\supSat = (\cInfty - \cBaseOut)/\cBaseOut$ far away from droplets is now created by chemical reactions.
In the typical case where the dilute phase is large compared to the length scale~$\DiffLenOut = (\DOut/\abs{\kOut})^{\frac12}$, the chemical reactions equilibrate far away from droplets and the composition thus reaches the value $\cInfty = \cBaseOut + \sBaseOut/\kOut$, so $\sLoc(\cInfty) = 0$; see \figref{fig:4_concentration_profile}(b).
The associated supersaturation reads $\supSat = \sBaseOut/(\kOut\cBaseOut)$ and is thus positive if $\sBaseOut > 0$ since the reaction is auto-inhibitory outside droplets ($\kOut > 0$); see \secref{sec:coarse_grained_active_droplets}.
The resulting transport of droplet material can be quantified by the flux outside the droplet interface, which reads
\begin{equation}
	\FluxOut \simeq 
	4 \pi  \DOut R \cBaseOut \left(\frac{\ell_\gamma}{R} - \supSat\right)
	\label{eqn:active_flux_outside}
\end{equation}
in the typical case $R \ll \DiffLenOut$.
The first term in the bracket captures the effect of surface tension, which is typically small.
Neglecting this term, we find that droplet material is transported toward the interface ($\FluxOut < 0$) if the dilute phase is  supersaturated ($\supSat > 0$), which is the case only in externally maintained droplets ($\sBaseOut>0$).

The droplet growth rate following from combining \Eqsref{eqn:droplet_growth_sym}--\eqref{eqn:active_flux_outside} reads
\begin{align}
	\difffrac{R}{t} \simeq \frac{\DOut\cBaseOut}{R(\cBaseIn - \cBaseOut)}\left(
		\supSat
		- \frac{\ell_\gamma}{R}
		+ \frac{R^2\sBaseIn}{3\DOut \cBaseOut}
	\right)
	\label{eqn:active_droplet_growth}
	\;.
\end{align}
The first term in the bracket describes droplet growth due to a supersaturated environment ($\supSat > 0$) or shrinkage in undersaturated environments ($\supSat < 0$).
The second term, which is only relevant for small droplets, captures the reduction of growth due to surface tension~$\gamma$.
The last term describes the growth of the droplet due to production of droplet material inside if $\sBaseIn>0$ or its shrinking when $\sBaseIn < 0$.
Note that \Eqref{eqn:active_droplet_growth} reduces to \Eqref{eq:dRdt} if chemical reactions are absent ($\sBaseIn=0$) and the supersaturation $\supSat$ is imposed.

\subsubsection{Single droplet in an infinite system.}
We begin by discussing the growth of a single droplet in an environment with constant supersaturation~$\supSat$.
This corresponds for instance to an infinite system where the supersaturation reaches its equilibrium value $\supSatEq=\sBaseOut/(\kOut\cBaseOut)$ far away from an isolated droplet.
The stationary states of this system can be determined from \Eqref{eqn:active_droplet_growth} and correspond to radii~$R$ for which the bracket vanishes.
The associated cubic equation in $R$ has at most three solutions, which we now classify.
If the chemical reactions are too strong, there are no physical solutions.
In particular, if $\abs\sBaseIn > \sBaseInMax$ with
\begin{align}
	\sBaseInMax &= \frac{4 \cBaseOut \DOut \abs{\supSat}^3}{9 \ell_\gamma^2}
	\;,
\end{align}
two solutions are complex while the third one is either negative (if $\sBaseIn<0$) or smaller than the interface width (if $\sBaseIn>0$), which is both unphysical.
Consequently, stable droplets can only exist for moderate chemical reactions, $\abs\sBaseIn < \sBaseInMax$.
In this case, the polynomial equation has three real solutions.
One of the solutions is always negative and thus unphysical, while the other two read
\begin{subequations}
\label{eqn:stationary_state_radii}
\begin{align}
	\RsA &\simeq \frac{\ell_\gamma}{\supSat} \, ,
\\
	\RsB &\simeq \left(\frac{3\DOut\supSat \cBaseOut}{-\sBaseIn}\right)^{\frac12}
		-\frac{\ell_\gamma}{2\supSat}
	\label{eqn:stationary_state_radiiB}
	\;,
\end{align}
\end{subequations}
which is correct up to linear order in $\ell_\gamma/R$.
The dynamics in the vicinity of these states follow from a linear stability analysis of \Eqref{eqn:active_droplet_growth}, where the droplet radius is perturbed away from the stationary state and we determine whether the dynamics will bring the droplet back to its stationary state or not~\cite{Zwicker2015}.
This gives us enough information to discuss the growth behavior of active droplets qualitatively.

\begin{figure}
	\includegraphics[width=\columnwidth]{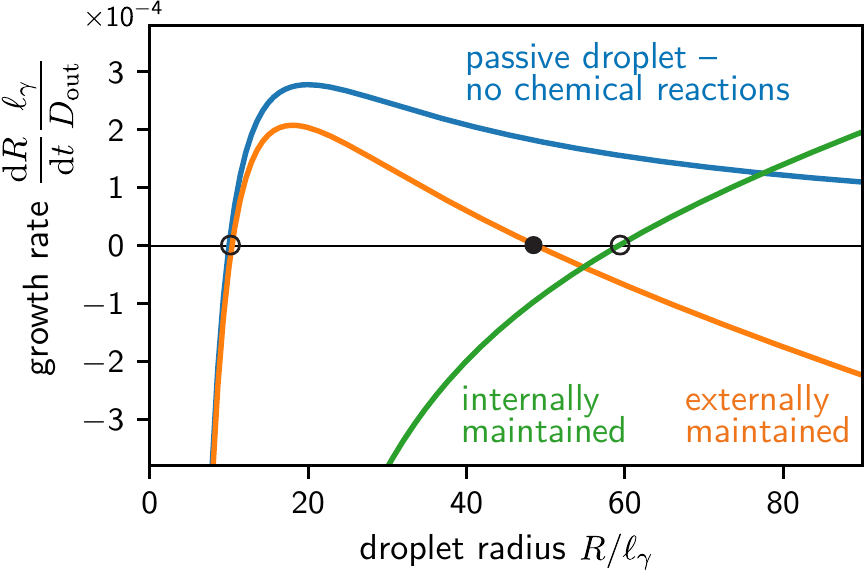} 
	\caption{
	Growth rate~$\diff R/\diff t$ of a single droplet in an infinite system as a function of the droplet radius~$R$.
	Shown are the values given by \Eqref{eqn:active_droplet_growth} for
	passive droplets (blue line; $\sBaseIn=0$, $\supSat=0.1$),
	externally maintained droplets (orange line; $\sBaseIn/\cBaseOut= -10^{-4} k_0$, $\supSat=0.1$), and
	internally maintained droplets (green line; $\sBaseIn/\cBaseOut= 10^{-4} k_0$, $\supSat=-0.1$) for $\cBaseIn/\cBaseOut=10$ where $k_0=\DOut/\ell_\gamma^2$.
	Unstable stationary states are marked with open circles, while the only stable one is marked with a black disk.
	}
	\label{fig:4_growth_rate}
\end{figure}

In the case of externally maintained droplets ($\sBaseIn< 0$, $\supSat>0$), we have $0 <  \RsA <  \RsB$, since $\abs\sBaseIn < \sBaseInMax$.
Here, the radius~$\RsA$ corresponds to an unstable state, while $\RsB$ is stable, see \figref{fig:4_growth_rate}.
Droplets smaller than $\RsA$ dissolve and disappear, so $\RsA$ is a critical radius  similar to the one discussed in section~\ref{sect:intro_growth_single_droplet}.
Externally maintained droplets that are larger than the critical radius~$\RsA$ grow until they reach the stable stationary state with radius~$\RsB$.
This state is not present for passive droplets in an infinite system and must thus be a consequence of the chemical reactions.
This can be seen by analyzing the two fluxes $\FluxIn$ and $\FluxOut$, which must be equal in the stationary state.
However, $\FluxIn$ scales with the droplet volume,
 while $\FluxOut$ scales with its radius, see \Eqref{eqn:active_flux_outside}.Consequently, if the droplet radius exceeds $\RsB$, the loss due to $\FluxIn$ dominates and the droplet shrinks back to the stationary state.
The chemical turnover inside the droplet thus stabilizes the stationary state.
Note that the two stationary states given in \Eqref{eqn:stationary_state_radii} only exist if the chemical reactions are not too strong ($\abs\sBaseIn < \sBaseInMax$).
In the limiting case $\sBaseIn = -\sBaseInMax$ the situation is degenerated and the two stationary states are identical.
The corresponding radius~$R_{\rm min}^{\rm ext}= (3 \ell_\gamma)/(2\supSat)$ can be determined from \Eqref{eqn:active_droplet_growth} and corresponds to the smallest externally maintained droplet that can be stable. 

In the case of internally maintained droplets ($\sBaseIn > 0$, $\supSat < 0$), the first solution given in \Eqref{eqn:stationary_state_radii} is negative and thus unphysical.
The second solution is always positive, but unstable to perturbations.
Consequently, $\RsB$ is the critical droplets size of internally maintained droplets, which can be significantly larger than critical sizes in externally maintained droplets.
Nucleation is thus typically suppressed efficiently, but it can be promoted by catalytically active particles, which catalyze the production of droplet material at their surface~\cite{Zwicker2014, Zwicker2015}.
Internally maintained droplets larger than the critical size grow up to the system size and there is no characteristic stable size.
This is because such droplets grow quicker if they become larger and this autocatalytic growth only stops when the dilute phase is depleted of material~$P$, which can only happen in a finite system~\cite{Zwicker2015}.

\subsubsection{Single droplet in a finite system.}
So far, we only considered systems that are large compared to the droplet size, so the supersaturation~$\supSat$ is effectively constant.
In contrast, in small systems a growing droplet can deplete the surrounding dilute phase significantly.
In particular, the average concentration~$\cOut$ of droplet material in the dilute phase evolves as
\begin{align}
	\difffrac{\cOut}{t} = \sLoc(\cOut) + \frac{\FluxOut}{\Vsys - \Vd}
	\label{eqn:supersaturation_dynamics_single}
	\;,
\end{align}
where the first term on the right hand side originates from the chemical reactions in the dilute phase and the last term accounts for the diffusive flux at the interface of the droplet of volume~$\Vd$.
In large systems ($\Vsys\gg \Vd$), the last term is negligible and $\cOut$ relaxes to $\cInfty$, where chemical equilibrium is obeyed, $\sLoc(\cInfty)=0$.
At this point, the supersaturation~$\supSat= (\cOut - \cBaseOut)/\cBaseOut$ attains its equilibrium value $\supSatEq=\sBaseOut/(\kOut\cBaseOut)$ and is thus independent of the droplet size, consistent with our assumptions in the previous subsection.
In small systems, however, the last term in \Eqref{eqn:supersaturation_dynamics_single} is not negligible and $\supSat$ depends on the droplet volume~$\Vd$.
For instance, in the case of externally maintained droplets, we have $\FluxOut < 0$ and the concentration~$\cOut$ outside the droplet is thus lower than $\cInfty$.
Consequently, the supersaturation is smaller for smaller systems and we would expect a reduced stationary droplet radius~$\RsB$, see \Eqref{eqn:stationary_state_radiiB}.
Indeed, when we solve for the stationary states of \Eqsref{eqn:active_droplet_growth} and \eqref{eqn:supersaturation_dynamics_single}, we find in the simple case of large diffusion and small surface tension that there is a stationary state with volume
\begin{align}
	\steady V &\simeq \frac{\sBaseOut\Vsys}{\sBaseOut - \sBaseIn}
	\label{eqn:steady_volume_small}
	\;,
\end{align}
which is stable for externally maintained droplets and unstable for internally maintained ones.
Consequently, the size of stable stationary droplets that are externally maintained depends on system size in small systems, while it is independent of the system size in large systems, see \Eqref{eqn:stationary_state_radii}.

Similar to passive droplets, we find that active droplets also have a critical radius, below which they shrink and disappear.
Consequently, droplets can only be nucleated spontaneously if a concentration fluctuation creates a large enough initial droplet.
In externally maintained droplets, this mechanism is similar to passive droplets, where the nucleation barrier is higher for larger surface tension and smaller supersaturation.
Conversely, the critical radius is generally larger for internally maintained droplets, but it becomes smaller for larger surface tension and smaller supersaturation.
This is because only a large enough droplet can produce enough droplet material to balance the efflux into the dilute phase.
This efflux is smaller for larger surface tension, so surface tension actually helps to nucleate internally maintained droplets.

Once droplets exceed their critical radius they grow spontaneously by absorbing droplet material from the surrounding (passive and externally maintained droplets) or by producing more droplet material (internally maintained droplets).
In the first case, the droplet growth rate is larger for smaller droplets and higher supersaturation.
However, the growth of externally maintained droplets comes to a stop at a finite size, at which the  loss of droplet material due to the chemical reactions inside is balanced by its influx over the surface.
Conversely, internally maintained droplets grow indefinitely in an infinite system, similar to passive droplets.
However, in contrast to passive droplets, this growth accelerates because of its autocatalytic nature.
Consequently, in the simple case of a binary fluid in an infinite system, only externally maintained droplets reach a finite droplet size, while both passive and internally maintained droplets grow to occupy the whole system.

\subsection{Arrest of droplet coarsening: Suppression of Ostwald ripening}\label{sect:Ostwald_supression}

\begin{figure*}
	\raggedleft
	\includegraphics[width=\textwidth]{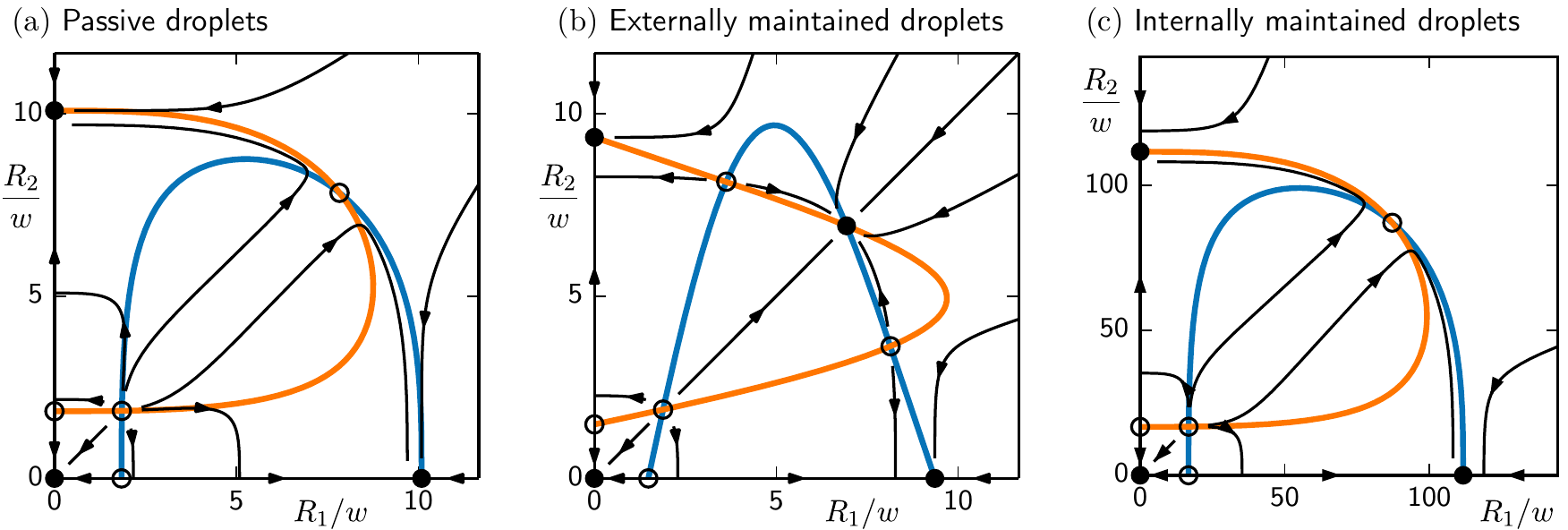}
	\caption{Behavior of two droplets as a function of their radii~$R_1$ and $R_2$, normalized with the interface width~$w$.
	The black arrows indicate the temporal evolution of the state variables $R_1$ and $R_2$ following from \Eqref{eqn:active_droplet_growth}, with $\supSat$ given by \Eqref{eqn:supersaturation_multiple}.
	The blue and orange lines are the nullclines, which indicate where the growth rate of droplets 1 and 2 vanish, respectively.
	Their intersections are stable (disks) or unstable fixed points (open circles).
	(a)~Passive droplets ($\sBaseIn = \sBaseOut = 0$)
	(b)~Externally maintained droplets ($\sBaseIn < 0$, $\sBaseOut > 0$)
	(c)~Internally maintained droplets ($\sBaseIn > 0$, $\sBaseOut < 0$).
	}
	\label{fig:4_ostwald_ripening}
\end{figure*}

If many active droplets are present in the same system, their dynamics will be coupled because they share the material in the dilute phase.
Intuitively, this coupling will be stronger if the droplets are closer together.
For instance, for externally maintained droplets, the length scale over which they deplete the surrounding dilute phase is given by $\DiffLenOut$ and we thus expect that such droplets do not interact significantly when they are further apart.
In particular, their stationary state radius~$\RsB$ given in \Eqref{eqn:stationary_state_radiiB} should not be affected much.
Conversely, if externally maintained droplets are close, they compete strongly for the material produced in the solvent and might thus only reach a smaller size.
This is in strong contrast to passive and internally maintained droplets that grow unbounded and will thus interact eventually, independent of the initial separation.

We study the interaction of multiple droplets in the simple case of sparse systems, where droplets are far apart from each other.
In this case, the growth dynamics of the droplets are coupled because they all exchange material with the same dilute phase, but direct interactions between droplets can be neglected.	
Considering $N$ droplets in a finite system of volume~$\Vsys$, the supersaturation~$\supSat$ in the dilute phase evolves as
\begin{align}
	\difffrac{\supSat}{t} \simeq \sBaseOut - \kOut\cBaseOut\supSat + \frac{1}{\Vsys} \sum_{i=1}^N J_{\mathrm{out}, i}
	\label{eqn:supersaturation_dynamics_multiple}
	\;,
\end{align}
which follows by generalizing \Eqref{eqn:supersaturation_dynamics_single} to many droplets in the limit that $\Vsys$ is large compared to the total volume of all droplets, $\Vsys \gg \sum_i V_i$.
Here, $R_i$ and $V_i=\frac{4\pi}{3}R_i^3$ are the respective radii and volumes of the droplets for $i=1,\ldots,N$ and $J_{\mathrm{out}, i}$ is the flux of droplet material right outside the interface of the $i$-th droplet integrated over its surface, which follows from \Eqref{eqn:active_flux_outside}.
For simplicity, we here consider the quasi-static case, where the concentration profile between droplets relaxes quickly compared to the growth of the droplets themselves, which is the typical situation~\cite{Zwicker2015}.
In this case, $\supSat$ will attain its stationary state value
\begin{align}
	\steady\supSat&=
		\frac{4 \pi N \ell_\gamma \DOut + \frac{\sBaseOut \Vsys}{\cBaseOut}}
			{\kOut\Vsys + 4 \pi \DOut \sum_i  R_i} 
		\;.
	\label{eqn:supersaturation_multiple}
\end{align}
Note that in the limit of a dilute system, $\Vsys/N \gg \DiffLenOut^3$, we recover the supersaturation at chemical equilibrium, $\supSatEq = \sBaseOut/(\kOut\cBaseOut)$.
Conversely, in passive or dense systems, the supersaturation is set by the equilibrium condition at the droplet surfaces, $\steady\supSat = \ell_\gamma / \steady R$, when all droplets have the same radius~$\steady R$.

The growth rate of each individual droplet is still described by \Eqref{eqn:active_droplet_growth}, but with the supersaturation now given by \Eqref{eqn:supersaturation_multiple}.
In the simple case of two droplets in the same system, the growth dynamics can be illustrated graphically.
\Figref{fig:4_ostwald_ripening} shows that passive and internally maintained droplets exhibit Ostwald ripening.
Conversely, a new stable stationary state (black dot) can emerge for externally maintained droplets, where both droplets coexist at the same size.
\Figref{fig:4_ostwald_ripening} thus indicates that Ostwald ripening can be suppressed in active droplets.

To understand when Ostwald ripening is suppressed, we next analyze the state  where all droplets have the same stationary radius~$\steady R$, which can be determined from the stationary state of \Eqref{eqn:active_droplet_growth}. 
Similar to the discussion of isolated droplets above, we can then use a linear stability analysis to discuss the qualitative dynamics in the vicinity of this state.
The detailed analysis given in \refcite{Zwicker2015} shows that there are two independent perturbation modes with qualitatively different dynamics:
The fast mode associated with the total droplet volume describes the fact that all droplets quickly take up excess material from the dilute phase until the stationary state of the total droplet volume is reached.
All other perturbation modes are associated with a slower exchange of material between droplets.
These modes all have the same perturbation growth rate~$\omega$ given by
\begin{equation}
	\omega = \frac{1}{\cBaseIn - \cBaseOut} \left(
		\frac{\ell_\gamma \DOut \cBaseOut}{\steady R^3} + \frac{2 \sBaseIn}{3}
	\right)
	\label{eqn:pertubration_rate_multiple}
	\;.
\end{equation}
If $\omega$ is positive, the associated mode is unstable and material will flow from smaller to larger droplets, \eg during Ostwald ripening.
Conversely, negative~$\omega$ indicates stable states, where this coarsening is suppressed.

Passive systems ($\sBaseIn=0$) are always unstable ($\omega>0$) and the droplets thus exhibit Ostwald ripening as discussed in section~\ref{sect:Ostwald_intro} and shown in \figref{fig:4_ostwald_ripening}(a).
The redistribution of material between droplets is driven by surface tension, which causes a larger concentration of droplet material right outside of smaller droplets, see \Eqref{eq:final_GT_relations}.
Moreover, the associated perturbation rate~$\omega$ is smaller for larger mean droplet size, such that exchange of material will be slower between larger droplets.
This implies that the coarsening of droplets slows down and only stops when a single droplet remains.

Internally maintained droplets ($\sBaseIn>0$) are also unstable and the associated growth rate is larger than that of passive droplets.
This is because the autocatalytic growth allows larger droplets to outcompete smaller ones, independent of surface tension effects.
Internally maintained droplets are thus more unstable than passive ones, but they can be stabilised by particles that catalyse the production of droplet material within the droplets~\cite{Zwicker2015,Zwicker2014}.

Multiple externally maintained droplets ($\sBaseIn<0$) can coexist when $\omega < 0$.
This is the case if their radius~$\steady R$ exceeds the critical value
\begin{align}
	\Rstab = \left(
			\frac{3\DOut \ell_\gamma \cBaseOut}{-2\sBaseIn}
		\right)^{\frac13}
	\;,
	\label{eqn:active_Rstab}
\end{align}
see \Eqref{eqn:pertubration_rate_multiple}.
This expression reveals that the stability originates from a competition of the destabilising effect of surface tension, which tends to increase $\Rstab$, and the stabilising effects induced by the diffusive fluxes driven by the chemical reactions.
This is similar to the isolated droplets that we discussed in the previous section: the influx toward a droplet scales at most with the droplet radius, see \Eqref{eqn:active_flux_outside}, while the material loss scales with the volume.
Consequently, multiple externally maintained droplets can stably coexist when the supersaturation in the dilute phase sustains the influx.

We showed that multiple active droplets interact because they compete for the same material from the dilute phase.
The associated diffusive fluxes between droplets are caused by surface tension effects and chemical reactions.
The flux due to surface tension is generally destabilising and causes the classical Ostwald ripening.
Conversely, the flux due to chemical reactions can be either destabilising (for internally maintained droplets) or stabilising (for externally maintained droplets).
If the stabilising contribution of the chemical reactions is stronger than the destabilising one due to surface tension, multiple active droplets can coexist in a stable state; see \figref{fig:4_ostwald_ripening}(b).

\subsection{Spontaneous division of active droplets}
\label{sect:spon_division}

\begin{figure*}
	\raggedleft
	\includegraphics[width=\textwidth]{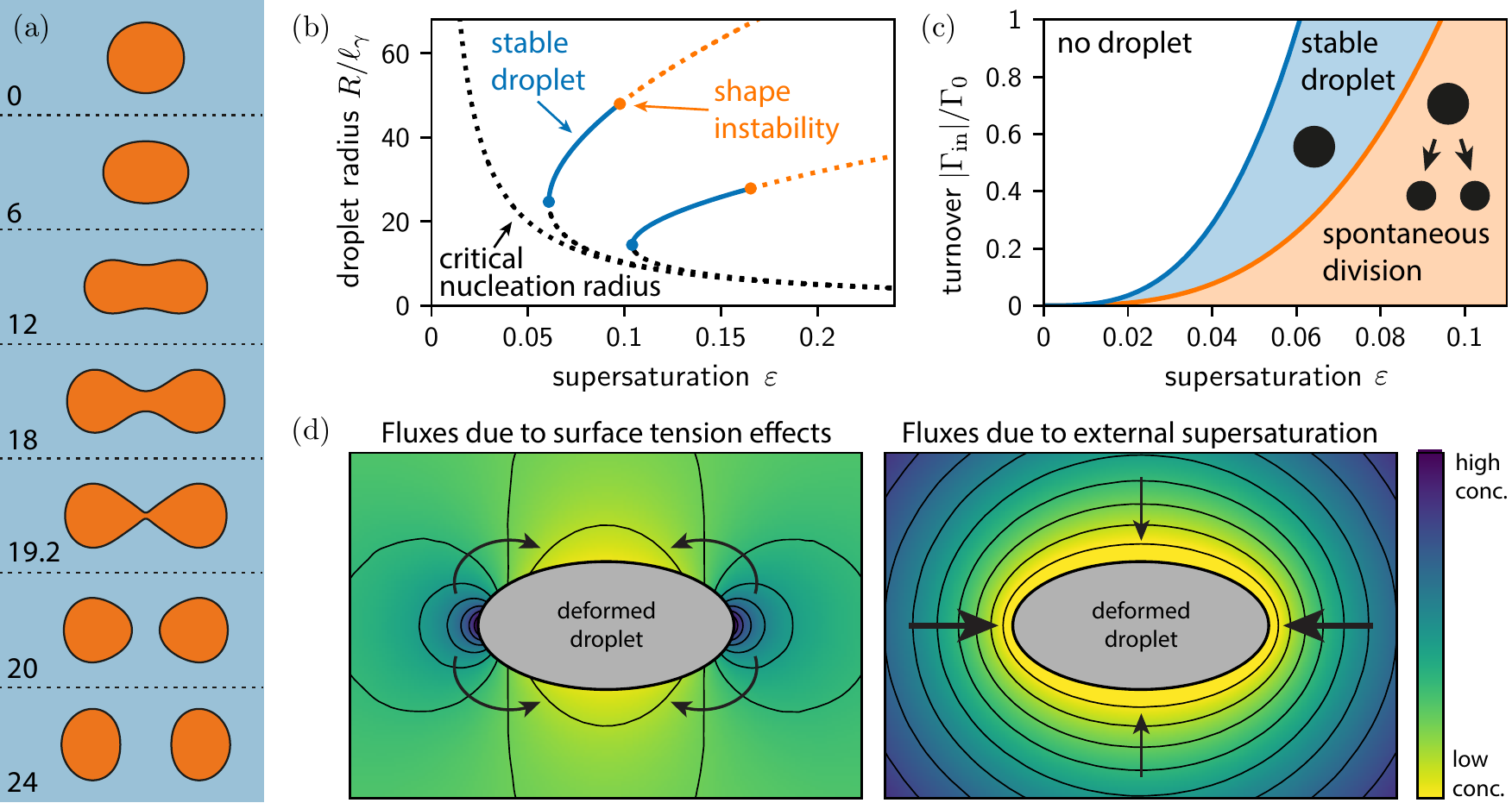}
	\caption{Spontaneous division of externally maintained droplets.
	(a) Sequence of shapes of a dividing droplet at different times as indicated (reproduced from \cite{Zwicker2017})
	(b) Stationary droplet radii~$R$ as a function of the supersaturation~$\supSat$ for three different turnovers ($\sBaseIn/\sBase_0 = 0, 1, 5$; from left to right, with $\Gamma_0 = -10^{-5}\cBaseIn \DOut \ell_\gamma^{-2}$) and $\cBaseIn/\cBaseOut=10$.
	Solid lines indicate stable stationary states, while dotted lines indicate states that are unstable with respect to size (black) or shape (orange). 
	The radii~$R$ were determined from \Eqref{eqn:active_droplet_growth}, assuming that droplets become unstable when $R>R_{\rm div}$ given by \Eqref{eqn:radius_shape_instability}.
	(c) Stability diagram of externally maintained droplets as a function of the supersaturation~$\supSat$ and the turnover~$\sBaseIn$ normalized by $\sBase_0 = \cBaseOut \DOut \ell_\gamma^{-2}$.
	Droplets dissolve and disappear (white region), are stable and attain a spherical shape (blue region), or undergo cycles of growth and division (orange region).
	The lines were determined analogously to (b) using the same parameters.
	(d) Schematics of deformed droplets and the surrounding concentration fields created by surface tension effects (left) and an external supersaturation (right).
	Material is transported from dark to light regions by diffusive fluxes (black arrows) perpendicular to the isocontours (black lines).
	Fluxes due to surface tension transport material from regions of high to low curvature, thus making droplets more circular (left).
	Conversely, the influx driven by an external supersaturation amplifies non-spherical shapes (right).
	The concentration fields have been obtained by solving the stationary diffusion \Eqref{eq:diff_equ_ch2}.
	The boundary condition \eqref{eq:final_GT_relations} at the surface of the deformed droplet accounts for surface tension effects (left) and a constant concentration~$c_\infty$ far away from the droplet represents the external supersaturation (right).
	}
	\label{fig:4_division}
\end{figure*}

So far, we analyzed the growth rate of active droplets assuming they maintain a spherical shape.
We found that the droplet growth dynamics are often determined by a competition between surface tension effects and diffusive fluxes toward the droplet interface.
Both effects depend on the droplet radius, or, more precisely, on the mean curvature of the droplet interface, which is given by $R^{-1}$ for spherical droplets.
In contrast, this mean curvature varies in non-spherical droplets, which suggests that the  competition between the two effects plays a role in the shape dynamics of active droplets.

The dynamics of a non-spherical droplet are described by the same physical principles that we discussed so far.
In particular, the interface velocity in its normal direction, given by \Eqref{eq:interface_vel}, depends on the local net flux of droplet material, which follows from the reaction-diffusion equations in the bulk phases.
Solving these equations numerically reveals that active droplets can show behaviours that are not present in passive droplets~\cite{Zwicker2017}.
\Figref{fig:4_division}a shows that externally maintained droplets can divide spontaneously, which can also happen multiple times~\cite{Zwicker2017}.
A linear stability analysis with respect to the droplet shape reveals that the  shape becomes unstable when the mean droplet radius~$R$ exceeds the critical value~\cite{Zwicker2017}
\begin{align}
	R_{\rm div} &\simeq \frac{11\ell_\gamma}{\supSat}
	\label{eqn:radius_shape_instability}
	\;,
\end{align}
which is an approximation in the limit of large diffusive length scales ($\DiffLenIn, \DiffLenOut \gg R$) and equal diffusivities inside and outside the droplet ($\DIn=\DOut$).
\Figref{fig:4_division}b shows that there are typically three different regimes of externally maintained droplets for a given turnover~$\sBaseIn$ of droplet material inside the droplet.
If the supersaturation~$\supSat$ that is set by the chemical reactions in the dilute phase is too low, droplets cannot exists at all.
For intermediate values of $\supSat$, droplets larger than the critical radius grow to their stationary size as discussed in \secref{sec:active_single}.
For even larger $\supSat$, droplets at the stationary size become unstable ($\steady R > R_{\rm div}$) and growing droplets start dividing before they reach a stationary size.
After division, these droplets can grow further and divide again.
This proliferation continues until the system is depleted of droplet material and the supersaturation~$\supSat$ decreases to a point where the stationary state is stable with respect to shape changes.
Whether externally maintained droplets divide or not depends on the balance between the availability of droplet material (supersaturation $\supSat$) and the turnover inside the droplet (reaction flux~$\sBaseIn$), see \figref{fig:4_division}(c).

The instability of the shape of externally maintained droplets can be qualitatively explained by a competition of surface tension effects with diffusive fluxes~\cite{Golestanian2017}, similar to the multiple droplets discussed in the previous section.
Because of surface tension, interface regions of larger mean curvature exhibit a larger concentration of droplet material right outside the interface, see \Eqref{eq:final_GT_relations}.
This reduces the influx of droplet material and thus attenuates growth at regions of high curvature, stabilising the spherical shape, see \figref{fig:4_division}(d).
Conversely, an externally driven material influx enhances the growth at regions of high curvature where the influx is larger, which is evident from the closer isocontour lines  in \figref{fig:4_division}(d).
This effect generally destabilises the spherical shape and has been described for phase separating systems by Mullins and Sekerka~\cite{Mullins1963,Mullins1964}.
In the presence of internal turnover~$\sBaseIn$ of active droplets, the degradation of 
building blocks inside may balance their influx  leading to a roughly maintained droplet volume.  Such a droplet that is fastest growing at its high curvature edges but of roughly maintained volumed may exhibit  a narrowing of its waistline.  
However, whether the instability can dominate the stabilising surface tension effects and lead to a pinching off and thereby the division into two drops  depends on the parameters, as shown in \figref{fig:4_division}(c).

In internally maintained droplets the diffusive fluxes are reversed.
Consequently, both the surface tension effect and the closer isocontour lines enhance the efflux of material at regions of higher curvature, which thus stabilizes the spherical shape of internally maintained droplets.
We thus expect that internally maintained droplets are more stable  with respect to shape perturbations than passive droplets.


\section{Active emulsions: relevance to biology with versatile applications}

\begin{figure*}
	\includegraphics[width=\textwidth]{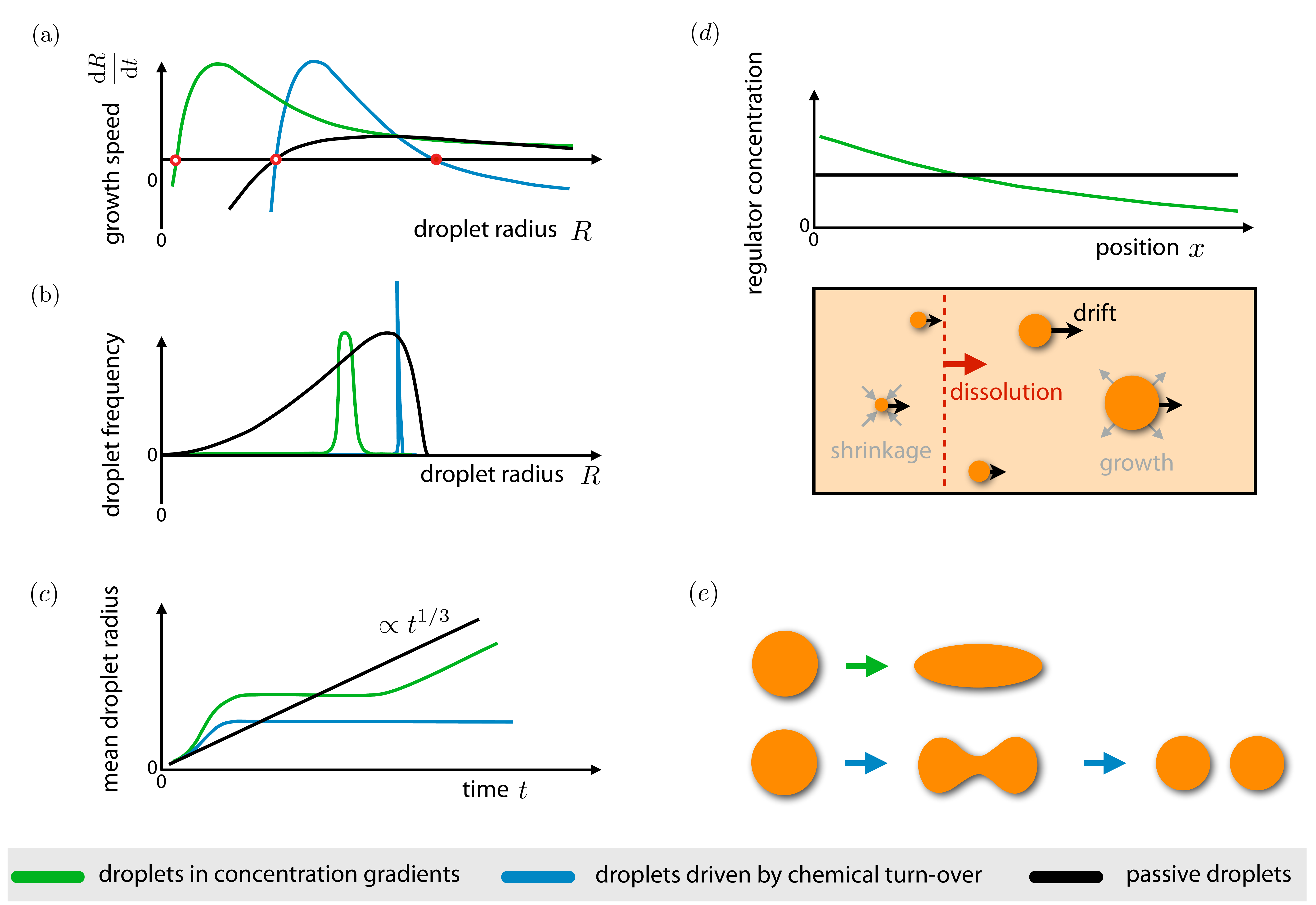}
	\caption{Overview of the novel behaviour and phenomena that occur in {\rm active emulsions}. 
	Curves in green correspond to droplets in a gradient of molecules which affect phase separation 
while blue curves indicate droplets in the presence of non-equilibrium chemical reactions. Results for passive systems are depicted as black lines.
(a) Droplet growth speed and (b) droplet frequency as a function of droplet radius $R$.
The (unstable) critical radius is indicated by an open red circle, while there can be a stable stationary radius in the case of non-equilibrium chemical reactions (shown as bold and red circle). 
(c) Mean droplet radius as a function of time $t$. The radius of passive droplets grows $\propto t^{1/3}$, while active emulsion break this scaling law. 
(d) Positioning of droplets to one boundary of the system (bottom) in the presence of a regulator gradient (top) by droplet drift and position-dependent dissolution and growth. 	(e) 	In the presence of a regulator gradient droplet can deform (top), while in the presence of non-equilibrium chemical reactions droplets can even undergo a shape instability and divide (bottom). 
}
	\label{fig:5_overview}
\end{figure*}

In this review we have discussed the 
physics of phase separation and the dynamics of droplets under conditions that deviate from passive systems.
In particular, we discussed demixing systems in the presence of a concentration gradient of a component that affects phase separation and 
droplets in the presence of chemical reactions that are driven away from thermal equilibrium, see  \figref{fig:5_overview}.
In both systems the dynamics of phase separation is significantly affected.
The systems favour non-equilibrium stationary states and exhibit novel phenomena that are not present in phase separating systems at thermal equilibrium.
We therefore refer to this novel class of physical systems as {\it active emulsions}.

The difference to passive systems becomes apparent already at the level of a single droplet, which can exhibit a qualitatively different growth speed in case of active droplets, see \figref{fig:5_overview}(a). 
For passive systems, the growth speed  as a function of the radius of the droplet has 
 an unstable fixed point. The associated critical radius increases over time and drives the coarsening in emulsions.  
Actively spending energy to create and maintain a regulator gradient that affects phase separation can reduce the critical radius in a position-dependent manner. 
Droplets at one end of the regulator gradient grow faster than droplets at the other end.
Eventually, the larger droplets outcompete the smaller ones, much like in Ostwald ripening, but the positional bias by the gradient effectively positions the surviving droplets toward one end.
Concomitantly, these larger droplets all grow to a similar size, i.e., the droplet size distribution narrows during the positing dynamics (figure~\ref{fig:5_overview}(b)).
This narrowing  is in stark contrast to the universal size distribution of passive droplets undergoing Ostwald ripening. 
As most droplets are positioned to one end, the ripening arrests for a certain time period, thus
breaking the universal growth law in passive systems where droplet size increases proportional to $t^{1/3}$ (figure~\ref{fig:5_overview}(c)).
Besides this position-dependent dissolution and growth process, droplets also drift along the gradient (figure~\ref{fig:5_overview}(d)).
During this drift, droplets can deform (figure~\ref{fig:5_overview}(e)).
In contrast, in passive systems, droplets maintain their spherical shape and do not drift. However, when all droplets are at one end of the gradient, the environment becomes locally homogeneous  and the dynamics slowly returns to classical Ostwald ripening.

If the regulator concentration gradient is created by an inhomogeneous external potential,
a phase separated system can favour novel equilibrium states. 
The condensed phase either sits at the minimum or the maximum of the external potential, i.e., the concentration profile describing the condensed phase is either correlated or 
anti-correlated with the regulator profile.
The system can also undergo a discontinuous phase transition between these two states when the interactions between the molecules are changed.
In the absence of an external potential there is no positional bias and each position of the condensed phase corresponds to the same free energy.
 
Droplet dynamics can also be affected by spending energy to drive chemical reactions that involve the droplet material away from equilibrium.
In this case, single droplets can exhibit a stable stationary state at a finite size, see \figref{fig:5_overview}(a).
Consequently, all droplets larger than the critical size tend toward this stationary radius, leading to a suppression of the Ostwald ripening that would occur in passive systems.
These dynamics leads to an infinitely sharp distribution in the absence of fluctuations (figure~\ref{fig:5_overview}(b)) with a mean droplet size that saturates at this stable radius (figure~\ref{fig:5_overview}(c)). 
Chemical reactions can thus modify the dynamics of phase separation such that the universal broadening is replaced by an evolution to a non-equilibrium stationary state of a monodisperse, active emulsion.
Moreover, in the presence of non-equilibrium chemical reactions, droplets can also undergo shape instabilities triggering the division of droplets, see \figref{fig:5_overview}(e). 
In contrast, passive droplets only exhibit the reverse process of droplet fusion, which is driven by surface tension effects.
In active droplets, this stabilising tendency of surface tension can be overcome by the influx of droplet material sustained by the non-equilibrium chemical reactions.

Demixing systems in the presence of external forces, like a maintained regulator gradient or non-equilibrium chemical reactions, share
some inherent similarities. 
Both systems break the universal coarsening dynamics of Ostwald ripening and modify the stationary state of the system.
While a regulator gradient causes a bias in the position of the single stationary droplet, 
the non-equilibrium chemical reactions can select a state composed of multiple droplets of equal size. 
The breaking of the universal dynamics occurs via additional fluxes that are absent in classical phase separation. These fluxes are generated by the dissipation of energy, either to assemble and maintain a concentration gradient that in turn interacts with the demixing components or to drive chemical reactions of the demixing components away from their equilibrium.

Active fluxes are ubiquitous in living systems to keep them away from thermal equilibrium.
In fact, if such systems were to reach equilibrium, they would be non-living, passive matter.
At the same time, it has been shown that liquid phase separated compartments are important for the spatial organisation inside a large variety of cells  (see  reference~\cite{banani2017biomolecular} and references therein). 
Such compartments can organise biomolecules in space and time to control chemical reactions, which is an essential building block for biological function.
However, how these compartments are controlled by cells is poorly understood.
Studying the physics of phase separation in non-equilibrium environments is thus highly relevant for cell biology~\cite{Cliffs_last_rev}.

Although several novel phenomena arising from active fluxes have been discovered in phase separating systems, many aspects have not been addressed, yet. 
For example, in order to realise an experimental system in which non-equilibrium reactions suppress Ostwald ripening or drive the division of droplets, a better understanding of  the couplings of the chemical pathways to phase separation is necessary. 
To this end, one major challenge is to find a proper choice of chemicals.
Promising candidates are synthetically designed reaction cycles including a fuel-driven reaction from a precursor component toward a building block capable to form liquid-like assemblies~\cite{boekhoven2015transient}.
Specifically, what is the minimal system that can be realised experimentally that shows dividing droplets for example?
If such experimental conditions were found, such systems could be used to set the droplet size in microfluidic devices in the context of chemical engineering~\cite{morel1972numerical} or support the potential role of phase separated liquid-like compartments for the origin of life~\cite{vieregg2016polynucleotides}.

Droplets can also carry information encoded in their chemical composition,
i.e., a specific set of molecules dissolved inside. 
Controlling such droplets with non-equilibrium chemical reactions not only allows  the change of the droplet size but also the information content.
Such a system offers the opportunity to perform aqueous computing at larger-than-molecular scales~\cite{Head1999,head2002aqueous}.
In addition, concentration gradients can be used to position these droplets.
As droplets have reached their target position this chemical information can be released by droplet dissolution, which can be induced by modifying the interactions between the droplet material and the solvent.
Such a liquid system could represent a first building block for aqueous computing, which allows to process chemical information in space. 

Controlled active emulsions could also be used to physically seal or open junctions in microfluidic devices.
The control over the droplet position also enables to modify the physical properties of target surfaces inside devices by wetting.
All these tasks require a solid theoretical understanding of how the formation, position and composition of droplets can be controlled and how physical parameters should be chosen.

In summary, we have discussed a new class of physical systems which we refer to as \textit{active emulsions}.
These emulsions are relevant to cell biology.
They may allow to develop novel applications in the field of chemical engineering or aqueous computing.
Maybe, active emulsions could be relevant in the research of how life could have emerged from an inanimate mixture composed of set of simple chemically active molecules.
However, the class of active emulsions also challenge our theoretical understanding of spatially heterogeneous systems driven far away from thermal equilibrium and can be used to refine existing theoretical concepts. 
In particular, \textit{active emulsions} are characterised by non-equilibrium fluxes that maintain these system away from thermal equilibrium.
The physics of phase separation in the presence of non-equilibrium 
chemical reactions pose several theoretical challenges and questions,
 such as the role of fluctuations, the couplings of diffusive and chemical fluxes, and what are the minimally required ingredients necessary for the phenomena discussed in this review.

\ack
We would like to thank 
  Omar Adame Arana,
   Giacomo Bartolucci,  
       Jonathan Bauermann,
    Siheng Chen,
Erwin Frey,
 Elisabeth Fischer-Friedrich, 
  David Fronk,
  Lars Hubatsch,
  Jacqueline Janssen,
    Jan Kirschbaum, 
  Samuel Kr\"uger,
    Matthias Merkel, 
     Nirbhay Patil, 
      Payam Payamyar, 
    Orit Peleg,
 Wolfram P\"onisch,  
 Rabea Seyboldt, 
 and
 Jean David Wurtz, 
 for helpful comments and discussions on our review.
We thank Samuel Kr\"uger for proving the data for Fig.~\ref{fig:Fig_3_profiles} and Fig.~\ref{fig:Fig_3_phasetransition}.
Finally, we would  like to acknowledge 
Simon Alberti,
Martin Z. Bazant, 
Edgar Boczek, 
Clifford P. Brangwynne, 
Andres Diaz, 
Titus Franzmann,
Anatol Fritsch,
Alf Honigmann,
Anthony A. Hyman, 
Louise Jawerth, 
Jose Alberto Morin Lantero,
Mark Leaver,
Avinash Patel,
Shambaditya Saha, 
Jeffrey B. Woodruff, 
 and 
Oliver Wueseke 
for stimulating scientific discussions about  active emulsions and droplet-like compartment inside cells.
C.A.W. and D.Z. also thank the German Research Foundation (DFG) for financial support.

\appendix

\section{Stability of the interfacial profile}
\label{sec:appendix_stability}

The stability of the inhomogeneous solution~$c_{\rm I}(x)$ given in \Eqref{eq:tanh} can be assessed by considering the change in free energy $\Delta F [c_{\rm I}, \epsilon]$ due to an infinitesimal perturbation~$\epsilon(x)$.
Using \Eqref{eq:triF}, we have
\begin{align}
	\Delta F [c_{\rm I}, \epsilon] &= \int \diff^3 \vect{r} \left[
		\frac{b\epsilon^2}{2}\left( 3 \tanh^2\!\left(\frac{x}{w}\right) - 1 \right)
		+\frac{\kappa}{2} \left(\partial_x \epsilon \right)^2
	\right]
	\label{eqn:free_energy_second_variation}
	\;.
\end{align}
The reference state $c_{\rm I}(x)$ is stable if all perturbations increase the free energy, \ie if $\Delta F[c_{\rm I}, \epsilon] > 0$ for all permissible functions $\epsilon(x)$.
Note that without chemical reactions the mass of the individual components is conserved, which implies that only those $\epsilon(x)$ are permissible that obey the constraint $\int \epsilon \, \diff x=0$.

To see whether all perturbations lead to positive~$\Delta F$, we determine the perturbation~$\epsilon_*(x)$ with the minimal~$\Delta F$.
If this value is positive, all other perturbations also increase the free energy and the base state~$c_{\rm I}(x)$ is stable.
A necessary condition for $\epsilon_*(x)$ is that it satisfies the Euler-Lagrange equations corresponding to \Eqref{eqn:free_energy_second_variation}.
Defining the linear operator $\mathcal A(\epsilon) = \delta \Delta F/\delta \epsilon$,
\begin{align}
	\mathcal A(\epsilon) 
	 = b\epsilon \left[ 3 \tanh^2\!\left(\frac{x}{w}\right) - 1 \right] - \kappa \pp_x^2 \epsilon
	\;,
	\label{eqn:appendixA_operator}
\end{align}
the Euler-Lagrange equations can be expressed as
\begin{align}
	\mathcal A(\epsilon_*) = \tilde\lambda
	\;.
	\label{eqn:appendixA_euler_lagrange}
\end{align}
Here, $\tilde\lambda$ is a Lagrange multiplier that ensures mass conservation in the case without chemical reactions, while $\tilde\lambda=0$ in the case with chemical reactions.
Using partial integration on \Eqref{eqn:free_energy_second_variation}, we also have
\begin{align}
	\Delta F [c_{\rm I}, \epsilon] &= \frac12 \int \diff^3\vect{r} \, \epsilon \, \mathcal A(\epsilon)
	\label{eqn:appendixA_energy_operator}
	\;,
\end{align}
where we neglected the boundary terms assuming that the perturbation vanishes at the boundary or the system exhibits periodic boundary conditions.
To evaluate~$\Delta F$, it is convenient to express $\epsilon_*(x)$ in terms of the eigenfunctions~$\epsilon_n(x)$ of the operator~$\mathcal A$.
The associated eigenvalue problem $\zeta_n \epsilon_n = \mathcal A(\epsilon_n)$ has already been considered in the context of Schr\"{o}dinger's equation with a potential similar to the first term in \Eqref{eqn:appendixA_operator}~\cite{landau_quantum}.
In particular, it has been shown that the discrete part of the spectrum consists of only two eigenvalues, $\zeta_0=0$ and $\zeta_1 = \frac32b$, while the continuous spectrum obeys $\zeta \ge 2b$.
In the following, the sum symbol will denote both summation over the discrete and integration over the continuous part of the spectrum. 
The solution~$\epsilon_*(x)$ can then be expressed as
\begin{align}
	\epsilon_*(x) &= \sum_{n=0}^{\infty} a_n \epsilon_n(x)
	\label{eqn:eigenfunction_decomposition}
	\;,
\end{align}
where $a_n$ are the corresponding series coefficients, which have to obey $\sum_{n=0}^{\infty} a_n \zeta_n = \tilde\lambda$.
Using this in \Eqref{eqn:appendixA_energy_operator}, we find
\begin{align}
	\Delta F[c_{\rm I}, \epsilon_*] &=
	\frac{A}{2}   \sum_{n=1}^\infty \zeta_n \int \dd x \, a_n^2 \bigl\{\epsilon_n(x) \bigr\}^2
	\ge 0
	\label{eqn:appendixA_energy_final}
	\;,
\end{align}
where $A$ is the cross-sectional area $A=\int \diff y \diff z$.
Equation~\eqref{eqn:appendixA_energy_final} implies $\Delta F > 0$ if any $a_n \neq 0$ for $n\ge1$, so the base state is stable with respect to these perturbations.
In particular, the state $c_{\rm I}$  would only be unstable if there are perturbations with $a_0 \neq 0$ and $a_n = 0$ for $n\ge1$.
Note that the term for $n=0$ does not appear in \Eqref{eqn:appendixA_energy_final} since the eigenvalue~$\zeta_0$ vanishes.
The associated eigenfunction is $\epsilon_0(x) = \partial_x c_{\rm I}(x)$, which does not conserve the mass of the individual components since $\int\epsilon_0(x)\diff x = 2(b/a)^{1/2}$.
Consequently, this mode is forbidden if chemical reactions are absent, so that all perturbations increase the free energy in this case.
Conversely, with chemical reactions, $a_0\neq0$ is allowed and this mode is marginal since it does not change the free energy, $\Delta F(c_{\rm I}, \epsilon_0)=0$.

Taken together, we showed that the interfacial profile~$c_{\rm I}(x)$ given in \Eqref{eq:tanh} is stable in the case without chemical reactions.
Chemical reactions introduce a marginal mode, which corresponds to a translation of the interface.
However, for finite systems, the boundary conditions \eqref{eq:interface_bcs}, and thus the interface profile~$c_{\rm I}(x)$, are only approximate and all inhomogeneous states might be unstable in very small systems~\cite{krapivsky_b10}.

\section{Entropy production of a system with phase separation and chemical reactions}
\label{sec:appendix_entropy_production}

To derive dynamical equations for a binary fluid exhibiting phase separation and chemical reactions, we here consider the associated entropy production.
For a closed, isothermal system of constant volume, the entropy production is related to the change of free energy, $\text{d}S/\text{d}t=-T^{-1} \text{d}F/\text{d}t$.
For simplicity, we here consider the free energy $F[c_A,c_B]=\int \diff^3 r \,  f(c_A, c_B)$, neglecting the contributions proportional to the gradients of the concentration fields.
Hence,
\begin{align}
	\label{eqn:free_energy_change}
	\frac{\text{d}F}{\text{d}t} &= \int \diff^3 r \left( \mu_A \partial_t c_A + \mu_B \partial_t c_B \right) \\
	\notag
	&=  - \int \diff^3 r \left[
		\mu_A \left( \nabla \cdot \vect{j}_A - s \right) +
		\mu_B \left( \nabla \cdot \vect{j}_B + s \right)
	\right]\\
	\notag
	&=  \int \diff^3 r \left[
		\vect{j}_A  \cdot \nabla \mu_A 
		+ \vect{j}_B \cdot \nabla \mu_B 
		+ s \left( \mu_A -\mu_B \right)\right] \\
	\notag
	& \quad + \text{boundary terms}
	\;,
\end{align}
where we have used $\mu_A=\partial f/\partial c_A$, $\mu_B=\partial f/\partial c_B$,
and the conservations laws $\partial_t c_A=- \nabla \cdot \vect{j}_A + s$ and $\partial_t c_B=- \nabla \cdot \vect{j}_B - s$.
Note that the boundary terms originating from partial integration can be neglected when appropriate boundary conditions are applied or an infinite system is considered. 
  
For each component~$i$, the flux~$\vect j_i$ can be split into a convective part with velocity $\vect{v}$ and an exchange current $\vect{j}$.
 As discussed in section~\ref{sec:dynamical_equation},
 incompressibility and equal molecular volumes then imply $\vect{j}_A=\vect{v} c_A +\vect{j}$ and $\vect{j}_B=\vect{v} c_B -\vect{j}$.
 Moreover,  changes of the intensive thermodynamic quantities  are coupled, implying the Gibbs-Duhem relationship $c_A \text{d}\mu_A+c_B \text{d}\mu_B=\text{d}p$, where $p$ is the pressure.
Hence,
\begin{align}
	\frac{\text{d}F}{\text{d}t} &= \int \diff^3 r \left(
		\vect{j}\cdot \nabla \bar{\mu} +
		p \nabla \cdot \vect{v} + 
		s \,  \bar{\mu}
	\right) 	
	\;,
\end{align}
where $\bar{\mu}=\mu_A-\mu_B$ is the exchange chemical potential.
In the case of equal molecular volumes of $A$ and $B$, incompressibility implies $\nabla \cdot \vect{v}=0$~\cite{julicher2009generic}, leading to
\begin{align}
	\frac{\text{d}S}{\text{d}t} &= \frac1T \int \diff^3 r \left( -\vect{j}\cdot \nabla \bar{\mu}  - s \, \bar{\mu} \right) 	
	\;.
	\label{eqn:entropy_change}
\end{align}
This expression reveals that the thermodynamic fluxes $\vect j$ and $s$ are coupled to the thermodynamic forces $\nabla \bar\mu$ and $\bar\mu$, respectively.
Expressing the fluxes as linear functions of their forces, we obtain $\vect j= - \mob \nabla \bar{\mu}$ and $s=- \mobR \bar{\mu}$, where the Onsager coefficients $\mob$ and $\mobR$ must be positive to obey the second law of thermodynamics ($\diff S/\diff t \ge 0$).
The entropy production only vanishes at equilibrium, where the equilibrium conditions $ \nabla \bar{\mu}=0$ and $\bar{\mu}=0$ are obeyed and the fluxes vanish.

\section{Thermodynamic constraints on chemical reaction rates}
\label{sec:appendix_detailled_balance}
 
\newcommand{\Peq}{P^\mathrm{eq}}


We consider a binary system consisting of particles $A$ and $B$ with a simple conversion reaction $B \rightleftharpoons A$.
Assuming local equilibrium, the system is described by the concentrations~$c_A$ and $c_B$ of the $A$ and $B$-particles.
Since the particles occupy all the space, both concentrations are connected via the relationship  $c_A = \nu^{-1} - c_B$, where $\nu$ is the molecular volume of both species.
The concentration of $A$ particles obeys
\begin{align}
	\partial_t c_A &= \sF - \sB
	\label{eqn:appendixB_rate_law}
	\;,
\end{align}
where $\sF$ and $\sB$ denote the respective forward and backward reaction flux.
To derive the condition on the ratio of these fluxes given in \Eqref{eqn:foward_backward_ratio}, we here analyze the associated lattice model
introduced in section~\ref{sec:stat_mech}.

We consider a system of $M$ lattice sites and focus on the reactions occurring at the arbitrary lattice site $n$. 
The probabilities to find an $A$ or $B$ particle at this site are given by $P({\vect{\sigma}}_n,t)$ 
or  $P(\bar{\vect{\sigma}}_n,t)$, respectively. 
Here, 
the configuration of all particles on the lattice where lattice site $n$ is occupied by an $A$ particle ($\sigma_n=1$) is denoted as 
${\vect{\sigma}}_n=\sigma_1,..., \sigma_n=1,...,\sigma_M$.
Conversely, $\bar{\vect{\sigma}}_n=\sigma_1,..., \sigma_n=0,..., \sigma_M$, is a configuration where the $n$-th site is occupied by $B$.
These probability functions are related to the volume fraction~$\phi=\nu c_A$ of $A$ particles and concentration fields by 
\begin{align}
	\phi &
	=  \sum_{\Omega_n}  P({\vect{\sigma}}_n,t)
	= 1 - \sum_{\bar{\Omega}_n}  P(\bar{\vect{\sigma}}_n,t)
	\label{eqn:AppendixB_concA}
	\;,
\end{align}
where $\Omega_n$ or $\bar{\Omega}_n$ denote the set of all possible configurations 
with an $A$ or $B$ particle at lattice site $n$, respectively.
The time evolution of the probabilities is captured by the master equation
\begin{align}
\label{eq_master}
\partial_t P({\vect{\sigma}}_n,t) &= -\partial_t P(\bar{\vect{\sigma}}_n,t)
\\&=
\notag
k_\rightarrow^n(\bar{\vect{\sigma}}_n) P(\bar{\vect{\sigma}}_n,t) 
-
k_\leftarrow^n({\vect{\sigma}}_n) P({\vect{\sigma}}_n,t) 
	\; ,
\end{align}
where $k_\rightarrow^n(\bar{\vect{\sigma}}_n) $
and $k_\leftarrow^n({\vect{\sigma}}_n)$ denote the
forward and backward rates, which generally depend on the configuration.
The equations discussed so far also hold when the reaction is not in equilibrium.

When the chemical reactions are equilibrated, the probability distribution is time independent, leading to 
the condition of detailed balance:
\begin{align}
\label{eq_detailled_balance}
	k_\rightarrow^n(\bar{\vect{\sigma}}_n) \Peq(\bar{\vect{\sigma}}_n)
	=
	 k_\leftarrow^n({\vect{\sigma}}_n) \Peq({\vect{\sigma}}_n)
	\;.
\end{align}
 Here,  the equilibrium distribution to find a specific configuration, $\sigma_1,...,\sigma_M$, is given as 
\begin{equation}
	\label{eq:P_specific}
	\Peq(\sigma_1,...,\sigma_M)= \frac{1}{Z}  \exp\left(-\frac{H(\sigma_1,...,\sigma_M)}{\kb T}\right)
	\;,
\end{equation}
with the Hamiltonian $H(\sigma_1,...,\sigma_M)$ defined in \Eqref{eq:H} and the partition function $Z$ given by \Eqref{eq:Z}.
Using \Eqref{eq_detailled_balance}, we have
\begin{multline}
\label{eq_detailled_balance_micro_rate}
	\frac{k_\rightarrow^n(\bar{\vect{\sigma}}_n)}
		{ k_\leftarrow^n({\vect{\sigma}}_n) } 
= \frac{ \Peq({\vect{\sigma}}_n)}
	{\Peq(\bar{\vect{\sigma}}_n) }
\\
= \exp\left(-\frac{H({\vect{\sigma}}_n)-H(\bar{\vect{\sigma}}_n)}{\kb T} \right)
	\;.
\end{multline}
For the considered case of a single reaction step and for short ranged interactions, 
 the forward and backward rates  do not depend on the full configuration~$\set\sigma$. 
 Instead, the rates solely 
depend on the particle configurations in the vicinity of  site $n$, which is determined by the characteristic length scale of the interaction.
 In particular, using a mean field approximation, 
 the rates solely depend on the local volume fraction
  $\phi$. 
In this case, the energy difference appearing in \Eqref{eq_detailled_balance_micro_rate} simplifies to
\begin{equation}
\label{eq_ham_diff}
	H({\vect{\sigma}}_n)-H(\bar{\vect{\sigma}}_n)  
	=\frac1M \difffrac{E(\phi)}{\phi} 
	\;,
\end{equation}
where $E(\phi)$ is the mean field energy given by \Eqref{eq_mf_energy}, such that
\begin{equation}
\label{eq_ham_diff2}
	\difffrac{E(\phi)}{\phi} = z M \left[ 
 e_{AA} \phi + e_{AB} (1-2\phi) -  e_{BB} (1-\phi) \right] \, .
 \end{equation} 
Combining \Eqsref{eq_detailled_balance_micro_rate} and  \eqref{eq_ham_diff}, we find that the  rates solely depend on the local volume fraction $\phi$, \ie
 $k_\rightarrow^n(\bar{\vect{\sigma}}_n)=k_\rightarrow(\phi)$ and
 $k_\leftarrow^n({\vect{\sigma}}_n)=k_\leftarrow(\phi)$.




We now return to the case where chemical reactions are not equilibrated.
Taking the time derivative of \Eqref{eqn:AppendixB_concA} and using \Eqref{eq_master}, we write the time evolution of the composition as
\begin{align}
\label{eq_local_volume_fraction}
	\partial_t \phi
	&=\sum_{\Omega_n}  \Big[
	 k_\rightarrow^n(\phi) P(\bar{\vect{\sigma}}_n,t) 
 -	k_\leftarrow^n(\phi) P({\vect{\sigma}}_n,t) \Big]
\notag\\
& = 
  k_\rightarrow(\phi) \left( 1 - \phi \right) -
	k_\leftarrow(\phi) \phi
	\; ,
\end{align}
where we have used that the forward and backward rate solely depend on $\phi$ within the mean field approximation.
Comparing with \Eqref{eqn:appendixB_rate_law}, we can identify the forward and backward reaction flux
as $\sF=k_\rightarrow(\phi) \left( \nu^{-1} - c_A \right)$ and $\sB=k_\leftarrow(\phi) c_A$, since $\phi = \nu c_A$.
Hence, the ratio of the reaction fluxes reads
\begin{align}
\label{eq_reaction_flux}
	\frac{\sF}{\sB}
	= \frac{ k_\rightarrow(\phi)}{k_\leftarrow(\phi)}   \frac{\left(1 - \phi \right)}{\phi}
	= \exp{\left(-\frac{ \bar{\mu}}{\kb T} \right) } \, .
\end{align}
where we defined
\begin{align}
	\bar\mu &= z \left[  e_{AA} \phi +e_{AB} (1-2\phi) - e_{BB}  (1-\phi)  \right] 
	\notag
	\\
& \quad + \kb T \ln{\left( \frac{\phi}{1-\phi} \right)}
	\;,
	\label{eqn:appendixB_chem_pot}
\end{align}
which can be identified with the exchange chemical potential $\bar{\mu} = \mu_A - \mu_B  = \nu \frac{\partial f}{\partial \phi}$ associated with the free energy density~$f(\phi)$ given in \Eqref{eq:free_en}.
Note that \Eqref{eqn:appendixB_chem_pot} does not contain any gradient term since we here focused on the simple case of homogeneous mean field.
However, \Eqref{eq_reaction_flux} also holds for more general situations and is referred to as detailed balance of the rates~\cite{Julicher1997}.

\section*{References}


\end{document}